\documentclass{aa}

\usepackage[varg]{txfonts}
\usepackage{sidecap}
\pdfoutput=1
\usepackage{etoolbox}
\makeatletter
\usepackage{adjustbox}
\usepackage{hyperref}
\pdfoutput=1
\usepackage{amssymb}
\usepackage{graphicx}
\usepackage{pdflscape}
\usepackage{tikz}

\usepackage{natbib}
\usepackage{xcolor}

\pdfoutput=1
\begin{document}

   \title{Spatially resolved star-formation relations of dense molecular gas in NGC 1068}


   \author{M. S\'anchez-Garc\'ia\inst{1,2}
	\and
	  S. Garc\'ia-Burillo\inst{2}
	\and
           M. Pereira-Santaella\inst{1}
         \and  
           L. Colina\inst{1}
         \and  
           A. Usero\inst{2}
         \and   
           M. Querejeta\inst{2}
        \and   
           A. Alonso-Herrero\inst{3} 
           \and
	A. Fuente\inst{2}
          }

   \institute{\inst{1}Centro de Astrobiolog\'{\i}a (CSIC/INTA), Ctra de Torrej\'on a Ajalvir, km 4, 28850 Torrej\'on de Ardoz, Madrid, Spain \\
              \email{mariasg@cab.inta-csic.es}\\
             \inst{2}Observatorio Astron\'omico Nacional (OAN-IGN)-Observatorio de Madrid, Alfonso XII, 3, 28014 Madrid, Spain \\
             \inst{3}Centro de Astrobiolog\'{\i}a (CAB, CSIC-INTA), ESAC Campus, E-28692 Villanueva de la Ca\~nada, Madrid, Spain \\
             }

   \date{}

 
  \abstract
{The current understanding of star formation (SF) contemplates that the regulation of this phenomenon in galaxy disks reflects a complex balance between processes that operate in molecular gas on local cloud-scales but  also on global disk-scales. }
{We analyse the influence of the dynamical environment on the SF relations of the dense molecular gas in the starburst (SB) ring of the Seyfert 2 galaxy NGC~1068.}
{We used ALMA to image the emission of the 1--0 transitions of HCN and HCO$^{+}$, which 
trace dense molecular gas 
 in the $r\sim1.3$~kpc SB ring of NGC 1068, with a 
resolution of 56~pc. We also used ancillary data of CO(1--0), as well as CO(3--2) and its underlying continuum 
emission at the resolutions of $\sim100$~pc and $\sim40$~pc, respectively. These observations allow us 
to probe a wide range of molecular gas densities ($n_{\rm H_2}\sim10^{3-5}$cm$^{-3}$). The SF rate (SFR)  in the SB ring of 
NGC 1068 is derived from Pa$\alpha$ line emission imaged by HST/NICMOS. We analysed how different 
formulations of SF relations change depending on the adopted aperture sizes and on the 
choice of molecular gas tracer.}
{The scatter in the Kennicutt-Schmidt  relation, linking the SFR density ($\Sigma_{\rm SFR}$) with the (dense) molecular gas surface density ($\Sigma_{{\rm dense}}$),  is about a factor of two to three lower for the HCN and HCO$^{+}$ lines compared to that derived from CO(1--0) for a 
common aperture. Correlations lose statistical significance below a critical spatial scale $\approx$ 300-400~pc for all gas tracers. The efficiency of SF of the dense molecular gas, defined as SFE$_{\rm 
dense}\equiv\Sigma_{\rm SFR} / \Sigma_{{\rm dense}}$, shows a scattered distribution as a function of the HCN luminosity ($L'$(HCN)) around a mean value of $\simeq0.01$Myr$^{-1}$. An alternative prescription for SF relations, which includes the 
dependence of SFE$_{\rm dense}$ on the 
combination of $\Sigma_{\rm dense}$ and the velocity dispersion ($\sigma$), resolves the degeneracy associated with the 
 SFE$_{\rm dense}$-$L'$(\rm HCN) plot. The SFE$_{\rm dense}$ values show a positive trend with the boundedness 
of the gas, measured by the parameter $b$ $\equiv$ $\Sigma_{\rm dense}$/$\sigma^{2}$. 
We identify two branches in the SFE$_{\rm dense}$--$b$ plot 
that correspond to two dynamical environments within the SB ring, which are defined by their proximity to 
the region where the spiral structure is connected to the  stellar bar. This region corresponds to the crossing of two overlapping density wave resonances, where an increased rate of cloud-cloud collisions would favour an enhanced compression of molecular gas.}
{These results suggest that galactic dynamics plays a major role in the efficiency of the gas conversion into stars. Our work adds 
supporting evidence that density-threshold star formation models, which argue that the SFE$_{\rm dense}$ should be roughly 
constant, fail to account for spatially resolved SF relations of dense gas in the SB ring of NGC~1068.}
   \maketitle
\section{Introduction}
The study of the processes that power star formation (SF) in galaxies is paramount to understanding how galaxies form and evolve.  If we assume that the gas scale-height is constant, the power law relating the gas volume density  and the SF rate (SFR) volume density, originally proposed by \citet{Schmidt59}, finds its equivalent  in terms of the corresponding surface densities of the SFR
($\Sigma_{\rm SFR}$) and the  gas  ($\Sigma_{\rm gas}$) in the expression:

 	\begin{equation}
	\Sigma_{\rm SFR}=A\Sigma^{N}_{\rm gas}.
	\end{equation}\label{Eq1}
	
In the above equation $A$ is a normalization constant and $N$ is the power-law index.  Under the hypothesis that  the  relevant  time  scale  for  SF is  the local free-fall  time for the gas, the theoretically predicted value for $N$ is 1.5.
This relation is known as the Kennicutt-Schmidt (KS) law \citep{Schmidt59, Kennicutt98}.  In addition to the empirical power law  relation between $\Sigma_{\rm gas}$ and $\Sigma_{\rm SFR}$, another key parameter in SF studies is the star formation efficiency (SFE) defined as: 
	\begin{equation}
	{\rm SFE}=\Sigma_{\rm SFR}/\Sigma_{\rm gas},
	\end{equation}\label{Eq2}
which represents the inverse of the depletion time ($T_{\rm dep}$) of the gas that is consumed by SF.

Different  proxies  for  $\Sigma_{\rm SFR}$ and $\Sigma_{\rm gas}$ have been chosen in observations carried out during the last decades on different galaxy populations.
First, from spatially unresolved galaxy-scale global measurements, which used CO and HI as neutral gas tracers in different galaxy samples, observers found a single law with a range of indexes $\simeq1.2-1.7$ \citep{Kennicutt98,Yao2003, bouche2007, Daddi10, Genzel10, Liu15, Kennicutt21}.  However, there is mounting evidence that global KS laws show signs of bimodality or multimodality, which tend to separate the branches of normal galaxies  from that of more extreme merger systems \citep{Daddi10, Genzel10, Liu15, Kennicutt21}. 

 Furthermore, high resolution imaging of neutral gas in galaxies has allowed the analysis of the KS relation at kpc and sub-kpc scales  in a growing number of galaxies \citep[e.g.,][]{Kennicutt2007, Bigiel2008, Leroy2008, Blanc2009, Casasola15, Leroy2017}.  In particular, \citet{Bigiel2008} used CO(2--1) as a tracer of molecular gas in a sample of  18 nearby star forming galaxies and found a single linear KS relation ($N\simeq1$), which holds  for  gas surface densities $\Sigma_{\rm gas}$ $\geq$ 10 M$_{\odot}$pc$^{-2}$. This dividing line identifies the transition from atomic to molecular gas. \cite{Leroy2008} derived a radial dependence in $N$ with a decreasing SFE at larger radius within individual galaxies.  Moreover, \cite{Casasola15}  derived KS relations in the nuclear regions of four low luminosity AGN using interferometric CO images on spatial scales between 20 to 200~pc. The KS relations were found to be sublinear, but also superlinear, with a wide range of slopes $\sim[0.5-1.3]$.  \citet{Leroy2017} studied the local dynamical state of molecular gas in M51 and found that the gas with stronger self-gravity forms stars at a higher rate. The variability in resolved KS relations also suggests higher SFE in lower mass low metallicity galaxies \citep[e.g.,][]{Schruba11, Leroy13} and in late Hubble types \citep[e.g.,][]{Colombo18, Ellison21}.

The high resolution CO observations of M~33 published by \citet{Onodera2010} found that at spatial scales similar to those of giant molecular clouds (GMCs) ($\sim50-100$~pc)  the correlation between $\Sigma_{\rm SFR}$ and $\Sigma_{\rm gas}$ is lost. In qualitative agreement with this picture, \citet{Schruba2010} observed a breakdown of the SF relation at scales $\leq300$~pc in M~33. Moreover, the recent work of \citet{Williams2018} found significant correlations  in M~33  down  to  scales  of  100~pc, while  the  measured Schmidt  index  shows  a  marked dependence on the spatial scale. The breakdown of the KS relation observed below a "critical" spatial scale can be attributed to the need of averaging over sufficiently large scales in order to have a statistical sampling of star forming sites at different evolutionary stages \citep[e.g.,][]{Kruijssen14b}.

The most recent  studies of the SF relation reported above, which use low-J CO lines (sometimes in combination with HI data) as tracers of the bulk of neutral gas, cast doubts on the existence of a "universal" or "unimodal" KS relation at all spatial scales.
Dense molecular gas, namely gas with volume densities typically exceeding $10^{4-5}$ cm$^{-3}$, is believed to condense into GMCs and be therefore more directly related to recent and massive star formation \citep{Lada2010, Andre2010}. In particular, observations of HCN(1--0) and HCO$^{+}$(1--0)  lines, which have associated critical densities of n$_{\rm crit}$[HCN(1--0)] $\sim$ 1.7 $\times$ 10$^{5}$ cm$^{-3}$ and n$_{\rm crit}$[HCO$^{+}$(1--0)] $\sim$ 2.9 $\times$ 10$^{4}$ cm$^{-3}$ \citep{Shirley2015}, are well suited to fairly trace the dense molecular gas mass ($M_{\rm dense}$) in galaxies. Dense gas probes have been used to study  "galaxy-scale" KS relations \citep[e.g.,][]{GaoSolomon2004a, GaoSolomon2004b, SolomonVan2005, Gracia-carpio2006, Gracia-carpio2008, Burillo2012, Liu15}. 
As expected, the SF relations derived for the dense gas  
show a less scattered linear correlation (i.e., with $N\simeq1$), compared to the global KS laws obtained from low-J CO lines. However, the residual but nevertheless significant scatter present in  the  SFR-$M_{\rm dense}$  plane has been interpreted as indicative of different average physical properties of the dense gas in normal SF galaxies and mergers \citep{Gracia-carpio2008, Burillo2012}. 

High resolution (kpc and sub-kpc) single-galaxy studies of the SF relations of the dense molecular gas have started to resolve the 
degeneracy in the SFR-$M_{\rm dense}$ parameter space  by showing how scaling laws change for different dynamical 
environments within a galaxy, including the Milky Way \citep{Lon13, Kru14, Murphy2015, Usero2015, Bigiel2015, Bigiel2016, Chen2017, Viaene2018, Querejeta2019, Jim19, Beslic21}. 
In particular, \cite{Usero2015} observed HCN(1--0) and CO(1--0) lines at several positions in the disks of 29 SF galaxies and found 
that SFE$_{\rm dense}$, derived from the IR/HCN ratio, is $\sim6-8$ times lower near galaxy centres than in the outer regions of 
the disks. Furthermore, \cite{Querejeta2019} found that SFE$_{\rm dense}$ values measured on $\sim100$~pc scales from radio 
continuum-to-HCN ratios  vary by more than 1~dex among the different dynamical environments of the disk of M~51.
More recently, \cite{Beslic21} found significant differences in the $\sim100$~pc-scale SFE$_{\rm dense}$ values between the 
galaxy centre, bar, and bar-end regions of the nearby barred galaxy NGC~3627. These  results contradict models that rely on a 
universal gas density threshold for star formation \citep[e.g.,][]{GaoSolomon2004b, Wu2005, Lada2010, Lada2012, Evans2014} 
and suggest instead that the dynamical environment of the dense molecular gas in GMCs can determine  its efficiency at forming 
stars. This is supported by models of turbulent SF \citep[e.g.][]{Krumholz2005, Krumholz2007,  Hennebelle2012, Meidt2013, Federrath2015, Meidt2016, Meidt2018, Meidt2020}.

In this paper we study the spatially resolved SF relations of the dense molecular gas in the starburst (SB) ring of the nearby \citep[$D\simeq14$~Mpc;][]{BH1997} Seyfert 2 barred galaxy NGC~1068, a target considered as an archetype of the composite starburst and active 
galactic nucleus (AGN) classification. Previous interferometer images have resolved the large-scale distribution of molecular gas in the disk of 
the galaxy \citep{Hel95, Schinnerer2000, Krips2011, Tsai2012, Burillo2014, Tak14, Viti2014, Burillo2017, Burillo2019, Sco20}. Molecular line and dust continuum emissions are 
detected from a $r\sim200$~pc circumnuclear disk (CND), from the  
$\sim2.6$~kpc-diameter stellar bar region, and from a 
ring, where molecular gas is accumulating and feeding a  SB episode. The SB ring is formed by a tightly wound 
two-arm spiral structure that starts from the ends of the stellar bar and unfolds in the disk over $\sim180^{\circ}$ in azimuth forming 
a pseudo-ring at $r\sim18\arcsec$(1.3~kpc). 


We used new images of the distribution of dense molecular gas ($n_{\rm H_{2}}$ $\geq$ 10$^{4-5}$cm$^{-3}$) obtained by the Atacama Large Millimeter Array (ALMA) in the 1--0 transitions of HCN and HCO$^{+}$ with a native
resolution of $\sim0\farcs8$ (56~pc). This spatial resolution is comparable to the typical size of GMCs. We also use high resolution ($\sim1\farcs5$=100~pc) CO (1--0) images of the galaxy obtained by the IRAM  array \citep{Schinnerer2000}, as well as available CO(3--2) and dust continuum images obtained by ALMA at a spatial resolution of $\sim0\farcs6$ (40~pc). The ensemble of these observations allows us 
to probe a wide range of molecular gas densities ($n_{\rm H_2}\simeq10^{3-5}$cm$^{-3}$) in the SB ring. To probe SF we use Pa$\alpha$ line images obtained by the {\it Hubble} Space Telescope (HST). We analyse how SF relations change depending on the adopted spatial resolution, on the choice of molecular gas tracer, and on the particular dynamical environment throughout the SB ring.


The paper is organized as follows. Sect.~\ref{observations} presents the new ALMA observations and accompanying ancillary data. We describe in Sect.~\ref{parameters} the conversion factors adopted. Sect.~\ref{maps} describes the molecular gas and Pa$\alpha$  images used in this work. We study the different KS relations derived in NGC~1068 in Sect.~\ref{SF}. Sect.~\ref{Trends} explores a different prescription of SF relations and analyses the environmental dependence of  SFE$_{\rm dense}$  as a function of a set of physical parameters in the different regions of the SB ring. We describe a scenario for the star formation in the SB ring in Sect.~\ref{discussion}. The main conclusions of this work are summarized in Sect.~\ref{summary}.


\section{Observations} \label{observations}
	In this section we present an overview of the different datasets used to probe the distribution of molecular gas (Sect.~\ref{tracers}) and the recent star formation (Sect~\ref{paalpha}) required 
	to derive the star-formation relations in the SB ring of NGC~1068. 
\subsection{Molecular gas tracers} \label{tracers}

\subsubsection{New ALMA data} \label{almadata}

We used ALMA to map the emission of HCN(1--0) and HCO$^{+}$(1--0) in the central $r\sim2.5$~kpc of the NGC~1068 disk.  Observations were executed during Cycle 2 in one track in August 2015 (project-ID: $\#$ 2013.1.00055.S, PI: S. Garc\'{\i}a-Burillo). We used band 3 receivers and a single pointing with a field of view (FOV) of $\sim$ 70$\arcsec$ ($\sim5$~kpc), covering the CND and the SB ring of the galaxy. Observations made use of 34 antennas of the array with projected baselines ranging from 12 m to 1430 m. The phase tracking centre was set to $\alpha_{2000}=02^{h}42^{m}40.771^{s}$,   $\delta_{2000}=-00^{\circ}00^{\prime}47.84\arcsec$, which is the centre of the  galaxy according to SIMBAD taken from the Two Micron All Sky Survey--2MASS survey \citep{Skrutskie2006}. The tracking centre is offset by $\leq$1$\arcsec$ relative to the AGN position: $\alpha_{2000}=02^{h}42^{m}40.71^{s}$, $\delta_{2000}=-00^{\circ}00^{\prime}47.94\arcsec$ \citep{Gallimore1996, Gallimore2004, Burillo2014, Burillo2016, Gal16, Imanishi2016}.
The galaxy has a systemic velocity of $v_{\rm sys}$(HEL)$\sim1130$~km s$^{-1}$ \citep{Burillo2014, Burillo2019}. 

Four spectral windows were placed, two in the 
lower sideband (LSB) and two in the upper sideband (USB). All the sub-bands have a spectral bandwidth of 1.875~GHz. The setup allowed us to simultaneously observe HCN($J=1-0$) (88.632~GHz at rest) and HCO$^+$($J=1-0$) (89.189~GHz at rest) in the higher frequency LSB band, as well as H$^{13}$CN($J=1-0$) (86.340~GHz at rest) and H$^{13}$CO$^+$($J=1-0$)  (86.754~GHz at rest) in the lower frequency LSB band. The two spectral windows in the USB band were centred around the  CS($J=2-1$) (97.981~GHz at rest) line and the continuum emission around 100~GHz, respectively. The CS(2--1) map was published by \citet{Sco20}. The 86.6~GHz-continuum  map of the galaxy was published by \citet{Burillo2017}.

The data were calibrated using the ALMA reduction package  {\tt CASA\footnote{http//casa.nrao.edu/}}. The calibrated uv-tables were exported to {\tt GILDAS\footnote{http://www.iram.fr/IRAMFR/GILDAS}}-readable format \citep{Guilloteau2000} in order to perform the mapping and cleaning steps as detailed below. We estimate that the absolute flux accuracy is about 5$\%$, which is in line with the goal of standard ALMA observations at these frequencies.   The synthesized beam obtained using natural weighting is $1\arcsec\times0\farcs6$ (70~pc~$\times$~42~pc) at a position angle PA~$=69\degr$. The line data cube was binned to a frequency resolution of $2.92$~MHz ($\sim10$~km~s$^{-1}$). We estimated a 1$\sigma$ sensitivity of 0.4~mJy~beam$^{-1}$ per channel of 10~km~s$^{-1}$  using line-free emission areas in the data. The conversion factor between Jy~beam$^{-1}$ and K is 247.6~K~Jy$^{-1}$ per beam in both emission lines. The spectral line maps were obtained after subtraction of the continuum emission performed in the $(u,v)$ plane using the {\tt GILDAS} tasks {\tt uv-average} and {\tt uv-subtract}. 
 
 We obtained the zeroth, first and second moment maps from the line data cubes using the {\tt GILDAS} task {\tt moments}, adopting a velocity window $\mid v-v_{\rm sys} \mid = 250$~km~s$^{-1}$, which is enough to  cover the  span of velocities due to rotation in the disk of the galaxy \citep{Burillo2014}. The total uncertainty on the velocity-integrated emission maps, $\Delta I $, was derived from:  
 
\begin{equation}
\Delta I  = \sqrt{\Delta I_{\rm noise}^{2} + \Delta  I^{2}_{\rm calib}}
\end{equation}
where
\begin{equation}
\Delta  I_{\rm noise}  =\sigma \times \delta V \times \sqrt{N_{\rm window}}
\end{equation}

%

$\Delta$I$_{\rm noise}$ is the velocity-integrated intensity error, which results from propagating the error of individual channels, $\sigma=0.4$~mJy~beam$^{-1}$, of width $\delta V=10$~km s$^{-1}$, 
to the number of channels considered in the integration window, $N_{\rm window}$ = 50. Furthermore, $\Delta$I$_{\rm calib}$ is the uncertainty due to the absolute flux calibration error, which is about $\simeq 5\%\times I$ for ALMA band 3 observations (see ALMA Technical Handbook \footnote{\href{http://almascience.eso.org/documents-and-tools/latest/documents-and-tools/cycle8/alma-technical-handbook}{http://almascience.eso.org/documents-and-tools/latest/documents-and-tools/cycle8/alma-technical-handbook}}). 
All in all, for a typical flux integrated value $\geq 5 \sigma$ characteristic of the regions in the SB ring of NGC~1068 studied in this work, the total uncertainty on $I$ as well as  on all the related parameters,  namely line luminosities and gas masses, amounts at most  to $\pm0.09$~dex 
in logarithmic units\footnote{The corresponding values for $\Delta$I$_{\rm calib}$ range respectively from $\simeq 10\%\times I$ to $\simeq 15\%\times I$ for the ALMA band 7 observations and the PdBI CO(1--0) data used in this paper 
(see Sect.~\ref{ancillary}). These values imply similar  typical uncertainties on $I\leq \pm0.09-0.10$~dex in either case for the regions of the SB ring of NGC~1068 examined in this work.}.

The largest angular scale (LAS) of our observations is $\sim4\arcsec$ ($\sim280$~pc). Since our observations do not contain short-spacing correction, the flux can start to be filtered out on scales larger than the LAS.
The HCN emission, and very likely also the HCO$^+$ emission, are expected to arise from a highly clumpy medium consisting of an ensemble of dense cloud cores. This particular hierarchy of the dense molecular gas probed by HCN and HCO$^+$ helped by the  velocity structure observed in the molecular disk of NGC~1068 \citep[e.g., see][]{Burillo2014} allows us to foresee that the amount of flux filtered on the spatial scales that are the most relevant for this paper is kept low in both lines. 

To validate this hypothesis we estimated the maximum percentage of missing flux in the HCN image of NGC~1068 by comparing the fluxes measured   by ALMA and by the IRAM 30m telescope (Usero et al.~private communication) at different locations of the disk. With this aim we derived the spatially-integrated fluxes using the single-dish aperture sizes  $\sim30\arcsec$ $\sim2$~kpc (see Appendix~\ref{app1} for details). The result of this comparison indicates that the maximum percentage of missing flux in the HCN ALMA map is about 25$\%$ on scales of 2~kpc.  As the spatial scales relevant for this paper are much smaller (40-700~pc), we can conclude that 25$\%$ is a conservative upper limit on the missing flux for HCN (and very likely also for HCO$^+$).

\subsubsection{Ancillary data} \label{ancillary}
We used the CO(3--2) line and 349 GHz (859 $\mu$m) continuum emission images of the galaxy obtained  by \citet{Burillo2014} with ALMA during the Cycle 0 of the array in band~7 
(project-ID: \# 2011.0.00083.S, PI: S. Garc\'{\i}a-Burillo).  The angular resolution of these data is $\sim0\farcs6\times0\farcs5$  at a position angle of $\sim60\degr$ (42~pc~$\times$~35~pc).
We refer to  \citet{Burillo2014} for a detailed description of the data reduction steps.
As the observations of  \citet{Burillo2014} do not contain short-spacing correction, we expect that a non-neligible amount of flux may start to be filtered out on scales beyond the reported LAS 
 $\geq6\arcsec$ (420~pc) for the continuum and CO(3--2) emission images.  Based on a comparison between the fluxes measured by ALMA and different single-dish telescopes using a set of apertures,
 \citet{Burillo2014} estimated  that their interferometer images may be filtering up to 20-30$\%$ and 65$\%$ of the total flux on spatial scales of about 1~kpc for the CO(3--2) and continuum emission, respectively.
 However, the clumpy distribution of the gas and also (in the case of the CO line) the velocity structure of the emission are expected to favour
the recovery of most of the flux in the line and continuum maps  on smaller  apertures ($\leq6\arcsec\sim420$~pc) centred on the brightest emission spots of the SB ring. 
 
 We also used the CO(1--0) line map of the galaxy obtained by the IRAM array on the Plateau de Bure Interferometer (PdBI), published by \citet{Schinnerer2000}. 
 The CO(1--0) line allows us to study the bulk of the molecular gas reservoir in the disk of NGC 1068. The  angular resolution of these data is $\sim1\farcs8\times1\farcs0$  at a position angle of $\sim24\degr$ (126~pc~$\times$~70~pc). \citet{Schinnerer2000} estimated  that  the  CO PdBI map misses about $25-30\%$  of  the  total flux on scales $\sim55\arcsec$ (3.8~kpc), based on the comparison between the CO flux measured by the PdBI and 
 the flux derived from the CO maps obtained through the combination of the Berkeley-Illinois-Maryland-Association (BIMA) array and the 12m Kitt Peak single-dish data published by \citet{Hel95}.
 We derived a new upper limit on the missing flux in the CO(1--0) map based on a comparison between the fluxes measured by PdBI and the  IRAM-30m telescope in 
 Appendix~\ref{app1} ($\sim$40-45$\%$ on scales $\sim 24\arcsec\sim1.7$~kpc).
As in this paper the relevant spatial scales used in our analysis  are smaller, we can therefore expect that the missing flux factors  reported above can be taken as strict upper limits.

\subsection{Star formation tracer} \label{paalpha}

We used the emission of the Pa$\alpha$ hydrogen recombination line at 1.875 $\mu$m to image the distribution of recent star formation in the disk of NGC~1068. We used the HST/NICMOS ($NIC3$) narrow-band ($F187N, F190N$) images of the galaxy retrieved from the {\it Hubble} Legacy Archive (HLA)\footnote{http://hla.stsci.edu/hlaview.html} to derive the continuum subtracted Pa$\alpha$ map, and followed the calibration and continuum subtraction steps detailed  in Sect.~2.2 of \citet{Burillo2014}. The pixel size of the HLA images is $0\farcs1\times0\farcs1$. The angular resolution (FWHM) of the Pa$\alpha$ image is $0\farcs26\times0\farcs26$ ($\sim18$~pc~$\times18$~ pc), as determined from the estimated size of the point spread function (PSF) in the observations. Uncertainties on flux calibration are at the 15-20$\%$ level \citep{Boker1999, AH2006}, which implies associated uncertainties $\leq \pm0.09-0.10$~dex for the regions of the SB ring of NGC~1068 examined in this work.
The Pa$\alpha$ line traces ionized gas produced by associations of massive 
($\gtrsim30M_{\odot}$) and young ($\lesssim8-10$~Myr) stars. The main advantage of the near infrared recombination  line compared to its optical counterpart, namely H$\alpha$, resides in the significantly lower extinction by dust of   Pa$\alpha$  \citep{Kennicutt1998bis, calzetti2007}. We may therefore neglect any dust-extinction correction when we derive the SFR from the Pa$\alpha$ fluxes. The validity of this hypothesis is  examined in Appendix~\ref{anexo-pa}, where we compare the Pa$\alpha$ fluxes measured by HST over a number of hot spots of the SB ring with those measured in H$\alpha$ using the ground-based image of the galaxy published by \citet{angeles2000}.  We estimate an overall low extinction correction at 1.875 $\mu$m for the SB ring knots:  A$_{\rm Pa\alpha}$ shows a median value $\sim0.03$~mag, compatible with optically thin emission (see Appendix~\ref{anexo-pa} for details). If we allow for a $\sim20\%$ uncertainty in the flux scales due to absolute calibration errors, we conclude that the Pa$\alpha$ map of the SB ring does not require any significant correction for dust extinction.

  \section{Conversion to physical parameters} \label{parameters}

 
 \begin{figure*}[htbp!]
   \centering
   \includegraphics[width=0.95\textwidth]{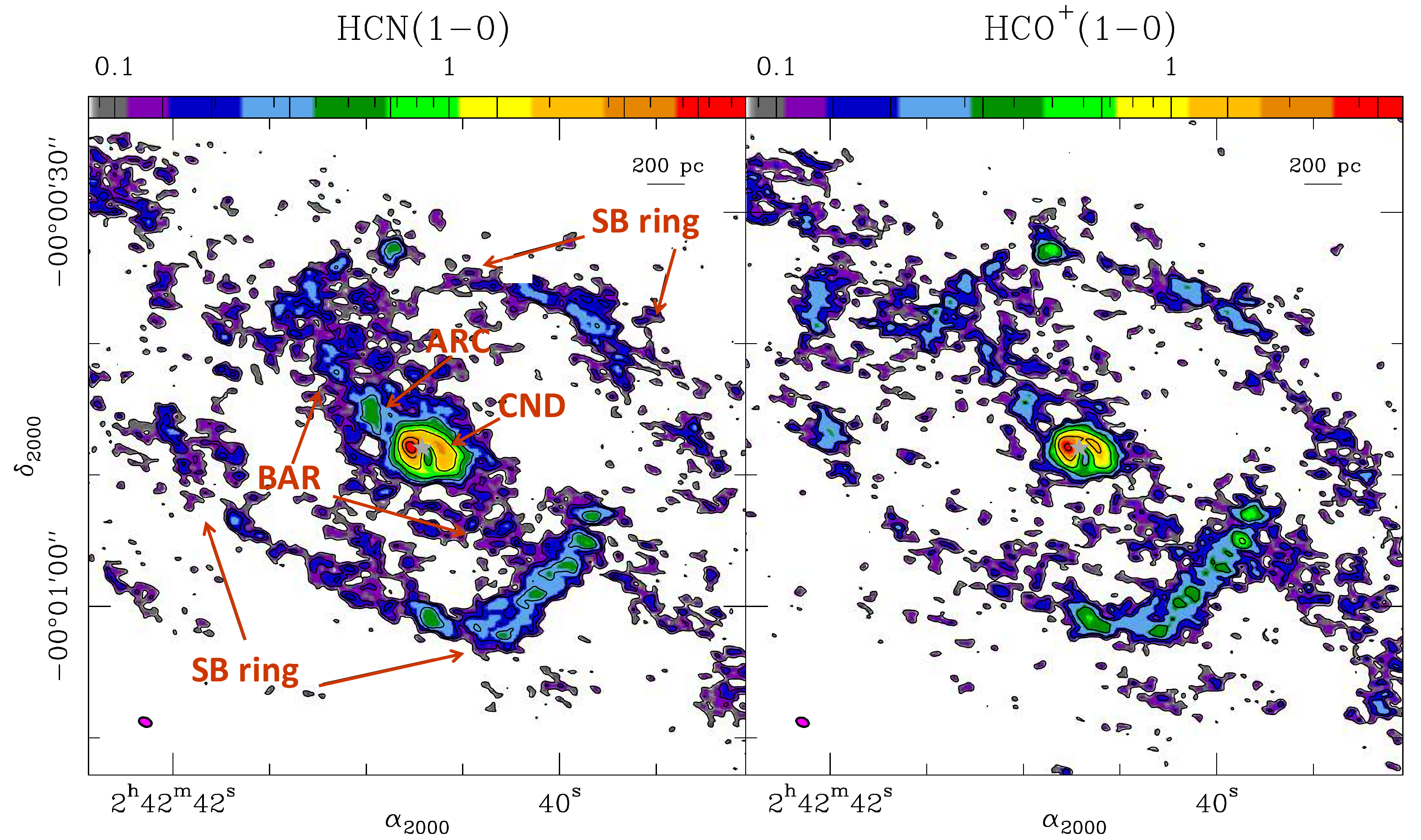}
   \caption{ {\it Left panel:}  velocity-integrated intensity map of HCN(1--0) of NGC~1068 obtained with ALMA. The map is shown in a logarithmic colour scale with contour levels: 3$\sigma$, 5$\sigma$, 7$\sigma$, 12$\sigma$, 24$\sigma$, 50$\sigma$, 120$\sigma$ and 200$\sigma$, where 1$\sigma$ = 0.028 Jy beam$^{-1}$km s$^{-1}$. We highlight the location of several representative regions of the emission, namely the CND, the bar, and the SB ring. {\it Right panel}: same as {\it left panel} but for the HCO$^{+}$(1--0) line map, with the same list of contour levels but here truncated at 120$\sigma$. The position of the AGN is identified by the dark grey star marker. The black bar in the top right corner of each panel shows the spatial scale in pc. The (magenta) filled ellipses at the bottom left corners of the panels represent the ALMA beam size (1$\arcsec$.0 $\times$ 0$\arcsec$.6 at PA=69$\degr$).} 
              \label{Fig_1}%
  \end{figure*}
   

\subsection{Molecular gas masses} \label{alphahcnhco}

To derive the distribution of "local" dense molecular gas mass  (M$_{\rm dense}$) from the HCN(1--0) velocity-integrated luminosities ($L_{\rm dense}'$) we assumed the standard conversion factor commonly applied  for "global" scales,  
$\alpha_{\rm HCN} = 10~M_{\odot}$~(K km s$^{-1}$pc$^{2}$)$^{-1}$ for HCN, following \citet{GaoSolomon2004a}, for the different "local" scales used in this work. This factor 
includes a correction for Helium. A canonical "constant" conversion factor is usually adopted in the literature for both lines, which are considered as reliable tracers of the dense molecular gas phase above densities n$\approx$10$^{4}$cm$^{-3}$ \citep[see, however,][]{Burillo2012, Eva20}.

This fixed conversion factor assumes that the HCN(1-0) emission is originated from gravitationally-bound "cores" or clumps with volume-averaged density n(H$_{2}$) $\sim$ 3 $\times$ 10$^{4}$ cm$^{-3}$ and a brightness temperature T$_{b}$ $\sim$ 35 K. However, under the hypothesis that the emission of the HCN line is mostly optically thick, if the volume-averaged density of the emitting clumps is lower than 3 $\times$ 10$^{4}$ cm$^{-3}$ or if the T$_{b}$ is larger than 35 K, the $\alpha_{\rm HCN}$ could be smaller than the factor suggested by \cite{GaoSolomon2004b}. In this context, \cite{Wu2005} estimated the value of $\alpha_{\rm HCN}$ in Galactic star-forming cores, finding a slightly lower conversion factor at smaller scales:   $\alpha_{\rm HCN}$= 7 $\pm$ 2 $M_{\odot}$~(K km s$^{-1}$pc$^{2}$)$^{-1}$. This value differs only by 30$\%$ from the "global" conversion factor used by \cite{GaoSolomon2004b}, which is the one adopted in this work. We nevertheless note that adopting a lower value of the conversion factor for HCN would result in slightly lower molecular gas surface density values, particularly in the regions of the SB ring that show comparatively higher  SF activity and SFE values (see discussion in Sect.~\ref{Trends}).

We therefore derived M$_{\rm dense}$ as:
\begin{equation}
M_{\rm dense}[M_{\odot}]=\alpha_{\rm HCN}L_{\rm dense}'[{\rm K~km s^{-1} pc^{2}}].
\end{equation}
The line luminosity L' is defined following \citet{Solomon1997} as: 
\begin{equation}
L'_{\rm dense}[{\rm K~km s^{-1} pc^{2}}]=3.25\times10^{7} \times S_{\rm dense}\Delta V \nu_{\rm obs}^{-2} D_{L}^{2}(1+z)^{-3},
\end{equation}
where the velocity-integrated fluxes $S_{\rm dense}\Delta~V$ are in Jy~km s$^{-1}$ particularized for each line, the observed frequency  $\nu_{\rm obs}$ is  in GHz and the luminosity distance $D_{L}$ is in Mpc units. 

We obtained face-on values of the dense molecular gas surface densities ($\Sigma_{\rm dense}$ in $M_{\odot}$~pc$^{-2}$ units) from:
\begin{equation}
\Sigma_{\rm dense}=\frac{M_{\rm dense}}{A_{\rm aperture}}\times cos(i), 
\end{equation}
where $i=40\degr$ is the inclination of the disk of NGC~1068 \citep{BH1997,Brinks1997,Burillo2014}
and $A_{\rm aperture}$ is the area of the aperture used in pc$^{2}$.

Similarly, we transformed the measured CO(1-0) luminosities into molecular gas masses and surface densities using the conversion prescription of \citet{Bolatto2013}, which assumes a standard Galactic 
conversion factor $\alpha_{\rm CO} = 4.4~M_{\odot}$~(K km s$^{-1}$pc$^{2}$)$^{-1}$ , which already includes a correction for Helium.

We used the  {\tt GILDAS} task {\tt gauss-smooth} to convolve the initial resolution versions of the molecular line data cubes 
with the appropriate Gaussian kernels adapted to generate all the image versions for the common set of spatial resolutions 
used in this work, which range from $\sim40-56$~pc ($\sim100$~pc for CO(1--0)) up to  $\sim700$~pc (see Sect.~\ref{KSn1068}).


\subsection{Star formation rates} \label{sfr}

We adopted the prescription proposed in \citet{Kennicutt:2012} regarding the conversion factor used to calculate the local SFR map from the Pa$\alpha$ line luminosities. In particular,  we
assumed a Kroupa initial mass function \citep{Kroupa:2001} as well as an intrinsic ratio for H$\alpha$/Pa$\alpha$~$\sim7.81$ \citep{HS1987}, which applies for the case B recombination at $T_{e}$= 5000~K and $n_{e}$= 10$^{3}$cm$^{-3}$. These conditions are found in starbursting galaxies \citep{Roy:2008, Rieke:2009}. 

 We obtain the SFR from the expression:
\begin{equation}
{\rm SFR}(M_{\odot} {\rm yr^{-1}})=4.15 \times 10^{-41}L({\rm Pa}\alpha, {\rm erg~s}^{-1})
\end{equation}

We derived the corresponding SFR surface densities ($\Sigma_{\rm SFR}$) in units of M$_{\odot}$~yr$^{-1}$pc$^{-2}$ from:
\begin{equation}
\Sigma_{\rm SFR}=\frac{\rm SFR}{A_{\rm aperture}}\times cos(i),
\end{equation}


We estimated the SFR for the range of spatial resolutions analysed in this work following the same procedure described in Sect.~\ref{alphahcnhco}, which 
uses  the  {\tt GILDAS} task {\tt gauss\_smooth}  to convolve the initial resolution images with the appropriate Gaussian kernels.

   
    \begin{figure}[tbhp]
   \centering
   \includegraphics[width=8.5cm]{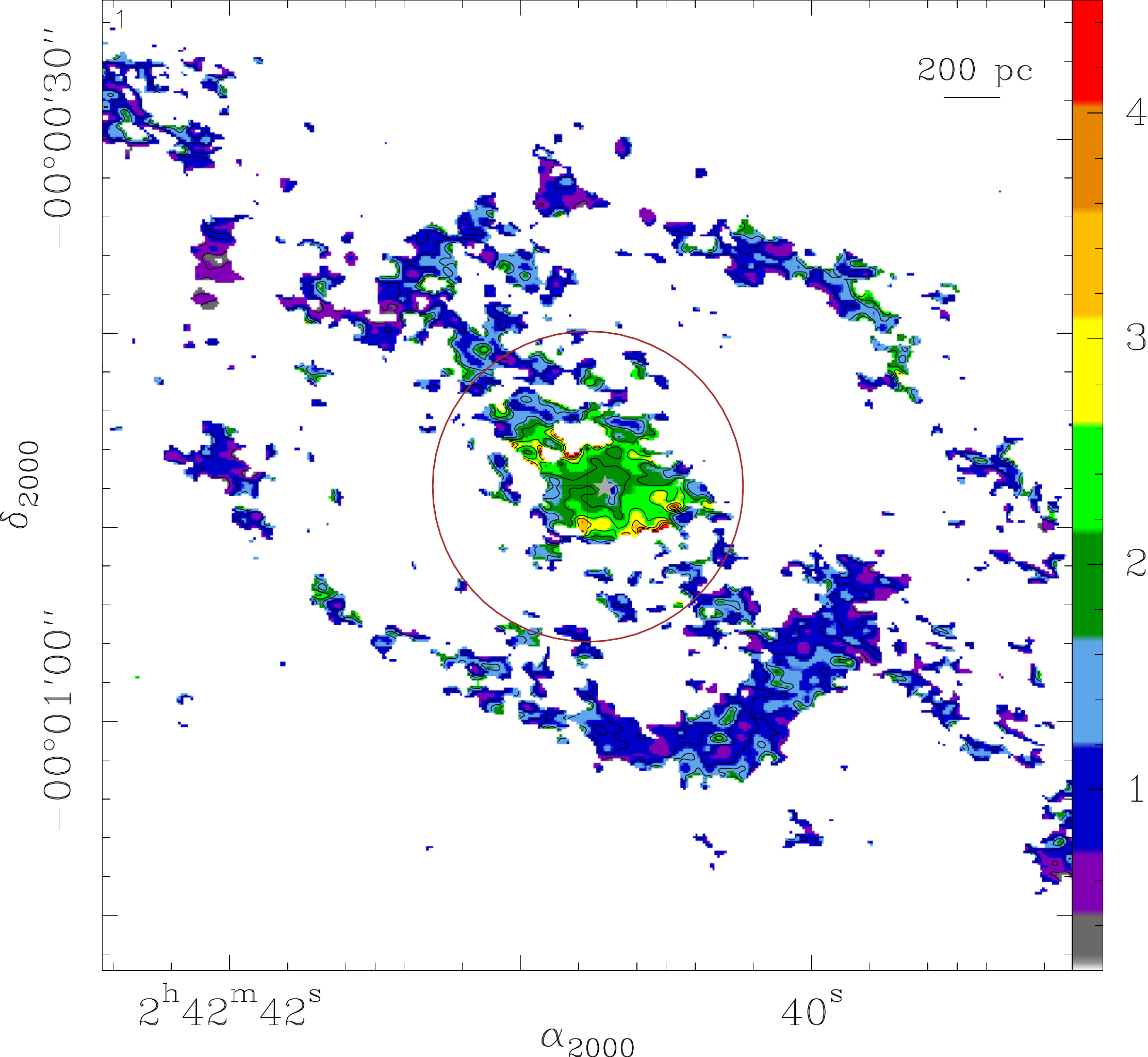}
   \caption{HCN(1--0)/HCO$^{+}$(1--0) brightness temperature ratio map ($R_{\rm dense}$) derived at the common spatial resolution of $\simeq$56~pc. The brown circle of 8$\arcsec$-radius ($\simeq$ 560 pc) locates the inner region of the galaxy disk where the molecular outflow signature has been identified in the kinematics of molecular gas \citep{Burillo2014, Burillo2019}. This region, purposely excluded from our analysis of SF relations, shows significantly higher line ratios ($R_{\rm dense}\simeq$~1.5-3.5) compared to the SB ring ($R_{\rm dense}\simeq$0.5-1.5). Other symbols as in Fig.~\ref{Fig_1}.}
              \label{Fig_2}%
   \end{figure}



   \begin{figure}[tbhp]
   \centering
   \includegraphics[width=8.5cm]{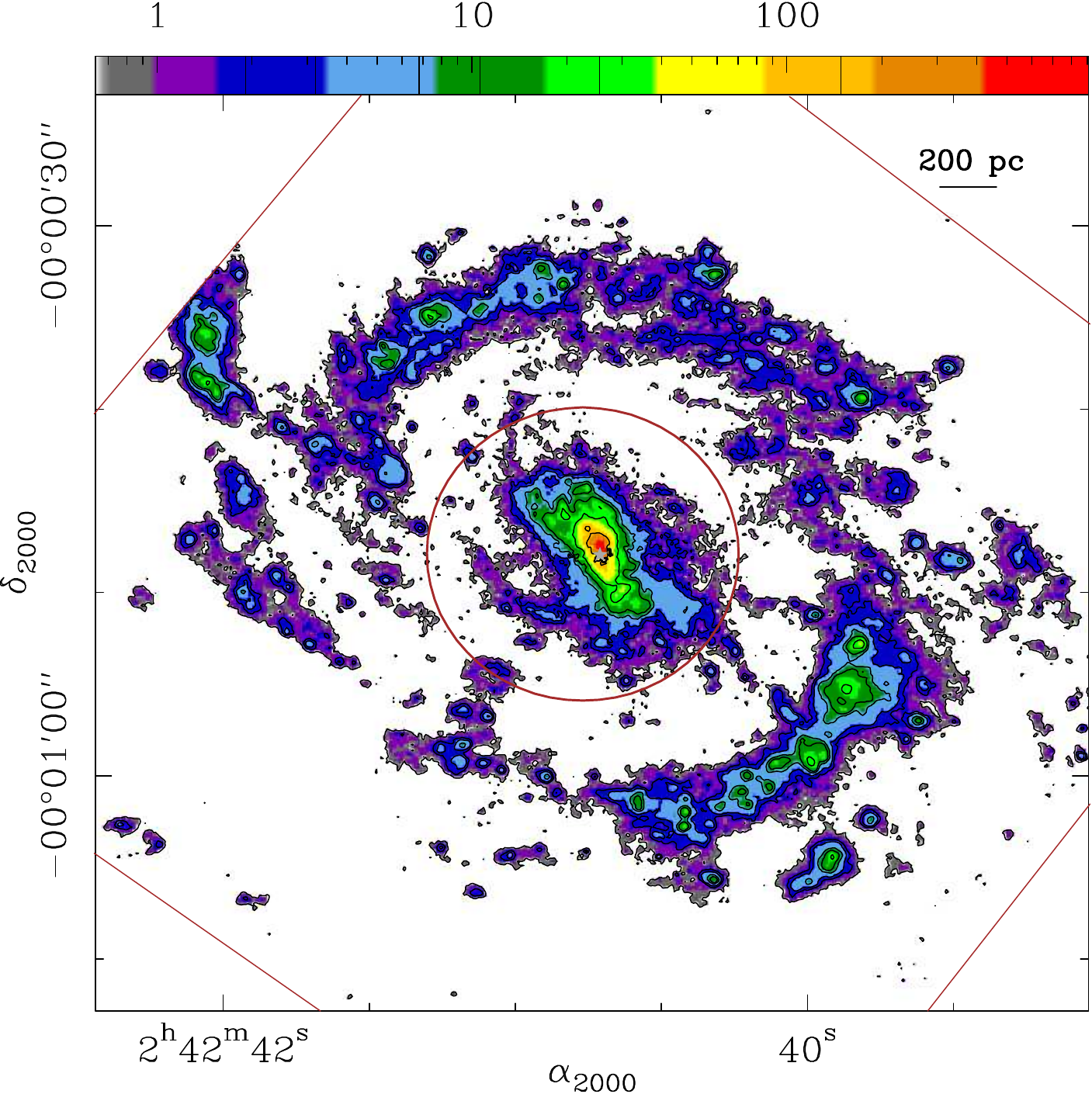}
   \caption{HST/NICMOS continuum-subtracted Pa$\alpha$ image of NGC~1068. The map is shown in colour scale with contour levels: 3$\sigma$, 9$\sigma$, 15$\sigma$, 32$\sigma$, 50$\sigma$, 120$\sigma$ and 700$\sigma$, where 1$\sigma$=0.212 10$^{-16}$erg s$^{-1}$cm$^{-2}$pixel$^{-1}$. The colour scale range is shown in units of 10$^{-16}$ erg s$^{-1}$ cm$^{-2}$ pixel$^{-1}$. The (truncated) square region identifies the outer edge of the HST/NICMOS field-of-view. The angular resolution of the image is 0$\farcs$26 $\times$ 0$\farcs$26. Other symbols as in Fig.~\ref{Fig_2}.}
              \label{Fig_3}%
	\end{figure}
	


 \begin{figure*}[hbtp]
   \centering
   \includegraphics[width=1\textwidth]{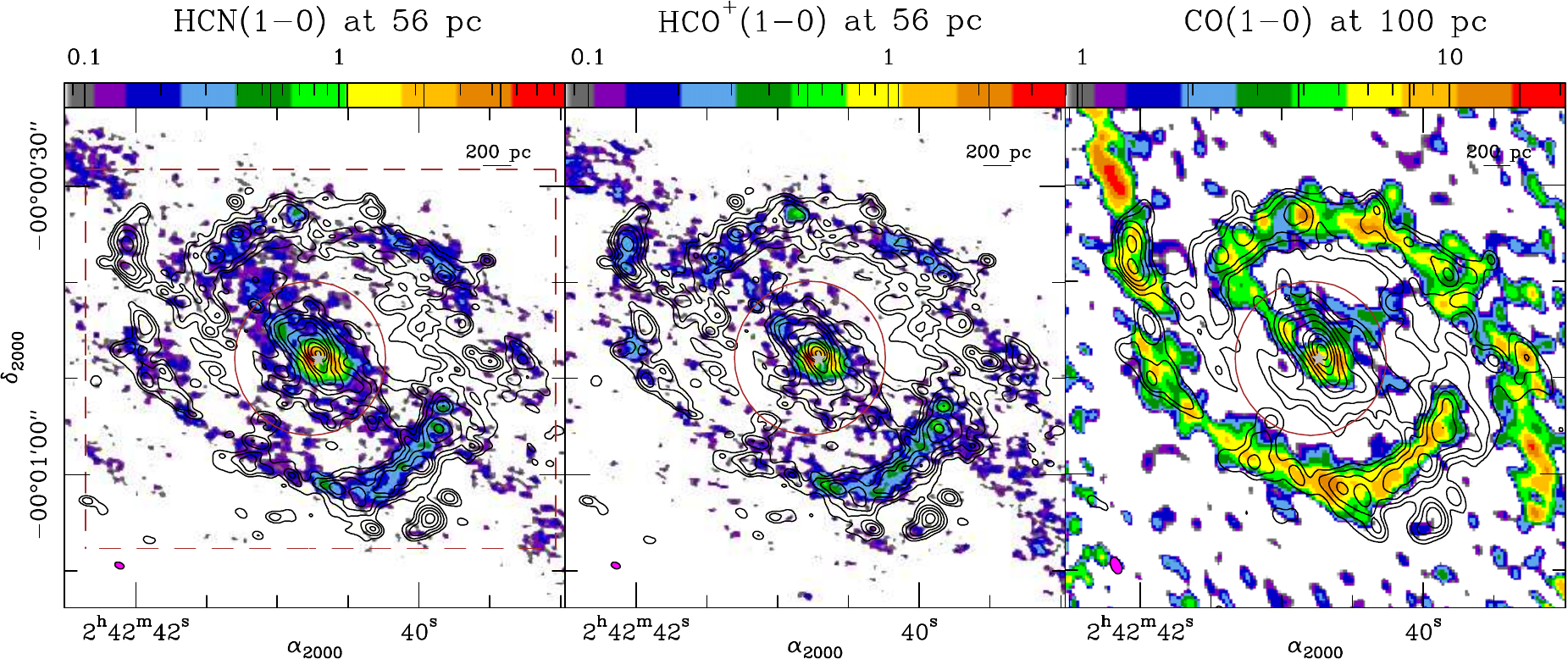}

   \caption{{\it Left panel:} overlay of the HST/NICMOS Pa$\alpha$ image (contours) on the ALMA HCN(1--0) map (colour scale). Contours with a logarithmic spacing from 3$\sigma$ to 1536$\sigma$, in steps of 0.3~dex, where 1$\sigma$=0.180$^{-16}$erg s$^{-1}$cm$^{-2}$pixel$^{-1}$. Colour scale is shown in units of Jy beam$^{-1}$km s$^{-1}$. {\it Middle panel:} same as  {\it left panel} but showing the overlay of the HST/NICMOS Pa$\alpha$ contours on the ALMA HCO$^{+}$(1--0) map (colour scale). {\it Right panel:} same as {\it left panel} but showing the overlay of the HST/NICMOS Pa$\alpha$ contours  on the PdBI CO(1--0) map of \citet{Schinnerer2000} (colour scale). Contour spacing same as in left and right panels, but with 1$\sigma$=0.162 10$^{-16}$erg s$^{-1}$cm$^{-2}$pixel$^{-1}$. The (magenta) filled ellipses at the bottom left corners in all panels represent the beam sizes of the molecular line maps: 1$\arcsec$.0 $\times$ 0$\arcsec$.6 at PA=69$\degr$ ({\it left}  and {\it middle panels}) and 1$\arcsec$.84 $\times$ 1$\arcsec$.09 at P.A. = 24$\degr$ (right panel). The HST/NICMOS Pa$\alpha$ images in each panel have been degraded to the corresponding spatial resolutions of the molecular gas tracers used in this comparison. Other symbols as in Fig.~\ref{Fig_2}.}
              \label{Fig_4}%
   \end{figure*}


\section{Dense molecular gas and SF maps} \label{maps}
\subsection{The HCN and HCO$^+$ maps} \label{mapsalma}
 
 Figure~\ref{Fig_1} shows the HCN(1--0) and HCO$^{+}$(1--0) velocity-integrated intensity maps of NGC 1068 obtained by ALMA in the central  $r\sim2.5$~kpc of the disk. Overall, the distribution of dense molecular gas is similar to that shown by other molecular gas tracers as seen in  previous interferometer images of 
the galaxy \citep{Schinnerer2000, Krips2011,Tsai2012, Burillo2014, Viti2014, Burillo2019, Sco20}. In particular,  the bulk of the  HCN(1--0) and HCO$^{+}$(1--0)  emission stems from three main regions:

{\it 1.~The CND}. Described as an asymmetric elliptical ring of  $6\arcsec \times4\arcsec$--size ($r\sim200$~pc), the CND shows two emission knots located $\sim 1\arcsec$ east and $\sim 1.5\arcsec$ west of the AGN. The CND ring is off-centred relative to the AGN locus. The two emission knots are bridged by weaker emission north and south of the AGN.   The morphology of the HCN and HCO$^+$ maps of the CND is to a large extent similar to that of the ALMA CO and CS maps \citep{Burillo2014, Burillo2019, Sco20}.

{\it 2.~The bar}.  There is HCN and HCO$^+$ emission in the  region occupied by the $\sim2.6$~kpc-diameter stellar bar, which is oriented along PA$=46^{\circ}\pm2^{\circ}$ \citep{Sco88,Schinnerer2000}. As shown by other molecular gas tracers, the emission in this region from both lines (especially for HCN), tends to accumulate along the leading edges of the bar. We also detect significant emission in the "bow-shock arc" feature identified in the CO(3--2) and continuum dust emission maps of \citet{Burillo2014} on the northeast side of the disk at $r\sim4 \arcsec-7\arcsec$ (300~pc--500~pc).  
 
{\it 3.~The SB ring}.  Most of the dense molecular gas in the disk concentrates in a ring of $r\sim18\arcsec$(1.3~kpc) formed by two tightly wound spiral arms, which unfold  over  $\sim 180^{\circ}$ in azimuth in the disk from the ends of the stellar 
bar. The SB ring concentrates also most of the massive star forming complexes in the disk identified in the Pa$\alpha$ image (see Sect~\ref{imagehst}). The emission of HCN and HCO$^+$ is unevenly distributed azimuthally over the SB ring: in 
both lines the emission is strongest around two regions, located  at $r\sim18\arcsec$ and PA~$\sim 15^{\circ}-75^{\circ} (\pm 180^{\circ}$), where the ring is connected to the stellar bar ends. The emission of HCN and HCO$^+$ in the SB ring is clumpy and it appears to be organized as coming from molecular cloud associations of $\geq50$~pc-size. The SB ring is connected at larger radii to  two emission lanes located  at the edge of the HCN and HCO$^+$ maps shown in Fig~\ref{Fig_1} along PA$\sim50^{\circ}\pm180^{\circ}$.


 \begin{figure*}[htb]
   \centering
    \includegraphics[width=18.6cm]{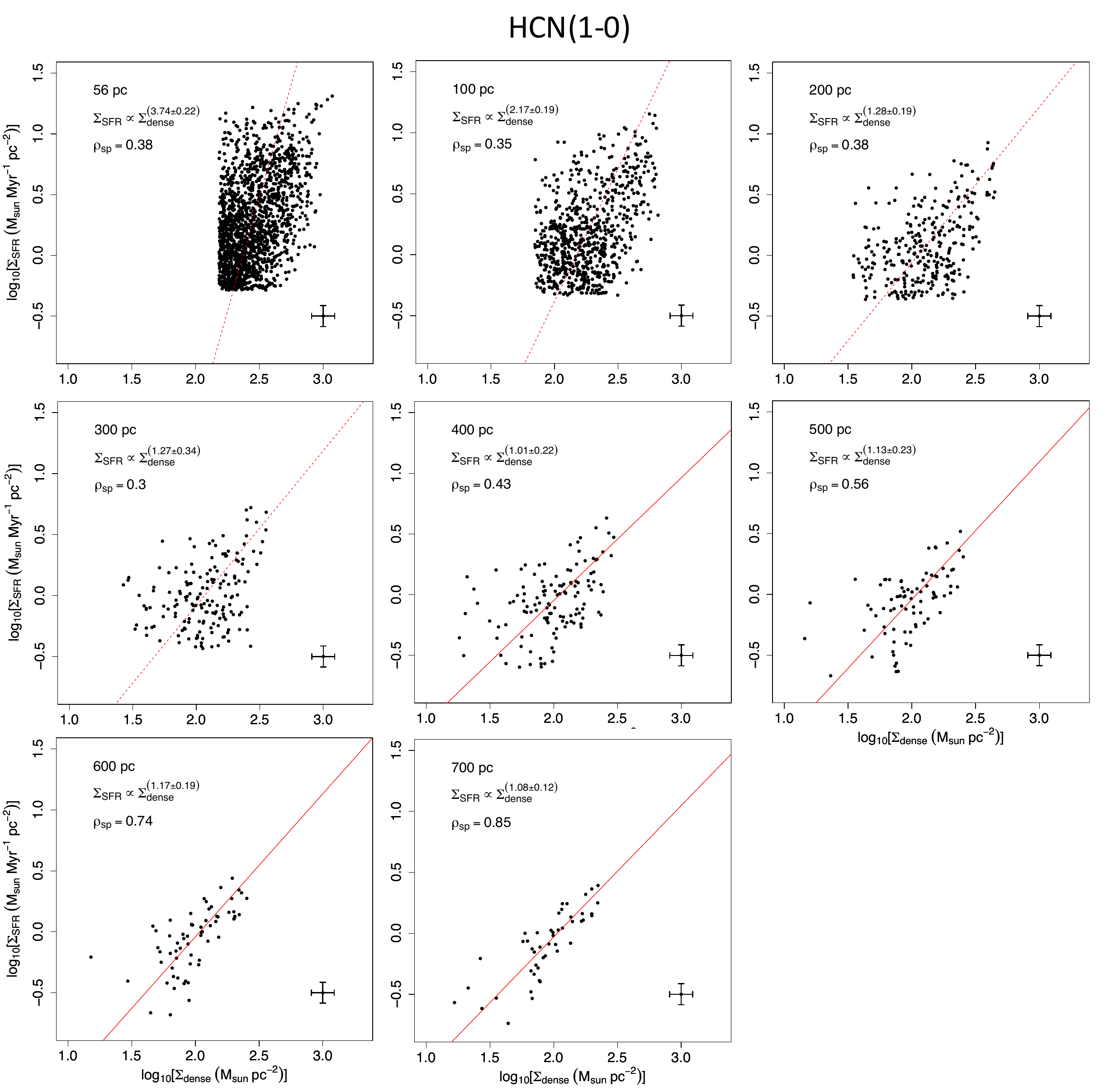}
   \caption{SFR surface density ($\Sigma_{\rm SFR}$) as a function of the dense molecular gas surface density ($\Sigma_{\rm dense}$) derived from HCN(1--0) for the different resolutions explored in this work: 56~pc, 100~pc, 200~pc, 300~pc, 400~pc, 500~pc, 600~pc and 700~pc. The red lines show the  orthogonal distance regression (ODR) fits to the data. The dashed lines identify specifically the correlations that are not found to be statistically significant.  We only show on both axes the data points above 3$\sigma$ which are considered in the fits. We indicate the Spearman's rank correlation coefficients ($\rho_{\rm sp}$) and the power-law indexes ($N$) of the best fits. Vertical and horizontal errorbars at the lower right corner of each panel account for the typical uncertainties, which amount to $\pm0.09$~dex on both axes.}
    \label{Fig_5}
   \end{figure*}


Figure \ref{Fig_2} shows the HCN-to-HCO$^{+}$ line brightness ratio  ($R_{\rm dense}$) in the disk of the galaxy. The $R_{\rm dense}$ map was obtained assuming a common 3$\sigma$ threshold on the integrated intensities of both lines. The $R_{\rm dense}$ ratio changes significantly across the different regions of the disk identified above. In particular,  $R_{\rm dense}\simeq 1.5-3.5$ in the CND and the bow-shock arc region, for which we estimated a mean value $\langle R_{\rm dense}\rangle\simeq$~2.2. On the other hand,  $R_{\rm dense}\simeq 0.5-1.5$ in the SB ring and the corresponding  mean value $\langle R_{\rm dense} \rangle \simeq$~1.1. 

The high $R_{\rm dense}$ values measured in the CND and the bow-shock arc are 
 related to the molecular outflow signature identified in the kinematics of 
molecular gas in these regions \citep{Burillo2014, Burillo2019}. The outflow is 
thought to be driven by the interaction of the AGN wind and the radio jet with the molecular 
gas in the disk in a mostly coplanar geometry. Besides leaving a distinct kinematic signature in the CND and the bow-shock arc, the 
outflow  has left its imprint on the excitation and the chemistry of 
molecular gas, which is under the influence of large-scale shocks and a strong UV 
irradiation in these regions \citep{Viti2014, Burillo2017}.

\subsection{The Pa$\alpha$ map} \label{imagehst}
Figure \ref{Fig_3} shows the  Pa$\alpha$ image of NGC~1068 obtained at the initial resolution of the HST/NICMOS camera: $0\farcs26\times0\farcs26$ ($\sim18$~pc~$\times18$~ pc). The central region of the image reveals strong
emission stemming from an asymmetric bipolar nebula of ionized gas. Both 
the morphology and the kinematics of the gas in this structure have been 
modelled in terms of an AGN-driven wind \citep[e.g.,][]{Cre00, Cec02, Das06, Mue11, Bar14, Miy20}. The AGN wind occupies a hollow bicone  which 
extends up to a radius $r\sim8\arcsec$ (550~pc) on its northern side.  The bicone feature is oriented
along PA~$\sim30^{\circ}$, and is characterised by a wide opening angle (
$\sim80^{\circ}$). In our subsequent analysis of the SF relations we will therefore screen the central  $r\leq8\arcsec$ region of the galaxy, where Pa$\alpha$ cannot be considered as a reliable tracer of SF.   

Outside the bright AGN bicone structure, most of the Pa$\alpha$ emission comes from the SB ring. Similar to the distribution of the dense molecular gas, SF traced by  Pa$\alpha$ is not uniformly distributed throughout the SB ring. As for HCN and HCO$^+$, the brightest SF complexes are located at the northeast section, and most particularly, at the southwest section of the SB ring (at PA~$\sim 195^{\circ}-255^{\circ}$). In either case these are the two regions where the ring is connected to the ends of the stellar bar. \citet{Rico-Vilas21} studied the 147~GHz free-free emission associated with SF and identified  a similar concentration of massive super star clusters in the bar-ring interface region.

Figure \ref{Fig_4} shows the overlay of the HST/NICMOS Pa$\alpha$ emission image on 
the HCN(1--0), HCO$^{+}$(1--0), and CO(1--0) maps. For a proper comparison we degraded the Pa$\alpha$ image to the common spatial resolution of HCN and 
HCO$^{+}$ ($\sim56$~pc), and to that of CO ($\sim100$~pc). A visual inspection of this figure illustrates that the Pa$\alpha$ maxima do not always coincide with the strongest emission peaks in
 HCN, HCO$^{+}$ or CO throughout the SB ring. 

\section{Star formation relations in NGC1068} \label{SF}

\subsection{Kennicutt-Schmidt laws} \label{KSn1068}

We use in this section the images described in Sects.~\ref{tracers} and ~\ref{paalpha} to obtain different versions of the pixel-wise spatially resolved 
KS relations  in the SB ring of NGC~1068 for a set of seven spatial resolutions, ranging from $\sim40-56$~pc ($\sim100$~pc for CO(1--0)) up to  $\sim700$~pc. 
To derive $\Sigma_{\rm SFR}$ and $\Sigma_{\rm gas}$ for each spatial scale we degraded our datasets to the selected resolutions using the {\tt GILDAS} task {\tt gauss-smooth}. In our analysis we only considered pixels with a signal $\geq3\sigma$ in $\Sigma_{\rm SFR}$ and $\Sigma_{\rm gas}$. Furthermore, to minimize the redundancy in the scatter plots we explored the  $\Sigma_{\rm SFR}$-$\Sigma_{\rm gas}$ plane using a grid with Nyquist sampling adapted for each spatial resolution.


 \begin{figure*}[hbtp]
   \centering
   \includegraphics[width=9cm]{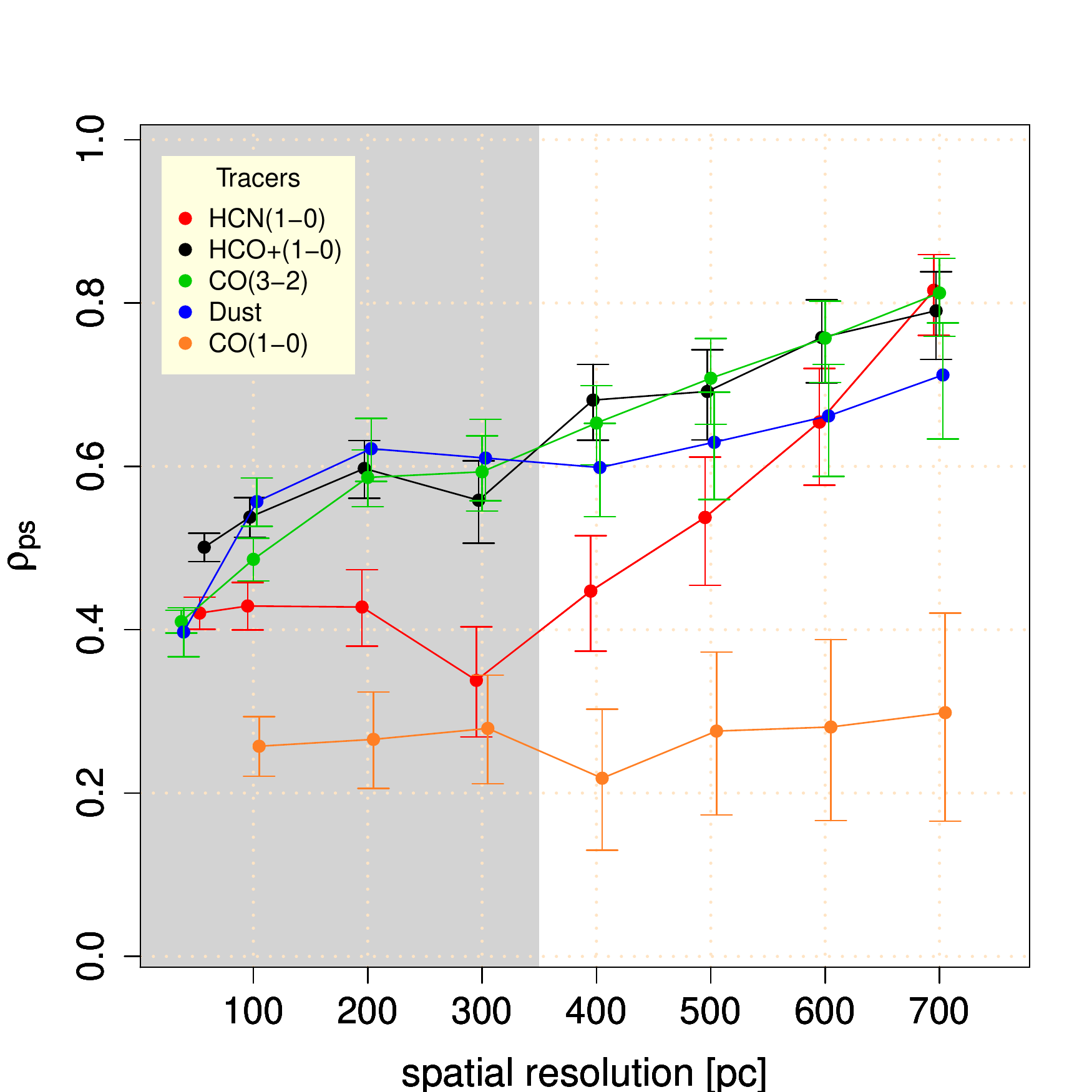}
    \includegraphics[width=9cm]{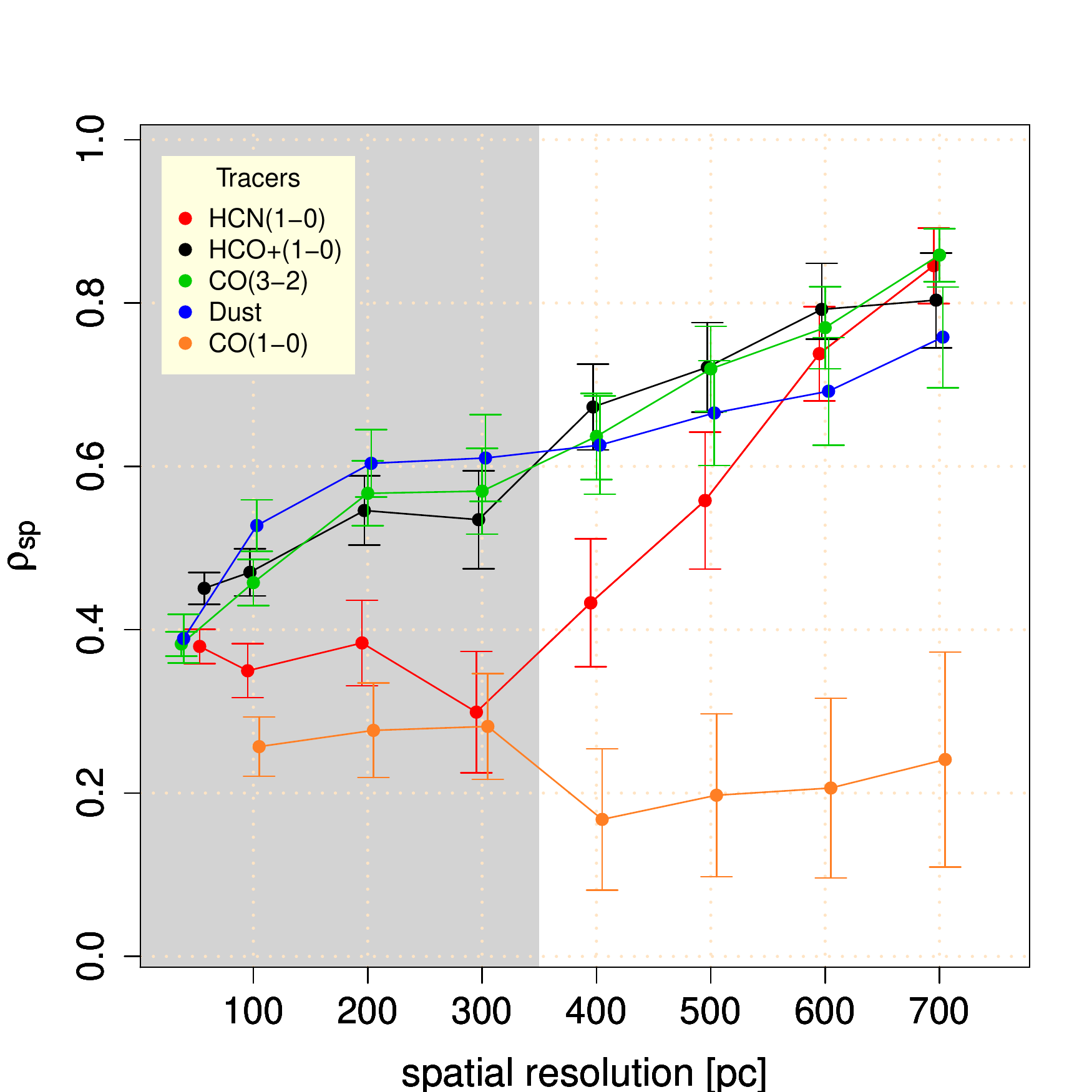}
      
   \caption{Pearson and Spearman correlation coefficients ($\rho_{\rm ps}$: {\it left panel};  $\rho_{\rm sp}$: {\it right panel}) of the KS laws derived  for the different gas tracers used in this work as a function of the spatial resolution. Error bars represent the 67 \% confidence interval around the mean values. For spatial scales larger than the range identified by the grey-coloured region ($\leq$350~pc)
   all the gas tracers (leaving aside CO(1--0)) show statistically significant correlations in their KS laws, defined by $p$-values $<$1$\%$ and $\rho_{\rm ps}$, $\rho_{\rm sp}>$0.4.}
              \label{Fig_6}
  \end{figure*}



\begin{figure}[tb!]
\centering
\includegraphics[width=1\linewidth]{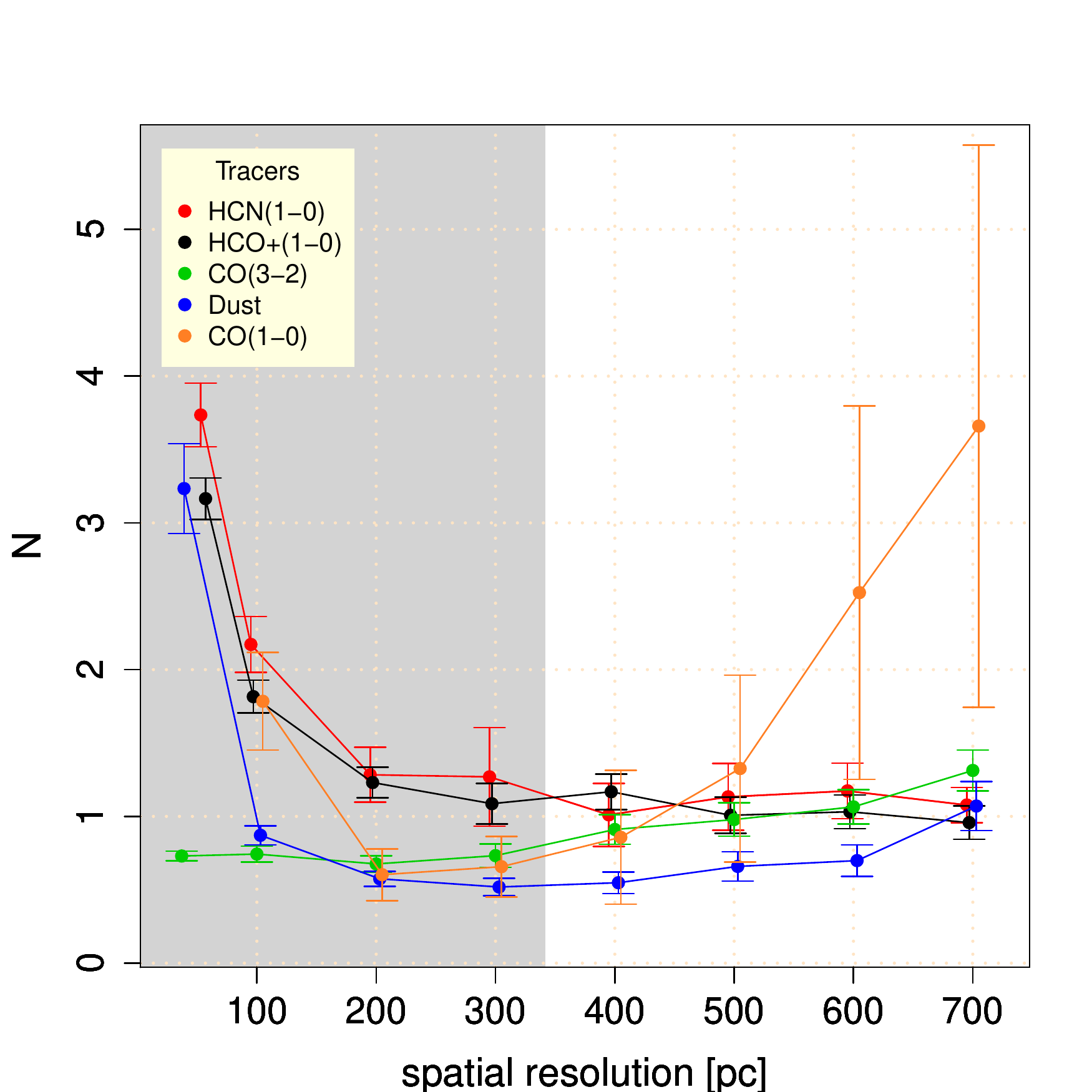}

\caption{Power-law indexes ($N$) of the KS relations obtained from the ODR fit to the data for the different gas tracers used in this work as a function of the  spatial resolution adopted. Symbols and colour codes as in Fig.~\ref{Fig_6}.}
   \label{Fig_7}
 \end{figure}
   


\begin{table*}[th!]

\centering
\caption{Correlation parameters obtained for the KS laws derived for different tracers and spatial scales.}

\resizebox{\linewidth}{!}{
 
\begin{tabular}{clllllllllllllll}
\hline

 & \multicolumn{3}{c}{\textbf{HCN(1--0)}} & \multicolumn{3}{c}{\textbf{HCO$^{+}$(1--0)}}   &   \multicolumn{3}{c}{\textbf{CO(1--0)}}            & \multicolumn{3}{c}{\textbf{CO(3--2)}}            &    \multicolumn{3}{c}{\textbf{Dust}}              \\
\textbf{scale (pc)} & \multicolumn{1}{c}{\textit{N}} & \multicolumn{1}{c}{\textit{$\rho_{ps}$}} & \multicolumn{1}{c}{\textit{$\rho_{sp}$}} & \multicolumn{1}{c}{\textit{N}} & \multicolumn{1}{c}{\textit{$\rho_{ps}$}} & \multicolumn{1}{c}{\textit{$\rho_{sp}$}} & \multicolumn{1}{c}{\textit{N}} & \multicolumn{1}{c}{\textit{$\rho_{ps}$}} & \multicolumn{1}{c}{\textit{$\rho_{sp}$}} &\multicolumn{1}{c}{\textit{N}} & \multicolumn{1}{c}{\textit{$\rho_{ps}$}} & \multicolumn{1}{c}{\textit{$\rho_{sp}$}} & \multicolumn{1}{c}{\textit{N}} & \multicolumn{1}{c}{\textit{$\rho_{ps}$}}  & \multicolumn{1}{c}{\textit{$\rho_{sp}$}} \\ \hline
40              & \multicolumn{1}{c}{-}           & \multicolumn{1}{c}{-}           & \multicolumn{1}{c}{-}           & \multicolumn{1}{c}{-}           & \multicolumn{1}{c}{-}           & \multicolumn{1}{c}{-}  & \multicolumn{1}{c}{-}  & \multicolumn{1}{c}{-}       & \multicolumn{1}{c}{-}      & \multicolumn{1}{c}{0.73 $\pm$0.03}           & \multicolumn{1}{c}{0.41}   & \multicolumn{1}{c}{0.38}         & \multicolumn{1}{c}{3.23 $\pm$ 0.31}           & \multicolumn{1}{c}{0.40}       & \multicolumn{1}{c}{0.39}     \\
56            & \multicolumn{1}{c}{3.74 $\pm$ 0.22 }           & \multicolumn{1}{c}{0.42}    & \multicolumn{1}{c}{0.38}        & \multicolumn{1}{c}{3.17 $\pm$ 0.14}           & \multicolumn{1}{c}{0.50}   & \multicolumn{1}{c}{0.45}          & \multicolumn{1}{c}{-}           & \multicolumn{1}{c}{-}           & \multicolumn{1}{c}{-}           & \multicolumn{1}{c}{-}           & \multicolumn{1}{c}{-}           & \multicolumn{1}{c}{-}          & \multicolumn{1}{c}{-}  & \multicolumn{1}{c}{-} & \multicolumn{1}{c}{-}  \\
100             & \multicolumn{1}{c}{2.17 $\pm$ 0.19}           & \multicolumn{1}{c}{0.43}     & \multicolumn{1}{c}{0.35}       & \multicolumn{1}{c}{1.82 $\pm$ 0.11 }        & \multicolumn{1}{c}{0.54}     & \multicolumn{1}{c}{0.47}       & \multicolumn{1}{c}{1.78 $\pm$ 0.33}           & \multicolumn{1}{c}{0.26}     & \multicolumn{1}{c}{0.26}       & \multicolumn{1}{c}{0.74 $\pm$ 0.05}           & \multicolumn{1}{c}{0.49}    & \multicolumn{1}{c}{0.46}        & \multicolumn{1}{c}{0.87 $\pm$ 0.07}           & \multicolumn{1}{c}{\textbf{0.56}}     & \multicolumn{1}{c}{\textbf{0.53}}       \\
200             &   \multicolumn{1}{c}{1.28 $\pm$ 0.19}                               &                               \multicolumn{1}{c}{0.43}  & \multicolumn{1}{c}{0.38}  &               \multicolumn{1}{c}{1.23 $\pm$ 0.11}                   &       \multicolumn{1}{c}{\textbf{0.60}}                         & \multicolumn{1}{c}{\textbf{0.55}}   &                               \multicolumn{1}{c}{0.61 $\pm$ 0.18}   &                \multicolumn{1}{c}{0.27}          & \multicolumn{1}{c}{0.28}         &                 \multicolumn{1}{c}{0.68 $\pm$ 0.06}                 &                               \multicolumn{1}{c}{0.59}  & \multicolumn{1}{c}{0.57}         &              \multicolumn{1}{c}{0.57 $\pm$ 0.05}                    &           \multicolumn{1}{c}{\textbf{0.62}      }       & \multicolumn{1}{c}{\textbf{0.61} }                 \\
300              &           \multicolumn{1}{c}{1.27 $\pm$ 0.34}                       &                               \multicolumn{1}{c}{0.34}   & \multicolumn{1}{c}{0.30}        &           \multicolumn{1}{c}{1.09 $\pm$ 0.14}                       &                  \multicolumn{1}{c}{0.56}     & \multicolumn{1}{c}{0.53}                   &                               \multicolumn{1}{c}{0.66 $\pm$ 0.21}   &                  \multicolumn{1}{c}{0.28}            & \multicolumn{1}{c}{0.28}            &         \multicolumn{1}{c}{0.73 $\pm$ 0.08}                         &                               \multicolumn{1}{c}{\textbf{0.59}} & \multicolumn{1}{c}{\textbf{0.57}}          &           \multicolumn{1}{c}{0.52 $\pm$ 0.06}                       &           \multicolumn{1}{c}{\textbf{0.61}}            & \multicolumn{1}{c}{\textbf{0.61}}        \\
400              &              \multicolumn{1}{c}{1.01 $\pm$ 0.22}                    &                               \multicolumn{1}{c}{\textbf{0.45}} & \multicolumn{1}{c}{\textbf{0.43}}          &                    \multicolumn{1}{c}{1.17 $\pm$ 0.12}              &                 \multicolumn{1}{c}{\textbf{0.68}}         & \multicolumn{1}{c}{\textbf{0.67}}                &                               \multicolumn{1}{c}{0.86 $\pm$ 0.46}   &                \multicolumn{1}{c}{0.22}             & \multicolumn{1}{c}{0.17}             &               \multicolumn{1}{c}{0.91 $\pm$ 0.11}                   &                               \multicolumn{1}{c}{\textbf{0.65}}  & \multicolumn{1}{c}{\textbf{0.69}}         &                  \multicolumn{1}{c}{0.55 $\pm$ 0.07}                &               \multicolumn{1}{c}{\textbf{0.60}}          & \multicolumn{1}{c}{\textbf{0.63}  }               \\
500           &               \multicolumn{1}{c}{1.13 $\pm$ 0.23}                   &                               \multicolumn{1}{c}{\textbf{0.54}}  & \multicolumn{1}{c}{\textbf{0.56}}         &              \multicolumn{1}{c}{1.01 $\pm$ 0.12}                    &                     \multicolumn{1}{c}{\textbf{0.69}}       & \multicolumn{1}{c}{\textbf{0.72}}              &                               \multicolumn{1}{c}{1.33 $\pm$ 0.64 }   &                        \multicolumn{1}{c}{0.28}       & \multicolumn{1}{c}{0.21}           &                    \multicolumn{1}{c}{0.98 $\pm$ 0.10 }              &                               \multicolumn{1}{c}{\textbf{0.71}}   & \multicolumn{1}{c}{\textbf{0.72}}        &                      \multicolumn{1}{c}{0.66 $\pm$ 0.11}            &           \multicolumn{1}{c}{\textbf{0.63}}           & \multicolumn{1}{c}{\textbf{0.67}}                    \\
600              &                  \multicolumn{1}{c}{1.17 $\pm$ 0.19}                &                               \multicolumn{1}{c}{\textbf{0.66}}  & \multicolumn{1}{c}{\textbf{0.74}}         &              \multicolumn{1}{c}{1.03 $\pm$ 0.11}                   &                \multicolumn{1}{c}{\textbf{0.76}}     & \multicolumn{1}{c}{\textbf{0.79}}                     &                               \multicolumn{1}{c}{2.16 $\pm$ 0.84}   &                \multicolumn{1}{c}{0.28}          & \multicolumn{1}{c}{0.23}                &             \multicolumn{1}{c}{1.06 $\pm$ 0.11}                     &                               \multicolumn{1}{c}{\textbf{0.76}}  & \multicolumn{1}{c}{\textbf{0.77}}         &                 \multicolumn{1}{c}{0.71 $\pm$ 0.11}                &                  \multicolumn{1}{c}{\textbf{0.67}}      & \multicolumn{1}{c}{\textbf{0.71}}                  \\
700              &                  \multicolumn{1}{c}{1.08 $\pm$ 0.12}                &                               \multicolumn{1}{c}{\textbf{0.82}}  & \multicolumn{1}{c}{\textbf{0.85}}         &              \multicolumn{1}{c}{0.96 $\pm$ 0.11}                    &                \multicolumn{1}{c}{\textbf{0.79}}     & \multicolumn{1}{c}{\textbf{0.81}}                     &                               \multicolumn{1}{c}{3.66 $\pm$ 1.92}   &                \multicolumn{1}{c}{0.30}          & \multicolumn{1}{c}{0.24}                &             \multicolumn{1}{c}{1.31 $\pm$ 0.14}                     &                               \multicolumn{1}{c}{\textbf{0.81}}  & \multicolumn{1}{c}{\textbf{0.86}}         &                 \multicolumn{1}{c}{1.07 $\pm$ 0.17}                 &                  \multicolumn{1}{c}{\textbf{0.71}}      & \multicolumn{1}{c}{\textbf{0.76}}                  \\

 \hline

\end{tabular}}
\tablefoot{We list for each gas tracer (HCN(1--0), HCO$^{+}$(1--0), CO(3--2), CO(1--0), and dust continuum emission) the power-law index ($N$), as well as the Pearson and Spearman coefficients of the log-space representation of the KS law ($\rho_{\rm ps}$ and $\rho_{\rm sp}$, respectively) obtained for a set of seven spatial resolutions ranging from 50 to 700~pc.
Values highlighted in boldface identify correlations that are noteworthy and statistically significant; these are characterised by $p$-values $<$1$\%$ and $\rho_{\rm ps}$, $\rho_{\rm sp}>$0.4 as derived for an equivalent number of pixels.}

\label{tab}
\end{table*}

Table~\ref{tab} lists the power-law indexes, as well as the Pearson correlation ($\rho_{\rm ps}$) and Spearman rank ($\rho_{\rm sp}$) parameters obtained for  the KS laws (in logarithmic space)  for the different spatial resolutions and  tracers used in this work.
We carried out the fits of log($\Sigma_{\rm SFR}$) versus log($\Sigma_{\rm gas}$) using the orthogonal distance regression (ODR) method.
We considered that a correlation is noteworthy and statistically significant when its two-sided p-value < 1$\%$ and both $\rho_{\rm ps}$ and $\rho_{\rm sp}$ are $\geq0.4$. Validated correlations are highlighted in boldface in Table~\ref{tab}. Furthermore, in order to counterbalance the dependence of the estimated $p$-values on the size of the sample, we derived these using a common number of randomly selected points for the different spatial scales. 

As an illustration of the wide variance of SF relations resulting from this analysis, we show in Fig.~\ref{Fig_5} the different versions of the KS law representing $\Sigma_{\rm SFR}$ as a function of $\Sigma_{\rm dense}$ (here derived from HCN) for all the spatial scales  explored in this work, namely from the "initial resolution" (56~pc) up to  700~pc. This figure shows that the correlation becomes looser with higher resolution, and it is hardly visible in the plot with a resolution of 56 pc. The KS relation for the dense gas at the "initial resolution" is highly scattered: $\Sigma_{\rm SFR}$ and $\Sigma_{\rm dense}$ span 1.5~dex each and show scarce evidence of correlation. We fitted a superlinear KS relation with a power-law slope $N=3.74\pm0.22$. The correlation parameters at the "initial resolution" have low values ($\rho_{\rm sp}=0.38$ and $\rho_{\rm ps}=0.42$) and their corresponding two-sided $p$-values $>1\%$. At scales of 100~pc  we obtained a power-law slope $N=2.17\pm0.19$ with correlation parameters of $\rho_{\rm sp}=0.35$ and $\rho_{\rm ps}=0.43$. In contrast, at 400~pc and 700~pc, the scatter in the KS relation is significantly reduced: $\Sigma_{\rm SFR}$ and $\Sigma_{\rm dense}$ span 1~dex each and the derived best-fit KS relation yields a power law index  $N=1.01\pm0.22$ and $N=1.08\pm0.12$,  with correlation parameters $\rho_{\rm sp}=0.43$, $\rho_{\rm ps}=0.45$ and $\rho_{\rm sp}$=0.85, $\rho_{\rm ps}$=0.82, with associated $p$-values $<1\%$, respectively.
 We represent the Kennicutt-Schmidt plots for all the tracers in Figures \ref{Fig_29},  \ref{Fig_30}, \ref{Fig_31} and \ref{Fig_32}.

Figure~\ref{Fig_6} shows how  $\rho_{\rm ps}$ and $\rho_{\rm sp}$ change as a function of the spatial resolution used for the different molecular gas and dust tracers used in this comparison. The grey-shaded region in Fig.~\ref{Fig_6} identifies the range of spatial scales where the correlation is judged not to be statistically significant. For spatial scales $\geq300-400$~pc the correlation improves monotonically as a function of the aperture size for all gas tracers excluding CO(1--0). The $\rho_{\rm ps}$ and $\rho_{\rm sp}$ parameters for CO(1--0) show values $\leq0.3$ for the entire range of spatial resolutions explored. We nevertheless find that for any spatial resolution the correlation parameters  derived from the high density tracers (CO(3--2), HCN(1--0) and HCO$^+$(1--0)) are about a factor of two to three larger than that derived from CO(1--0). Dust continuum emission shows a behaviour similar to that of the rest of high density tracers with the important particularity that the correlation is significant already at 100~pc scales. This result confirms that continuum emission is mostly sensitive to the column densities of dust which is being directly heated by recent SF activity in the SB ring.

The reported breakdown of the KS relations observed in the SB ring below a "critical" spatial scale of $\sim300-400$~pc is in qualitative agreement with the findings of  \citet{Onodera2010}, \citet{Schruba2010} in M~33 and \citet{Kreckel2018} in NGC~628. The existence of a "critical scale" for KS laws can be explained by the diverse evolutionary states of the GMC population, which can be singled out in high spatial resolution observations. The exact value of this "critical scale" may change from galaxy to galaxy. It may also depend on the criterion adopted to choose the data points used to generate the scatter plots: either a "blind" pixel-wise Nyquist sampling or a "biased" selection of apertures centred either around SF or gas emission peaks \citep[e.g., see discussion in ][]{Williams2018}. Different GMC states can reflect an ordered  time sequence of star formation determined by the large-scale dynamics in galaxy disks or a more stochastic pattern due to the local dispersal of molecular gas by stellar feedback.

Figure~\ref{Fig_7} shows how the power-law index $N$ obtained from the fit to the KS relation changes as a function of the spatial resolution  for the different molecular gas and dust tracers. 
As in Fig.~\ref{Fig_6}, the grey-shaded region in Fig.~\ref{Fig_7} identifies the range of spatial scales where the correlation is not statistically significant.
Overall, we find a strong scale dependence of $N$. The power-law index for HCN and HCO$^+$ shows a fairly systematic decrease with the spatial resolution from $N=3.5\pm0.3$ (at the "initial resolution") to $N=1.0\pm0.1$ (at 700~pc). The value of $N$ stays around $\sim1.0\pm0.1$ for the whole range of spatial scales where the correlation is significant ($\geq300-400$~pc). For CO(3--2) $N$ shows values marginally below unity in the 300~pc-500~pc range. Similarly, The value of $N$ for the dust continuum indicates a sublinear relation ($N<1$) within the range 100~pc--600~pc. The power-law becomes nevertheless linear at 700~pc as for most of the high density tracers. 
 The reported slightly different behaviour of CO(3--2) and dust continuum relative to HCN or HCO$^+$ can be attributed to the fact that, although all these tracers are sensitive to the presence of dense 
 molecular gas, CO(3--2) and dust continuum  are also to a large extent mostly sensitive to the presence
of comparatively hotter molecular gas, characterised by high kinetic temperatures \footnote{\citet{Tsai2012} studied the KS law in NGC 1068 using CO(3-2) and the FIR luminosity and also found a sublinear power-law for the L$_{\rm CO(3-2)}$-L$_{\rm FIR}$ relation  with $N\sim0.50$ at scales of $\sim300$~pc.}.

The linear behaviour of the power-law observed for all the dense gas tracers in NGC~1068 is in agreement with the results obtained in other galaxies \citep[e.g.,][]{GaoSolomon2004a, GaoSolomon2004b, Gracia-carpio2008, Wu2010, Burillo2012, Usero2015, Liu15, Chen2017, Williams2018, Querejeta2019}. 
The weak correlation found between Pa$\alpha$ and CO(1-0) at all scales indicates that the distribution of the general molecular gas traced by CO(1--0)  is not strongly correlated with the current location of recent star formation in the SB ring.  While we do not expect to have missed a high fraction of the CO(1--0) flux in the PdBI map on scales $\leq200-300$~pc (see Sect.~\ref{ancillary} and Appendix~\ref{app1}), the likely increasing fraction of flux filtered on scales close to the $\sim1$~kpc limit could be an explanation for the poor correlation shown by CO(1--0) in Fig.~\ref{Fig_6}.


%

\begin{table}[tbp]
\centering
\caption{Published work on SF relations}
\resizebox{\linewidth}{!}{
\begin{tabular}{llll}
\hline
\textbf{Reference}              & \textbf{Galaxies}    & \textbf{SFR
tracer}             & \textbf{Resolution (kpc)} \\ \hline
Garcia-Burillo+12 & SFG \& (U)LIRG & FIR             &
1.7-3.6             \\
Murphy+15         & NGC 3627                 & 33 GHz         &  0.3
                \\
Usero+15          & galaxy disks             & TIR             &
0.5-3.3             \\
Bigiel+16         & M51                      & TIR             & 1.1
                 \\
Chen+17           & M51                      & TIR             & 0.2
                 \\
Viaene+18         & M31                      & UV + 24 $\mu$m & 0.1
                \\
Querejeta+19      & M51                      & 33 GHz         & 0.1
                \\ \hline
\end{tabular}}

\tablefoot {List of references studying extragalactic SF relations of the dense molecular gas
traced by the HCN(1--0) line and using different SF tracers for a range of
spatial scales  ($\simeq0.1-3.6$~kpc).}
\label{tabinfinite}
\end{table}



\begin{figure*}[hbtp]
  \centering
   \includegraphics[width=18.75cm]{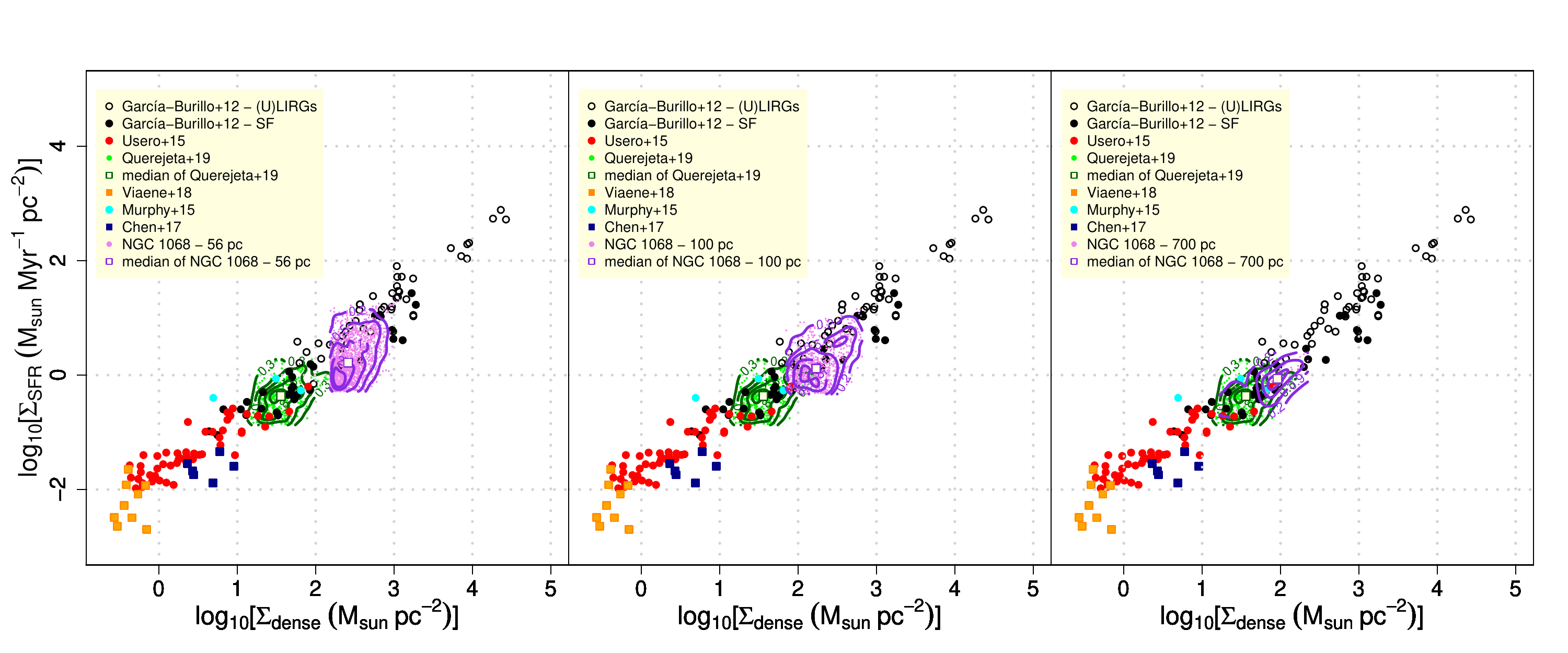}
   \caption{Comparison of the KS laws derived from HCN(1--0) in NGC\,1068 with those obtained in different populations of galaxies for a range of spatial scales: global ($\simeq$a few kpc) measurements of the disks of SF galaxies and (U)LIRGs \citep{Burillo2012}, $\simeq$0.3-1~kpc-scale regions of the disks of nearby SF galaxies \citep{Murphy2015, Usero2015, Bigiel2016}, and 0.1-0.2~kpc-size regions in the disk of M\,51\citep{Chen2017, Viaene2018, Querejeta2019}.  For the sake of a better comparison with the reference sample, the scatter-plots for NGC\,1068  (magenta circles and isodensity contours) are displayed for three spatial resolutions: 56~pc ({\it left panel}),  100~pc ({\it middle panel}), and  700~pc ({\it right panel}). The scatter-plot for M\,51 at 100~pc resolution is displayed by the green cricles and isodensity contours. The open squares stand for the median values of the distributions in 
   NGC\,1068 and M\,51.}
              \label{Fig_8}
\end{figure*}


\subsection{The Kennicutt-Schmidt laws of NGC~1068 in context} \label{context}

We compare in this section the KS relations derived from HCN(1--0) in the SB ring of NGC~1068 with those obtained by previous works in different populations of galaxies. Table \ref{tabinfinite} lists the references used in this  comparison. These works comprise galaxy-scale studies of SF galaxies (SFG) and (U)LIRG \citep[e.g., see compilation by][and references therein]{Burillo2012}, as well as
spatially resolved studies of individual galaxies  \citep{Usero2015, Murphy2015, Bigiel2016, Chen2017, Viaene2018, Querejeta2019}.
These references use the equations of Section~\ref{parameters} to convert HCN(1--0) intensities into (deprojected) surface densities of dense gas and adopt the same HCN conversion factor used in this paper. However, the SF tracers chosen in these works are different, as detailed in Table~\ref{tabinfinite}.

Figure~\ref{Fig_8} compares the different versions of the KS law derived from HCN(1--0) in NGC\,1068 at three spatial resolutions (56~pc, 100~pc, and 700~pc) with those obtained in the references listed in Table \ref{tabinfinite}.
NGC~1068 data lie within the linear power-law branch occupied by the rest of the galaxies shown in Fig.~\ref{Fig_8}. Galactic dense cores, at sub-pc scale, align along the same relationship \citep{Wu2005,Wu2010, Rosolowsky2011, Stephens2016, Shimajiri2017}. Taken at face value, this suggests that the star formation efficiency (SFR per unit dense molecular
mass) is  nearly constant on average. As expected, the internal scatter in the distribution NGC1068 data
points is lower as we move to larger apertures. Furthermore, NGC~1068 data points shift toward lower surface densities within the power-law as we move to larger apertures.

Compared at a common scale of 100~pc, the SB ring of NGC 1068 appears as a more extreme environment relative to the SF regions of M51 studied by \citet{Querejeta2019}. As illustrated in Fig.~\ref{Fig_8}, the median values of the distributions of $\Sigma_{\rm SFR}$ and $\Sigma_{\rm dense}$ are about a factor of three to five higher in NGC~1068: $\Sigma_{\rm SFR}${\small[NGC1068]}~$\simeq1.5~M_{\sun}$~Myr$^{-1}$pc$^{-2}\simeq3\times\Sigma_{\rm SFR}${\small [M51]} and
$\Sigma_{\rm dense}${\small [NGC1068]}~$\simeq174~M_{\sun}$~pc$^{-2}\simeq5\times\Sigma_{\rm dense}${\small [M51]}. Furthermore, when examined at scales of 700~pc, NGC\,1068 occupies in the KS plot a position intermediate between that of normal galaxies and (U)LIRGs, a result that hints at the relatively extreme conditions in the SB ring.

\subsection{Environmental dependence of the star formation efficiency of the dense gas}\label{SFE-deg}

The observed linear relation between $\Sigma_{\rm SFR}$ and $\Sigma_{\rm dense}$ shown in Fig.~\ref{Fig_8} has been considered as evidence of the validity of  density-threshold models of SF. For these models the SF 
efficiency of dense molecular gas, defined as SFE$_{\rm dense}$=$\Sigma_{\rm SFR}$/$\Sigma_{\rm dense}$ or its inverse, which represents  the depletion time of the dense gas ($T_{\rm dep}^{\rm dense}$=SFE$_{\rm dense}^{-1}$), are about constant in different populations of galaxies and also for different dynamical environments within galaxies  \citep{GaoSolomon2004b, Wu2005, Lada2010, Lada2012, Evans2014}.  In this section we use the NGC~1068 data to explore the existence of an environmental dependence of SFE$_{\rm dense}$ inside the SB ring. 

\begin{figure*}[htp]
   \centering
    \includegraphics[width=.98\linewidth]{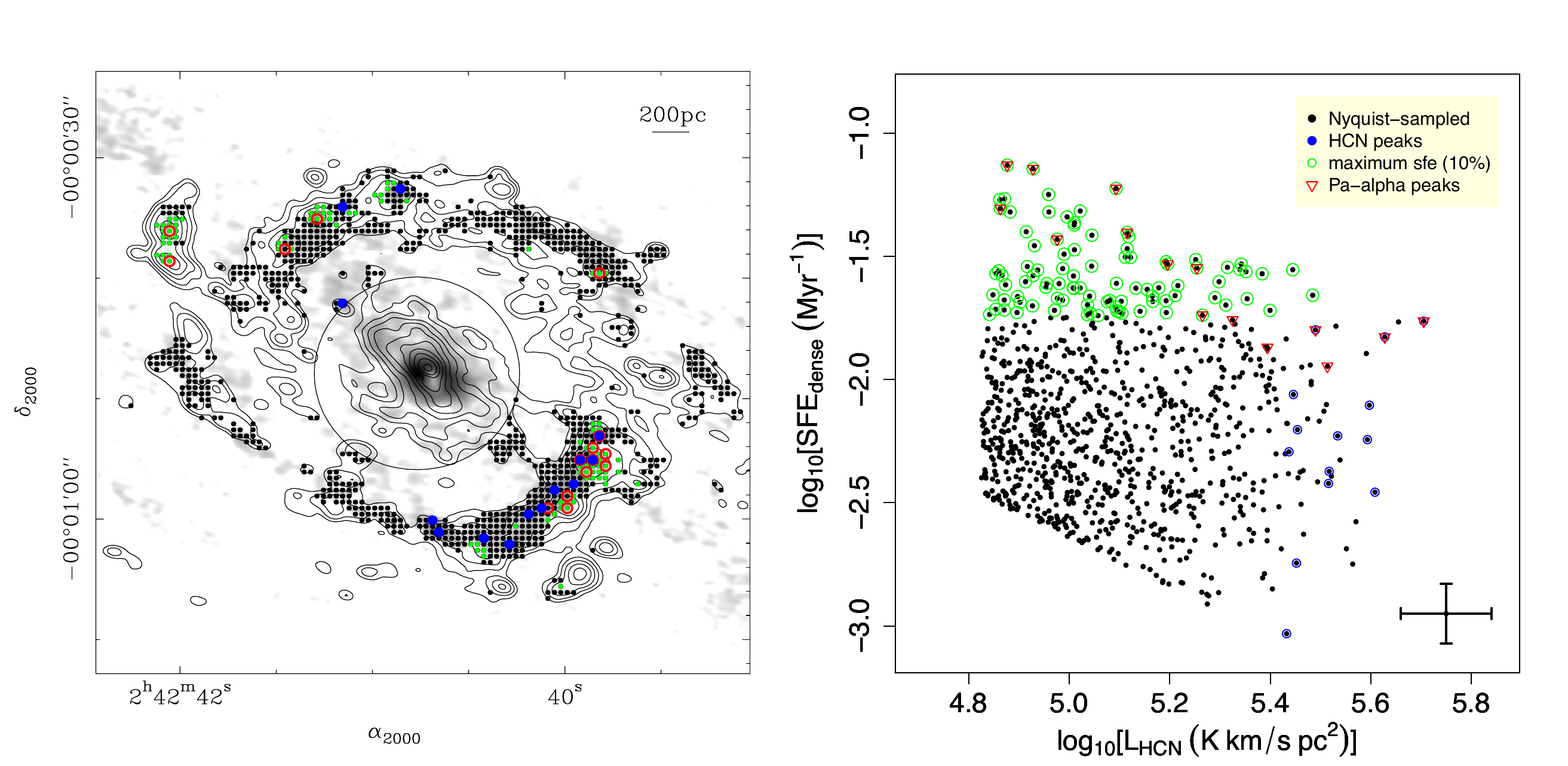}

   \caption{{\it Left panel}: overlay of the HST/NICMOS Pa$\alpha$  image (contours) on the HCN(1--0) map (grey scale). Contours and grey scale as in Fig.~\ref{Fig_4}.
 The set of  56~pc-size circular apertures used to extract the fluxes of  Pa$\alpha$ and HCN(1--0) are colour-coded differently, depending on whether they are centred on HCN peaks (in blue colour), on Pa$\alpha$ peaks (in red colour), or on  the $10\%$ of the apertures showing the highest values of SFE$_{\rm dense}$ (in green colour). The black dots identify the centres of all the Nyquist-sampled 56~pc-size apertures where  there are reliable estimates of SFE$_{\rm dense}$ across the SB ring. {\it Right panel}: distribution of SFE$_{\rm dense}$ values (in Myr$^{-1}$-units) as a function of L$_{\rm HCN}$ (in K km$^{-1}$pc$^{2}$-units) in the SB ring of NGC\,1068. Symbols as in {\it left panel}.  Vertical and horizontal errorbars at the lower right corner of the panel account for the typical uncertainties, which amount to $\pm0.13$~dex and $\pm0.09$~dex, respectively.}
              \label{Fig_9}
   \end{figure*}

 Figure~\ref{Fig_9} (left panel) overlays the HST/NICMOS Pa$\alpha$  image  on the HCN(1--0) map of NGC~1068 obtained at the  "initial resolution" of 56~pc. We identify the positions  used to extract the fluxes of  Pa$\alpha$ and HCN(1--0) over the SB ring
  from the Nyquist-sampled grid \footnote{The black dots in Fig.~\ref{Fig_9} single out the grid positions where the fluxes of HCN and Pa$\alpha$ are both $>3\sigma$.}. SFE$_{\rm dense}$ values  span almost 1.5~dex and show a highly scattered 
 distribution  as a function of $L'_{\rm HCN}$ around an "apparently" constant mean value of about 0.01~Myr$^{-1}$, equivalent to a $T_{\rm dep}^{\rm dense}\sim100$~Myr (see right panel of  Fig.~\ref{Fig_9}). This result is similar to the findings 
 of \citet{Gallagher2018}  and \citet{Querejeta2019} obtained from their high spatial resolution images of a sample of galaxies. Notwithstanding that we might attribute part of the scatter in the SFE$_{\rm dense}$--$L'_{\rm HCN}$ plot to possible small-scale variations of the $\alpha_{\rm HCN}$ conversion factor, any plausible range for these potential variations would nevertheless fall short of  accounting for the bulk of the 1.5~dex span shown in Fig.~\ref{Fig_9}.

 Although Figs.~\ref{Fig_8} and ~\ref{Fig_9} indicate that there is an overall relationship between the HCN luminosity and recent star formation in the SB ring,  we explore below the existence of systematic trends in SFE
 $_{\rm dense}$. With this aim, we selected within our initial grid a number of non-overlapping 56~pc-size apertures centred on local maxima either in the HCN or  
the Pa$\alpha$ maps, following the same procedure used by \citet{Querejeta2019} in their analysis of M~51 data. The local maxima were identified as the pixels with peak intensities within circular regions ("clumps"), obtained from successive cuts in the maps. We therefore started from a high threshold (a large multiple of the noise level) and iteratively explored lower values from isolated circular regions. To collect a similar number of points selected  from HCN and Pa$\alpha$ peaks we continued to identify new clumps down to  different threshold values of 12$\sigma$ and  44$\sigma$, respectively. We use different colours to identify in the two panels of Fig.~\ref{Fig_9} the apertures centred on HCN peaks (blue colour), on Pa$\alpha$ peaks  (red colour), and also on the $10\%$ of the apertures showing the highest values of SFE$_{\rm dense}$ (green colour).

 Figure~\ref{Fig_9} shows that the apertures showing the top $10\%$   SFE$_{\rm dense}$ values are not uniformly distributed throughout the SB ring.  High SFE$_{\rm dense}$ values are preferentially located at the southwest and northeast sections of the ring where the latter is connected  to the ends of the stellar bar. We also identify high SFE$_{\rm dense}$ values further out at  the northeast extreme of the SB ring. As expected, the clumps with the strongest 
  Pa$\alpha$ emission are also located  in the regions with the highest SFE$_{\rm dense}$. Furthermore, although the overall distribution of the highest (dense) gas column density apertures also corresponds to the ends of the stellar bar, on small-scales there is no one-to-one correspondence between Pa$\alpha$ and the HCN peaks.  This reflects the significant variance in the evolutionary states of the dense gas clumps in these regions. In particular, the gas ridge traced by HCN maxima tend to appear "upstream" relative to the Pa$\alpha$ maxima at the southwest section of the SB ring \footnote{We can assign an "upstream" location of HCN relative to Pa$\alpha$ after assuming that the sense of rotation of the gas in the disk is counterclockwise \citep[e.g.,][]{Burillo2014}.}.

\begin{figure*}[htp]
   \centering
    \includegraphics[width=.98\linewidth]{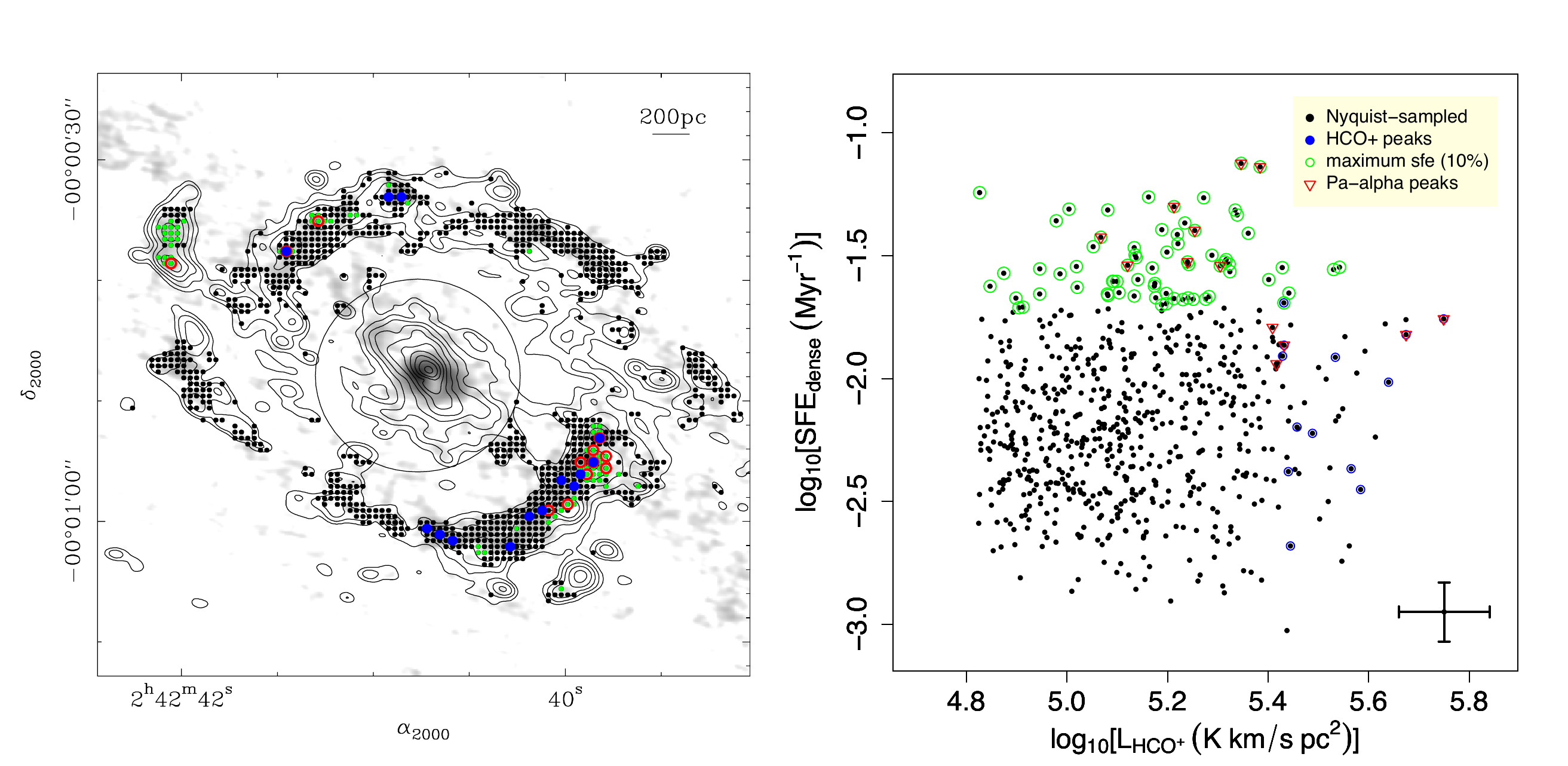}

   \caption{Same as Fig.~\ref{Fig_9} but replacing  HCN(1--0) by HCO$^+$(1--0)  along the x-axis of the {\it right panel}.}
              \label{Fig_10}
   \end{figure*}

In order to minimize the  bias  introduced due to the presence of HCN in the two axes of the right panel of Fig.~\ref{Fig_9}, we replaced HCN by HCO$^+$ luminosities along the x-axis by  selecting from our grid of points those positions satisfying  $L'_{\rm HCO^+}>3\sigma$.  Figure~\ref{Fig_10} shows the results obtained following the same steps leading to Fig.~\ref{Fig_9}.  As expected, Fig.~\ref{Fig_10} eliminates the slight anti-correlation trend identified in the SFE$_{\rm dense}$-$L'_{\rm HCN}$ plot of Fig.~\ref{Fig_9}, but the main results described above remain virtually unchanged.


\section{Star formation efficiency and dense gas fraction} \label{Trends}

The overall SF efficiency of molecular gas (SFE$_{\rm mol}$) can be expressed as the product of  SFE$_{\rm dense}$ and the dense gas fraction ($F_{\rm dense}$), namely: SFE$_{\rm mol} \equiv$ SFE$_{\rm dense} \times F_{\rm dense}$.
 There is mounting  evidence supporting the existence of significant variations in  SFE$_{\rm dense}$ and $F_{\rm dense}$ as a function of the galactic environment, based on observations of molecular clouds in the centre of our Galaxy  
 \citep{Lon13, Kru14} and in nearby galaxies for a range of spatial scales \citep{Usero2015, Bigiel2015, Bigiel2016, Gallagher2018, Querejeta2019, Jim19, Beslic21}. In the following sections we study the trends in   SFE$_{\rm dense}$ and $F_{\rm dense}$ as a function of a number of physical variables with the aim of resolving the degeneracy of SF laws in the SB ring of NGC~1068. Table~\ref{tabx} lists the  Spearman rank parameters for the different combinations of variables and spatial scales explored below.

\begin{table*}[htp]

\caption{Spearman rank parameters for different scaling relations.}

\centering
\resizebox{18cm}{!} {
\begin{tabular}{lccccccc}
\hline
& \multicolumn{1}{c}{\textbf{SFE$_{\rm dense}$} \textbf{vs.} \textbf{$\Sigma_{\rm star}$}} & \multicolumn{1}{c}{\textbf{F$_{\rm dense}$} \textbf{vs.} \textbf{$\Sigma_{\rm star}$}} & \multicolumn{1}{c}{\textbf{SFE$_{\rm dense}$} \textbf{vs.} \textbf{$\sigma_{\rm HCN}$}} & \multicolumn{1}{c}{\textbf{F$_{\rm dense}$} \textbf{vs.} \textbf{$\sigma_{\rm HCN}$}} & \multicolumn{1}{c}{\textbf{$\Sigma_{\rm SFR}$} vs. \textbf{F$_{\rm dense}$}} & \multicolumn{1}{c}{\textbf{T$_{\rm dep}$} \textbf{vs.} \textbf{b$_{\rm HCN}$}} & \multicolumn{1}{c}{\textbf{SFE$_{\rm dense}$} \textbf{vs.} \textbf{F$_{\rm dense}$}}\\
\hline
56 pc    & 0.10 &  --  & -0.22 &    --  &   --     &   -0.19  &  -- \\
100 pc  & 0.15 & \textbf{0.40}  & -0.28  & 0.10  & 0.22  &  \textbf{-0.43}  &  0.03 \\
400 pc   & \textbf{0.59}  & \textbf{0.67} & -0.06  & \textbf{0.50}  & \textbf{0.62} & \textbf{-0.66}   & 0.34 \\
 \hline                  
\end{tabular}}
\tablefoot{We list the Spearman rank parameters ($\rho_{\rm sp}$) for the different scaling relations and spatial scales studied in Sect.~\ref{Trends}. As in Table~\ref{tab}, values highlighted in boldface identify correlations that are statistically significant; these are characterised by $p$-values $<$1$\%$ and $\mid\rho_{\rm sp}\mid>$0.4}
\label{tabx}
\end{table*}


\begin{figure*}[h]
   \centering
    	\includegraphics[width=.61\linewidth]{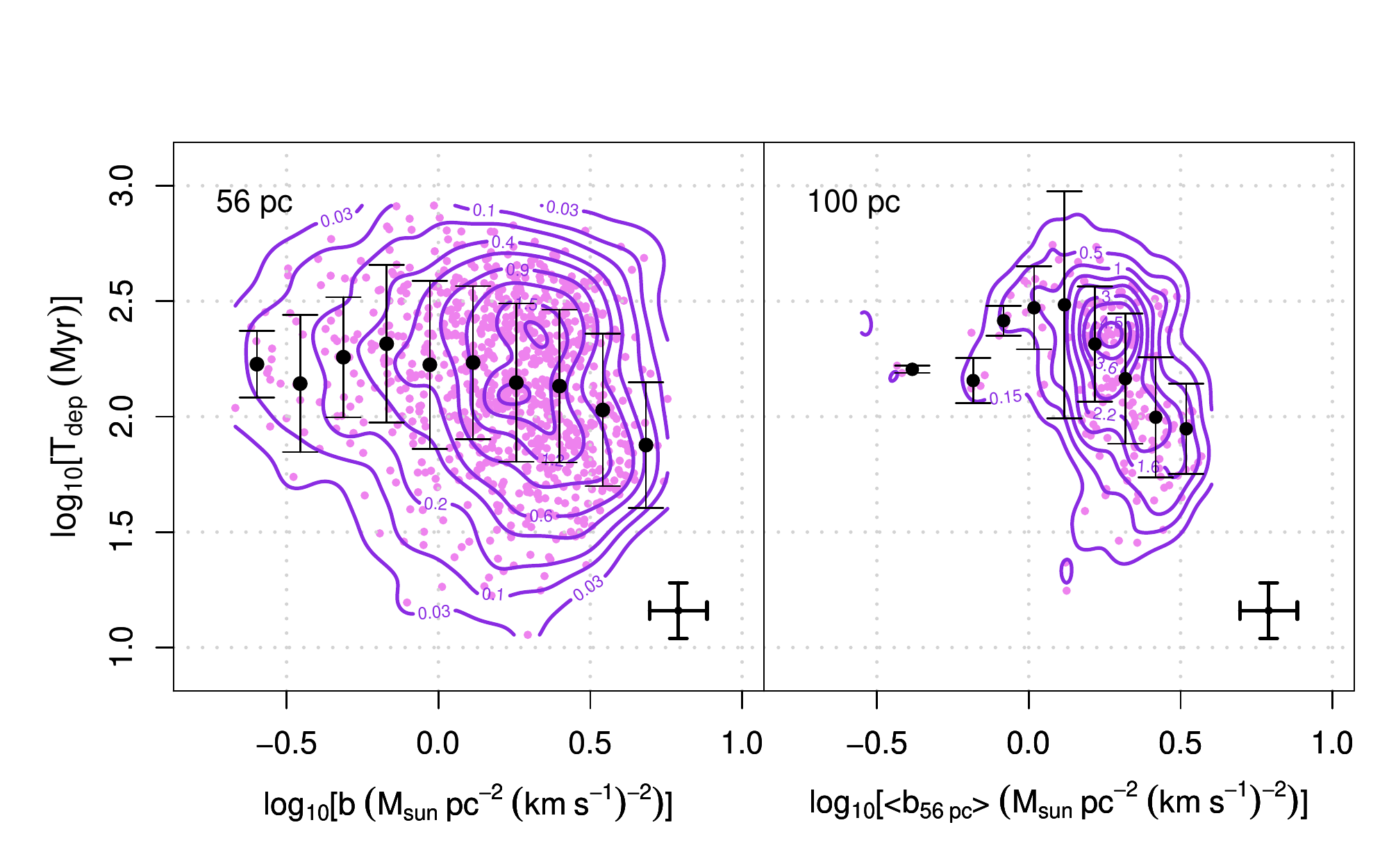}
    	\includegraphics[width=.38\linewidth]{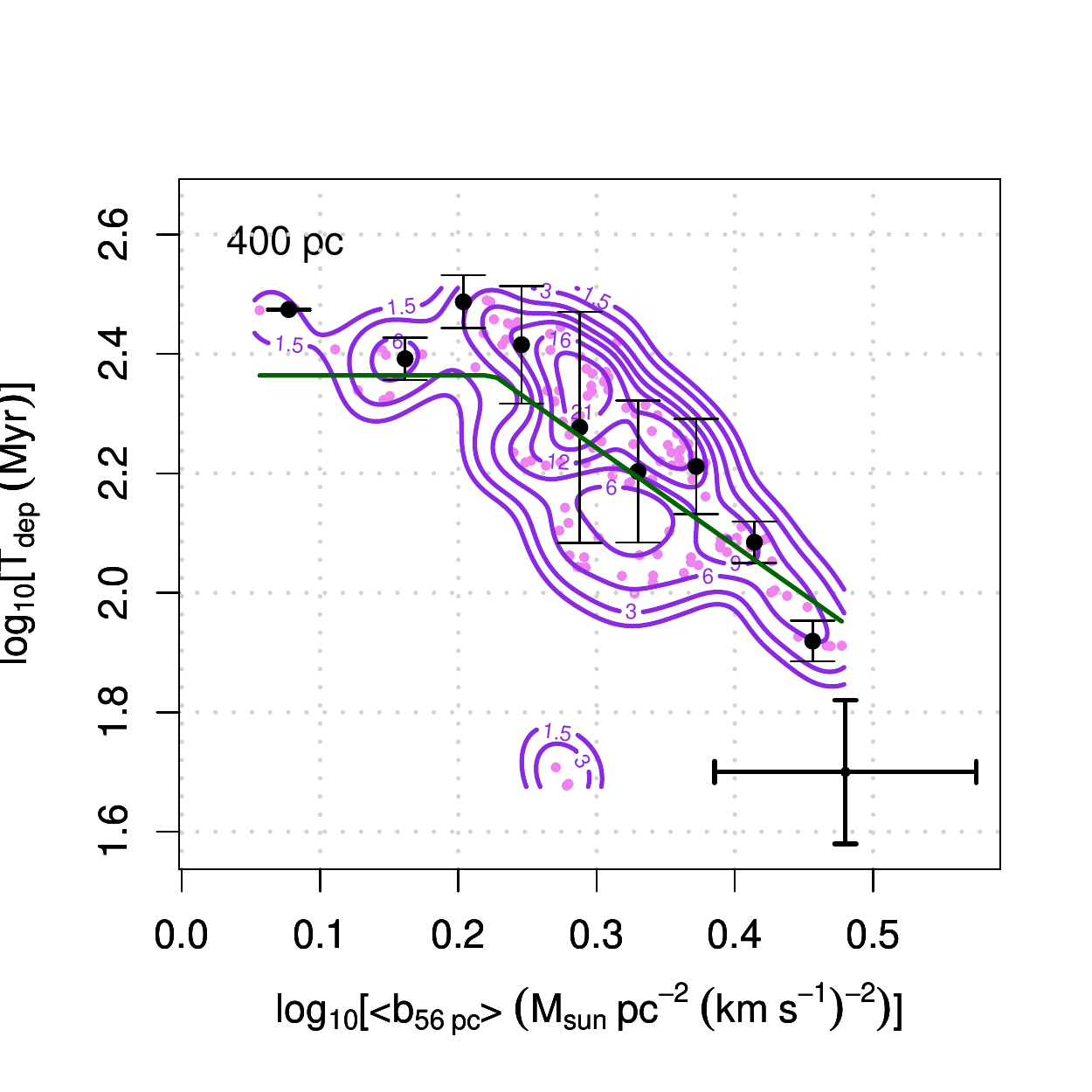}
   \caption{Depletion time of the dense molecular gas estimated from HCN, $T_{\rm dep}^{\rm dense}$ $\equiv$ $\Sigma_{\rm dense}$/$\Sigma_{\rm SFR}$, as a function of the self-gravity of the gas, measured by the parameter  \textit{b} $\equiv$ $\Sigma_{\rm dense}$/$\sigma^{2}$,  at different spatial scales (magenta circles and isodensity contours): 56~pc, namely the "initial resolution" of ALMA observations ({\it left panel}), 100~pc ({\it middle panel}), and 400~pc ({\it 
   right panel}). A significant trend in  $T_{\rm dep}^{\rm dense}$ as a function of $b$, identified in the middle and right panels, indicates a higher rate of star formation per unit mass of dense gas for regions with stronger self-gravity. 
 Values of $\langle b \rangle$ for the 100~pc and 400~pc apertures were derived from an intensity--weighted average of $b$  evaluated at the Nyquist-
   sampled grid of points for each aperture.  Values of  $T_{\rm dep}^{\rm dense}$  were derived using a Gaussian-weighted average. Black circles show median  $T_{\rm dep}^{\rm dense}$, and errorbars indicate the rms scatter in the 
   bins. The green lines at the right panel identify the two regimes in the  $T_{\rm dep}^{\rm dense}$-$\langle b \rangle$ plot found by the  MARS algorithm to fit the data using 400~pc as averaging scale. Vertical and horizontal errorbars at the lower right corner of each panel account for the typical uncertainties, which amount to $\pm0.13$~dex and $\pm0.09$~dex, respectively.}
              \label{Fig_11}
   \end{figure*}



 \begin{figure*}[htp]
   \centering
 \includegraphics[width=.495\textwidth]{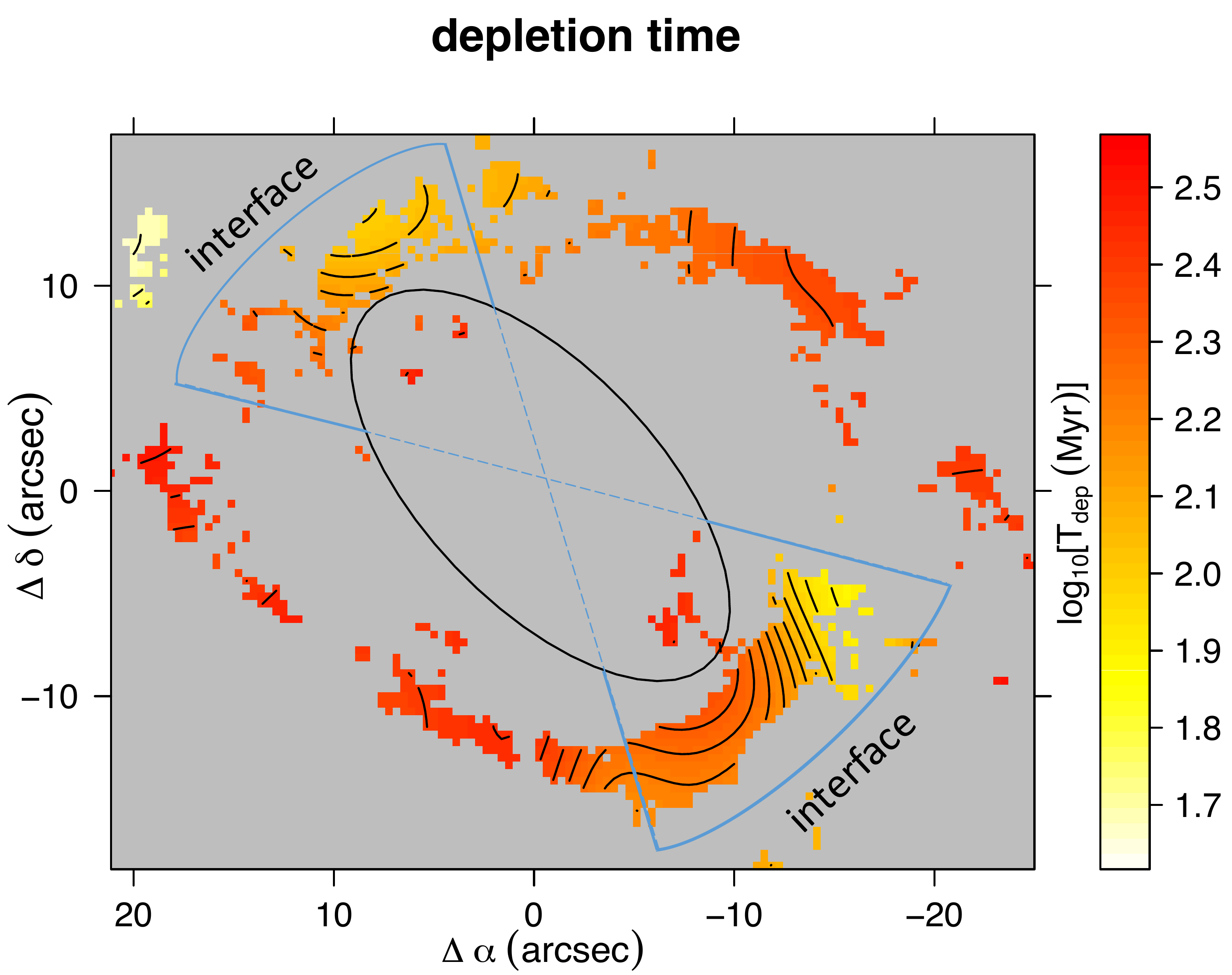}
\includegraphics[width=.495\textwidth]{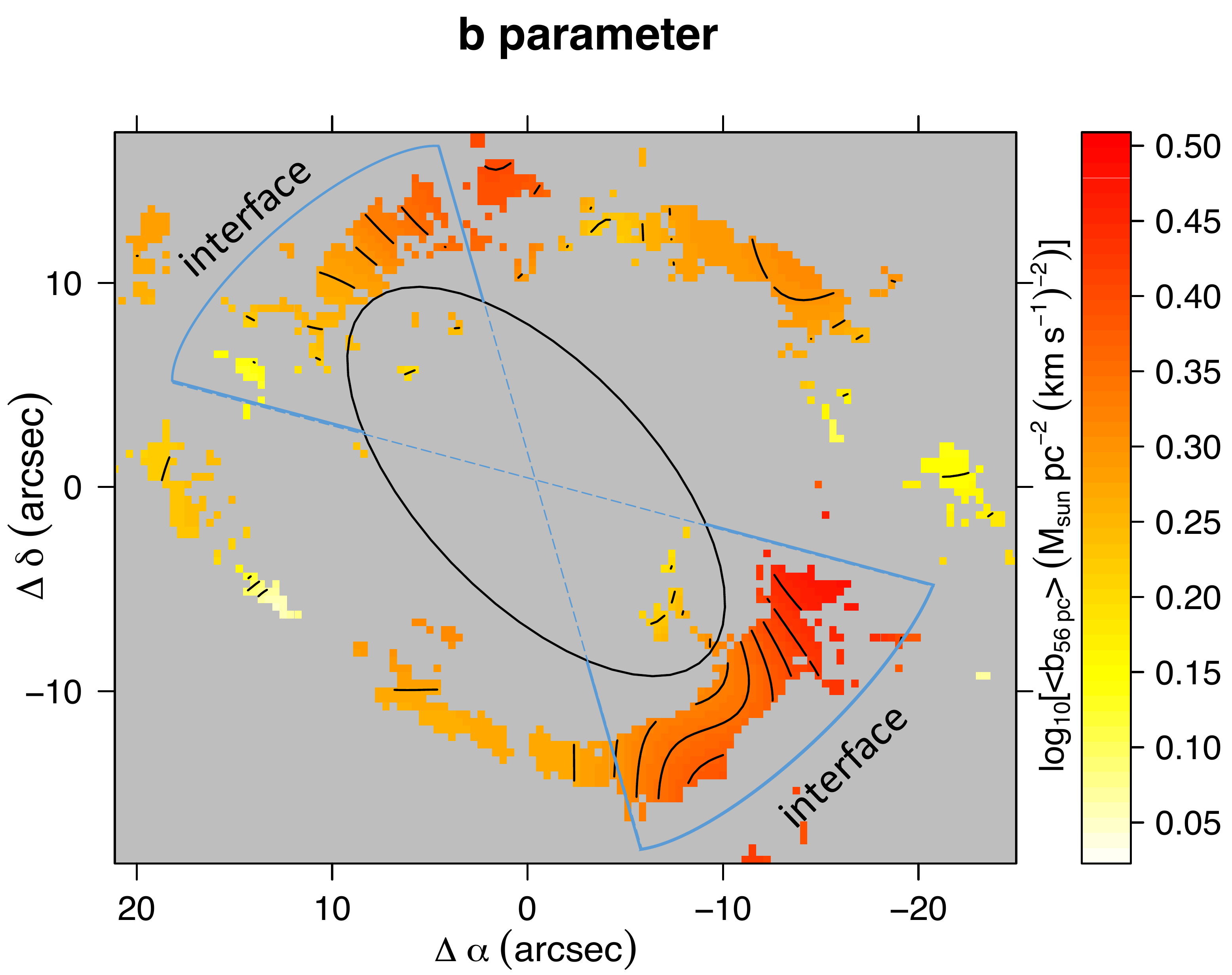}
     
   \caption{{\it Left panel:}~map of the depletion time of the dense molecular gas ( $T_{\rm dep}^{\rm dense}$) obtained after using 400~pc as averaging scale. Contour levels for log$_{\rm 10}$( $T_{\rm dep}^{\rm dense}$) go from 1.50 to 2.22 in steps of 0.045~(in Myr). {\it Right panel:}~map of the intensity-weighted average of the $b$ parameter ($\langle b \rangle$) in the SB ring of NGC\,1068 obtained after using 400~pc as averaging scale. Contour levels for log$_{\rm 10}$($\langle b\rangle$) go from 0.075 to 0.45 in steps of 0.025 (in $M_{\sun}$~pc$^{-2}$~(km~s$^{-1}$)$^{-2}$). $\Delta\alpha$  and $\Delta\delta$ offsets in arc seconds are relative to the phase tracking centre of ALMA. The ellipse identifies the extent of the nuclear stellar bar in the disk. We highlight the approximate extent of the bar-ring interface region in both panels.}
              \label{Fig_12}
   \end{figure*}


\subsection{Trends as a function of the boundedness of the gas}\label{Tdep-b}

In this section we use an alternative prescription for the SF relations of the SB ring, which includes explicitly the 
dependence of SFE$_{\rm dense}$ on a
combination of $\Sigma_{\rm dense}$ and the gas velocity dispersion ($\sigma$), in an attempt to resolve the degeneracy associated with the 
scatter in the SFE$_{\rm dense}$-$L'$(\rm HCN) plot of Fig.~\ref{Fig_9}. This approach was first used  by \citet{Leroy2017} in their analysis of SF relations in M~51 and also adopted by \citet{Kreckel2018} in a similar study
carried out in NGC~628. \citet{Leroy2017} and \citet{Kreckel2018} used CO(1--0) and CO(2--1) to trace  the bulk of the molecular gas on 40-50~pc-scales in M~51 and NGC~628, respectively. In the following analysis we use HCN(1--0) to specifically trace the dense molecular gas in the SB ring of NGC~1068 on spatial scales ($\sim56$~pc) comparable to those explored by \citet{Leroy2017} and \citet{Kreckel2018} \footnote{\citet{Querejeta2019}  applied the methodology of \citet{Leroy2017} to the HCN(1--0) data of M~51, yet at a spatial resolution of $\sim100$~pc,  that is,  about a factor of two lower than the one used in our work.}.

The degree of self-gravity or boundedness of molecular gas clouds  determines to a large extent their ability at forming stars. The virial parameter, defined as $\alpha_{\rm vir} \approx$ 2KE/UE, captures the balance of gravitational potential ($UE$) and kinetic energy ($KE$) and is commonly used in turbulent models of SF as a predictor of the efficiency of star formation \citep{Krumholz2005, padoan2012, padoan2017}.

 We define similarly to \citet{Leroy2017} the boundedness parameter, here particularized for the dense molecular gas phase traced by HCN, as $b_{\rm 56pc} \equiv \Sigma_{\rm dense}/\sigma^{2} \propto UE/KE \propto \alpha^{-1}_{\rm vir}$, where $\sigma$ is the velocity dispersion and $\Sigma_{\rm dense}$ is the column density of dense gas measured both at the "initial resolution" of 56~pc. We derived intensity-weighted averages of $b_{\rm 56pc}$ using two different apertures 
$\Delta A=100$~pc and $400$~pc over the Nyquist sampled grid, defined as follows:

  	\begin{equation}
		\langle b \rangle_{\rm \Delta A}(x_{0},y_{0})=\dfrac{\Sigma w(x,y) I_{\rm 56pc}(x,y) b_{\rm 56pc}(x,y)}{\Sigma w(x,y) I_{\rm 56pc}(x,y)}.
	\label{Eq8}		
	\end{equation} 	
The Gaussian weight $w(x,y)$ is defined as:

	\begin{equation}
		 w(x,y)=exp\left( \frac{-(\theta (x,y,x_{0},y_{0}))^{2}}{2 \sigma^{2}_{\rm \Delta A}} \right),
	\label{Eq9}	 
	\end{equation}

where $\theta (x,y,x_{0},y_{0})$ is the angular distance from the measurement point $(x_{0},y_{0})$, $\sigma_{\rm \Delta A}$ is the $\sigma$-width of the Gaussian averaging beam, and $\Delta A$ corresponds to the adopted averaging scale (100~pc and 400~pc in our case). In Eq.~\ref{Eq8}, $b_{\rm 56pc}$ and $I_{\rm 56pc}$ are the boundedness parameter and the HCN integrated intensity, respectively, measured at a generic position $(x,y)$ of the grid \footnote{\citet{Leroy2017} used a slightly different definition of the intensity-weighted average of the $b$ parameter: $\langle b \rangle \equiv \langle \Sigma_{\rm gas} \rangle /  \langle \sigma^{2} \rangle$. With this definition the averaging is performed separately for 
the numerator and the denominator of the $b$-parameter. As shown in Appendix~\ref{AppC}, the trends and statistical parameters derived following this definition are virtually identical to the ones obtained in Sect.~\ref{Tdep-b}.}.

Figure~\ref{Fig_11} shows how the depletion time of the dense molecular gas estimated from HCN, $T_{\rm dep}^{\rm dense}$ $\equiv$ SFE$_{\rm dense}^{-1}$, changes as a function of the self-gravity of the gas, measured by the $b$ parameter,  at three spatial scales:  56~pc (the "initial resolution"), 100~pc, and 400~pc. The values of $\langle b\rangle_{\rm \Delta A}$ for the 100~pc and 400~pc apertures were derived from Eq.~\ref{Eq8}. Furthermore, similarly to the rest of the physical parameters  analysed in Sects.~\ref{SFE-F} to \ref{SFE-sigma}, the average value of $T_{\rm dep}^{\rm dense}$ inside the two values of $\Delta A$ were derived using the Gaussian weighting function of Eq.~\ref{Eq9}.

Figure~\ref{Fig_11} shows  a significant monotonic decrease of  $T_{\rm dep}^{\rm dense}$ as function of  $\langle b\rangle_{\rm \Delta A}$ when we represent these parameters averaged over the two apertures. Specifically, we obtain (anti) correlation Spearman rank parameters $\rho_{\rm sp}=-0.43$ and $-0.66$ for  $\Delta A=100$~pc and $400$~pc, respectively, and associated two-sided $p$-values $<1\%$. Overall, this is indicative of a higher rate of star formation per unit mass of dense gas (lower $T_{\rm dep}^{\rm dense}$) for regions characterised by a stronger self-gravity (higher $b$ or lower $\alpha_{\rm vir}$ values).

The anti-correlation reported above is more pronounced in  the intensity-weighted version of the plot derived at 400~pc scales.  We identify a turnover in the $T_{\rm dep}-b$ scatter plot for  $\Delta A=400$~pc located around log$_{\rm 10}(\langle b \rangle) \simeq~0.2~M_{\sun}$pc$^{-2}$(km s$^{-1}$)$^{-2}$, as shown in the right panel of Fig.~\ref{Fig_11}. The observed change of tendency around this point defines two regimes in the  $T_{\rm dep}-b$ parameter space. To quantify the two-regime trend and the turnover,  we used the Multivariate Adaptive 
Regression Splines ({\tt MARS}) fit routine from the {\tt 
Rstudio} package\footnote{The {\tt MARS} algorithm creates a collection of so-called basis functions. In this procedure, the range of predictor values is partitioned in several groups. For each group, a separate linear regression is modelled, each 
with its own slope.}. We performed the {\tt MARS} fit on 100 realizations of the $T_{\rm dep}$ vs. $b$ relation at $\Delta A=400$~pc taking into account the data uncertainties.
From this Monte Carlo simulation, we obtained a turnover at log$_{\rm 10}(\langle b \rangle) =~0.23\pm0.02~M_{\sun}$pc$^{-2}$(km s$^{-1}$)$^{-2}$ which is indicated in the right panel of Fig.~\ref{Fig_11}.
Furthermore,  for $\langle b \rangle$ values below the turnover the trend is  approximately flat (slope = 0.02 $\pm$ 0.11). In contrast, the   {\tt MARS} routine fits a slope = -1.59 $\pm$ 0.15   
beyond the turnover. This slope is larger than the one obtained  by \citet{Leroy2017} ($\simeq-0.9$) in their analysis of the M~51 CO(1--0) data, derived using the same averaging spatial scales (400~pc). \citet{Kreckel2018} found nevertheless no
significant correlation in their analysis of the CO(2--1) data of NGC~628, which used an averaging  scale of 500~pc. Compared to the shallower or inexistent trends identified in M~51 and NGC~628, the steeper decline of $T_{\rm dep}^{\rm dense}$  with $\langle b \rangle$  in the SB ring of NGC~1068  reflects the tighter link between  the boundedness of dense molecular gas and star formation efficiency .

Figure~\ref{Fig_12} shows the spatial distribution  of $T_{\rm dep}^{\rm dense}$ and   $\langle b \rangle$ derived for $\Delta A=400$~pc in the SB ring of NGC~1068.  The   $T_{\rm dep}^{\rm dense}$ map of Fig.~\ref{Fig_12} confirms the picture drawn from the analysis of Sect.~\ref{SFE-deg}: the regions showing comparatively higher (lower) SFE$_{\rm dense}$ ($T_{\rm dep}^{\rm dense}$) values are preferentially located closer to the region where the SB ring is connected  to the ends of the stellar bar around  PA~$\sim 15^{\circ}-75^{\circ} (\pm 180^{\circ}$). We also see a similar spatial segregation in the  $\langle b \rangle$ map, which shows higher $\langle b \rangle$ values closer to the bar-ring interface region. Besides the azimuthal dependence of   $T_{\rm dep}^{\rm dense}$ and $\langle b \rangle$ within the SB ring, we also identify a radial dependence for both parameters especially in the southern section of the bar-ring interface: in particular, the highest (lowest)   $T_{\rm dep}^{\rm dense}$ ($\langle b \rangle$) values tend to appear "upstream" (smaller radii) along the gas circulation lines if we assume that the sense of rotation of the gas in the disk is counterclockwise.  The two branches in the   $T_{\rm dep}^{\rm dense}$-$\langle b \rangle$ plot of Fig.~\ref{Fig_11} correspond to a large extent to the two regions of the SB ring identified in Fig.~\ref{Fig_12}.
Similar results are found in Appendix~\ref{AppE} when we consider the HCO$^{+}$ as dense gas tracer. The trends and statistical parameters derived are practically identical to the ones obtained using HCN.

We can speculate if the reported trends of   $T_{\rm dep}^{\rm dense}$ as a function of  $\langle b \rangle$  for $\Delta A=400$~pc,  visualized in Figs.~\ref{Fig_11} and \ref{Fig_12},  could be entirely attributed to potential variations of the $\alpha_{\rm HCN}$ conversion factor on these spatial scales. In this context it is worth noting that the range explored by $T_{\rm dep}^{\rm dense}$ as a function of  $\langle b \rangle$, which amounts to 0.6~dex ($\simeq$a factor of four), is a significant factor of three larger than the differences found between the "global" kpc scale and the "small" pc scale HCN conversion factors reported by \citet{Wu2005}, as mentioned in Sect~\ref{alphahcnhco}. Although we have no way of confirming or refuting the existence of significantly larger changes of $\alpha_{\rm HCN}$ based on our data,  we note that adopting a 
lower value of the conversion factor for HCN would result in lower molecular gas surface densities, particularly in the regions of the 
SB ring of NGC1068 that happen to show comparatively higher (lower) SFE$_{\rm dense}$ ($T_{\rm dep}^{\rm dense}$)  values. These regions very likely comprise a collection of hot core-like
clouds akin to the Galactic SF cores studied by \citet{Wu2005} for which $\alpha_{\rm HCN}^{\rm cores}\simeq0.7 \times \alpha_{\rm HCN}^{\rm global}$. As a direct consequence, the trends shown in Figs.~\ref{Fig_11} and \ref{Fig_12} would be further enhanced rather than 
being suppressed.




\subsection{Trends as a function of the dense gas fraction} \label{SFE-F}

Observations of molecular gas in our Galaxy and in a number of nearby galaxies have found clear evidence of trends in SFE$_{\rm dense}$ as a function of $F_{\rm dense}$, which are indicative of anti-correlation \citep[e.g.,][]{Lon13, Chen2015, Murphy2015, Usero2015, Bigiel2016, Gallagher2018, Querejeta2019, Jim19}. In particular, \citet{Usero2015} found that  SFE$_{\rm dense}$ is about 6--8 times lower near the galaxy centres  than in the outer regions of the galaxy disks analysed in their IRAM-30m survey. This radial trend is reversed for $F_{\rm dense}$ in their sources. Furthermore,  \citet{Lon13} found  anomalously low SFR values in a large fraction of high-density molecular clouds in the centre of the Milky Way, suggestive of an anticorrelation between SFE$_{\rm dense}$ and $F_{\rm dense}$. A similar trend has been found recently by \citet{Querejeta2019} in their study of M~51, and by \citet{Jim19} from an analysis of the data obtained by the EMPIRE survey in nine spiral galaxies. We examine in this section the existence of a trend in SFE$_{\rm dense}$ as a function of $F_{\rm dense}$ in the SB ring of NGC~1068.

We define two proxies for the "dense gas fraction" in the SB ring. First,  the ratio of the dense molecular gas surface density derived from HCN (1--0)
to the bulk molecular gas surface density derived from from CO(1--0), that is, $F_{\rm dense}$=$\Sigma_{\rm dense}$/$\Sigma_{\rm mol}$ $\propto$ $I_{\rm HCN(1-0)}$/$I\rm _{CO(1-0)}$. We consider that CO line traces the bulk of the 
molecular gas in the galaxy ($n$(H$_2$)~$\gtrsim10^{2-3}$cm$^{-3}$), while HCN traces material with $n$(H$_2$)~$\gtrsim10^{4-5}$cm$^{-3}$. We therefore use the HCN(1--0)/CO(1--0)  line ratio (hereafter $R_{\rm HCN/CO}$) as a proxy for 
the dense gas fraction in the SB ring.  Secondly, we also use the CO(3--2)/CO(1--0) ratio (hereafter $R_{\rm 3-2/1-0}$) in the SB ring derived by \citet{Burillo2014} as an alternative  proxy for $F_{\rm dense}$.

   
   \begin{figure}[tb!]
   \centering
   \includegraphics[width=.94\linewidth]{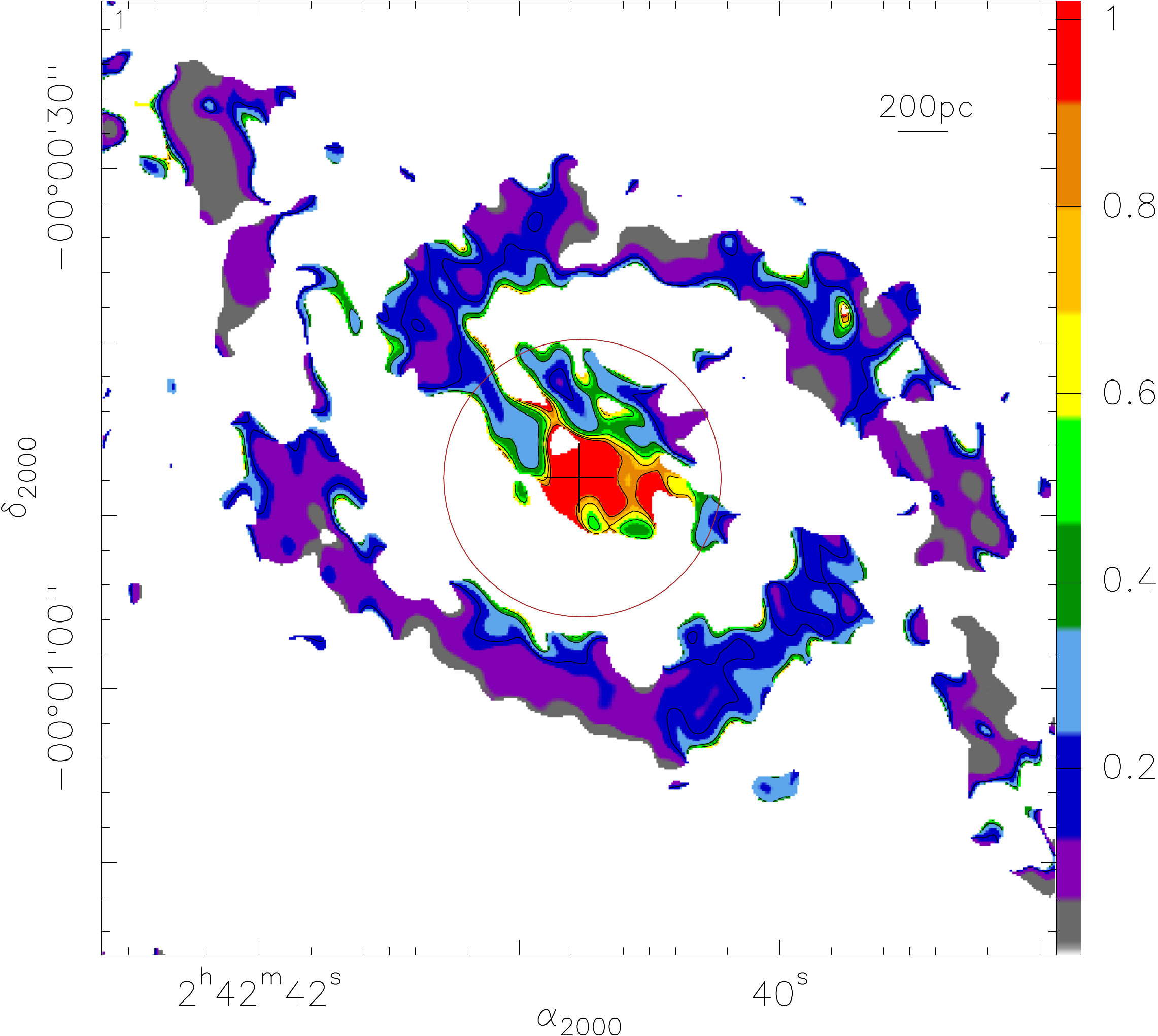}
   
   \caption{HCN(1--0)/CO(1--0) brightness temperature ratio map ($R_{\rm HCN/CO}$; colour scale and contours) derived at the common (lower) spatial resolution of the CO(1--0) observations of \citet{Schinnerer2000} ($\simeq$100~pc). The brown circle of 8$\arcsec$-radius ($\simeq$ 560 pc) locates the inner region excluded from our analysis of SF relations.} 
              \label{Fig_13}
   \end{figure}


   
      \begin{figure}[bt!]
   \centering
   \includegraphics[width=.9\linewidth]{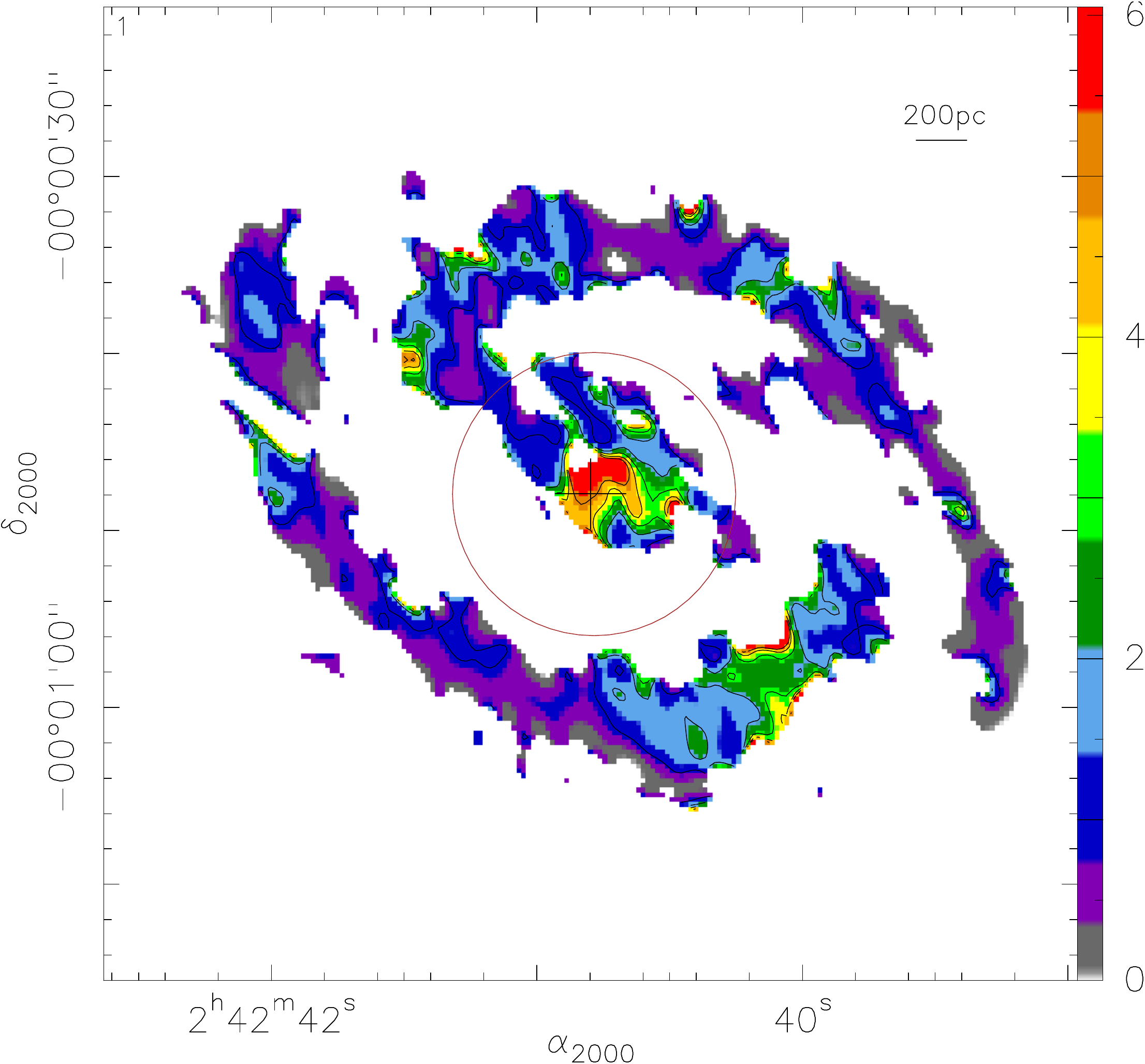}
   
    \caption{Same as Fig.~\ref{Fig_13} but showing the CO(3--2)/CO(1--0) brightness temperature ratio map ($R_{\rm 3-2/1-0}$; colour scale and contours) derived at the common (lower) spatial resolution of the CO(1--0) observations.} 
              \label{Fig_14}
   \end{figure}


 Figures~\ref{Fig_13} and \ref{Fig_14} show $R_{\rm HCN/CO}$ and $R_{\rm 3-2/1-0}$ in $T_{\rm mb}$ units. To derive the brightness temperature ratios, we degraded the HCN(1--0) and CO(3--2) maps to the spatial resolution of the CO(1--0) observations of \citet{Schinnerer2000}.
The $R_{\rm HCN/CO}$ ratio changes significantly across the disk of NGC~1068. The highest values of  $R_{\rm HCN/CO}$, $>0.4-1.2$, correspond to the CND. 
Although in the SB ring $R_{\rm HCN/CO}$ shows a wide range of values, this ratio is higher   in the bar-ring interface region  ($R_{\rm HCN/CO}\sim0.2-0.4$) than elsewhere in the ring ($R_{\rm HCN/CO}\sim0.05-0.2$). 
The $R_{\rm 3-2/1-0}$ ratio, shown in Fig.~\ref{Fig_14}, changes  also significantly across the disk. The $R_{\rm 3-2/1-0}$ ratio in the CND ($\sim2-6$) is higher 
 than in the SB ring, in agreement with previous estimates by  \citet{Krips2011} and \citet{Tsai2012}. While the average ratio is $\sim 1.2\pm0.02$ in the 
 SB ring , $R_{\rm 3-2/1-0}$ is higher in the bar-ring interface region ($R_{\rm 3-2/1-0}\sim0.6-3$) and comparatively lower elsewhere in the ring  ($R_{\rm 3-2/1-0}\sim0.1-0.8$).   

The $R_{\rm HCN/CO}$ ratio is hardly sensitive to kinetic temperature ($T_{\rm k}$), as the energy levels  giving rise to both rotational transitions are similar ($E_{\rm k}$[$J=1$, CO]~$\sim5.5$~K,  $E_{\rm 
k}$[$J=1$, HCN]~$\sim4.3$~K). However, as the excitation of both CO lines are  sensitive to both $n$(H$_2$) and $T_{\rm k}$,  $R_{\rm 3-2/1-0}$ is comparatively a more indirect and less straightforward tracer of    $F_{\rm dense}$ relative to  $R_{\rm HCN/CO}$. Leaving aside the uncertainties on the value of $\alpha_{\rm HCN}$ in SF regions, which may reflect a peculiar hot-core like chemistry, $R_{\rm HCN/CO}$ is therefore the most reliable proxy for the dense gas fraction and 
as such is widely used in extragalactic studies, and we therefore adopt it in the following analysis. In either case, we note that both line ratios suggest a higher excitation of HCN(1--0) and CO(3--2) lines relative to CO(1--0) in the bar-ring interface region.


   \begin{figure*}[htp]
   \centering
      \includegraphics[width=0.45\linewidth]{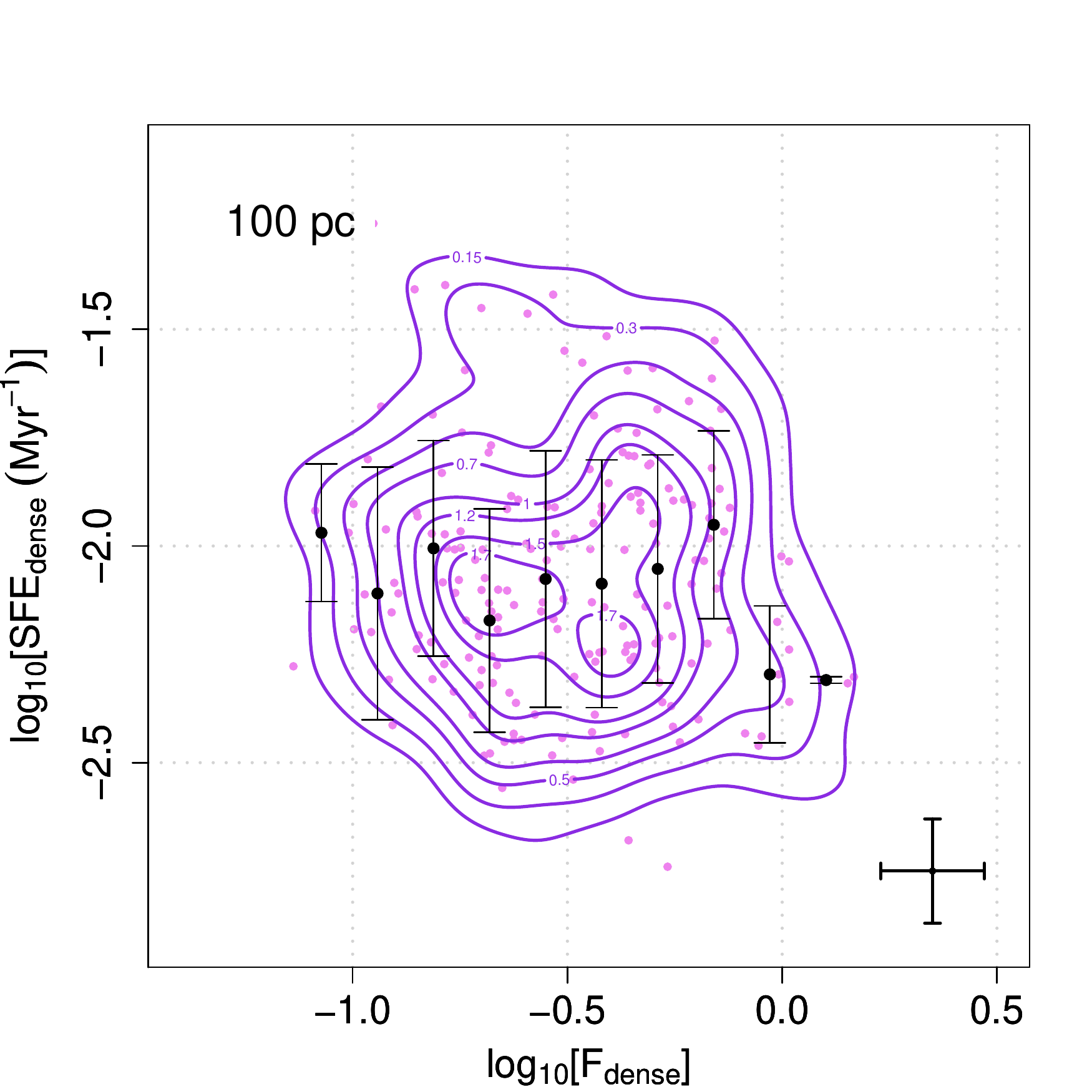}
   		\includegraphics[width=0.45\linewidth]{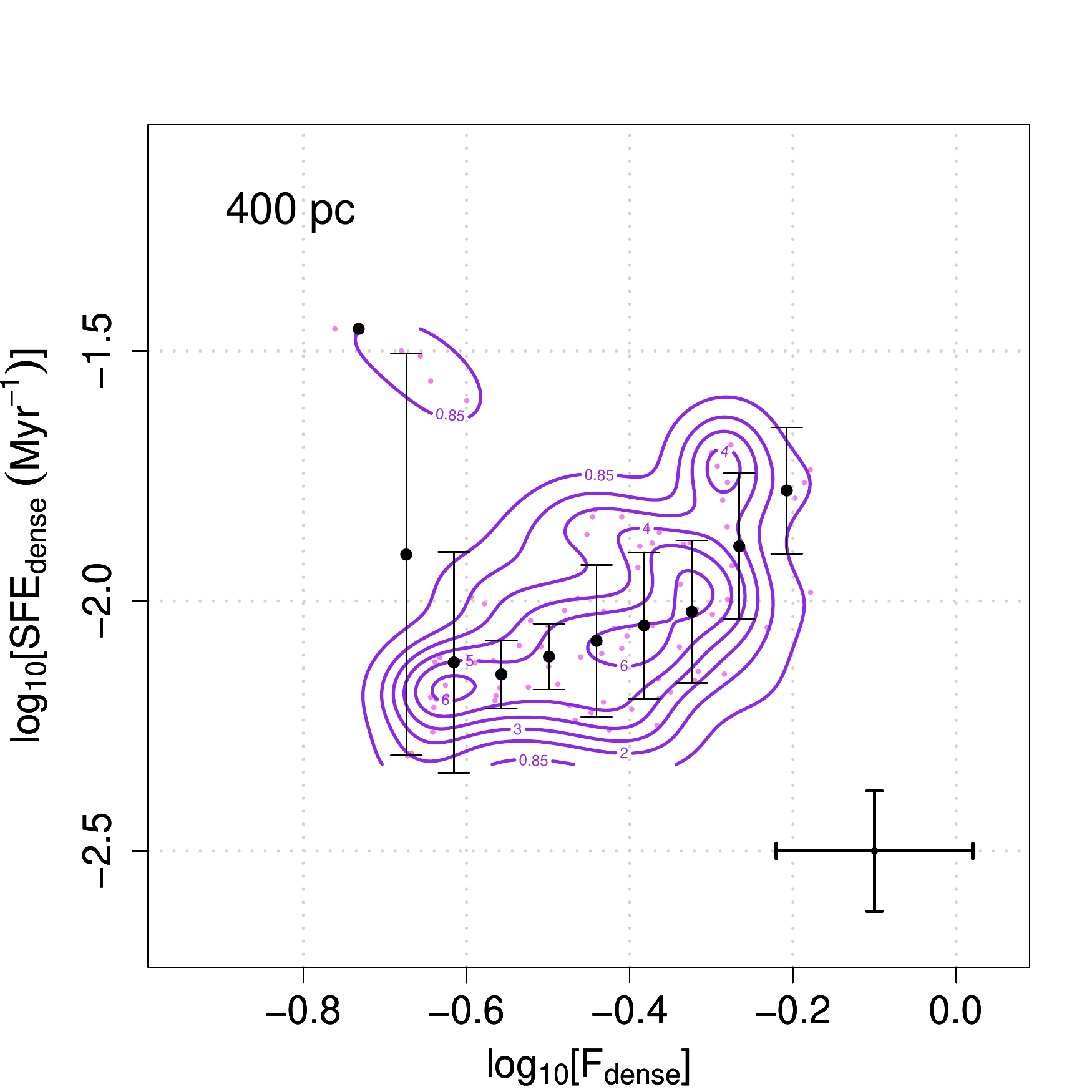}
   \caption{Star formation efficiency of the dense gas (SFE$_{\rm dense}$) as a function of dense gas fraction ($F_{\rm dense}$:  estimated from the HCN(1--0)-to-CO(1--0) line ratio of Fig.~\ref{Fig_13}) in NGC~1068 using two averaging scales: 100~pc ({\it left panel}) and 400~pc ({\it right panel}). Magenta circles correspond to data measurements evaluated at the Nyquist-sampled grid of points  after convolution with the appropriate Gaussian for each aperture.  Black circles show median SFE$_{\rm dense}$ values  and errorbars indicate the rms scatter in the bins. Vertical and horizontal errorbars at the lower right corner of both panels account for the typical uncertainties, which amount to $\pm0.13$~dex.}
              \label{Fig_15}
   \end{figure*}
   

Figure~\ref{Fig_15} represents SFE$_{\rm dense}$  as a function of $F_{\rm dense}$ estimated from $R_{\rm HCN/CO}$ in the SB ring. We adopted the same approach followed in Sect.~\ref{Tdep-b} to obtain estimates of both variables
over two averaging scales: $\Delta A$=100~pc and 400~pc.  Figure~\ref{Fig_15} shows that there is no significant trend in SFE$_{\rm dense}$ as a function of $F_{\rm dense}$ for $\Delta A$=100~pc: we estimate a Spearman rank parameter $\rho_{\rm sp}=+0.03$ with a
two-sided $p$-value~$>1\%$.  For $\Delta A=400$~pc there is  nevertheless a more significant positive correlation ($\rho_{\rm sp}=+0.34$, with a $p$-value~$<1\%$). This trend is a direct consequence of the spatial  distribution of SFE$_{\rm dense}^{-1}$ and  $F_{\rm dense}$ shown, respectively, in Figs.~\ref{Fig_12} and ~\ref{Fig_13}. The monotonic increase shown in the right panel of Fig.~\ref{Fig_15} 
suggests that the comparatively denser molecular gas of the bar-ring interface region inside the SB ring forms stars at higher rate per unit dense gas mass.

This result seems to be in contradiction with the anticorrelation trends found between  SFE$_{\rm dense}$  and $F_{\rm dense}$ in previous works in other galaxies. Specifically, \citet{Usero2015} found an index $N\sim-1.6$ for the SFE$_{\rm dense}\propto F_{\rm dense}^{N}$ power-law fitting their data. A similar index ($N\sim-1.5$) can be estimated from the data compiled by \citet{Querejeta2019}.

\subsection{Trends as a function of the stellar mass surface density} \label{SFE-star}

Several works studying the existence of trends in SFE$_{\rm dense}$  as a function of $\Sigma_{\rm star}$ on kpc-scales in a number of nearby galaxies have found that SFE$_{\rm dense}$ is seen to decrease in the central parts of galaxy disks, namely in regions characterised by high $\Sigma_{\rm star}$. These observations also showed that $F_{\rm dense}$ tends to increase systematically with $\Sigma_{\rm star}$ \citep{Chen2015, Usero2015, Bigiel2016, Gallagher2018, Jim19}. \citet{Querejeta2019} studied different regions in the disk of M~51 and found that similar correlations are recovered at 100~pc-scales. We investigate below the trends in SFE$_{\rm dense}$ and $F_{\rm dense}$ as a function of $\Sigma_{\rm star}$ in the SB ring of NGC~1068.

\begin{figure}[htp]
   \centering
   \includegraphics[width=1\linewidth]{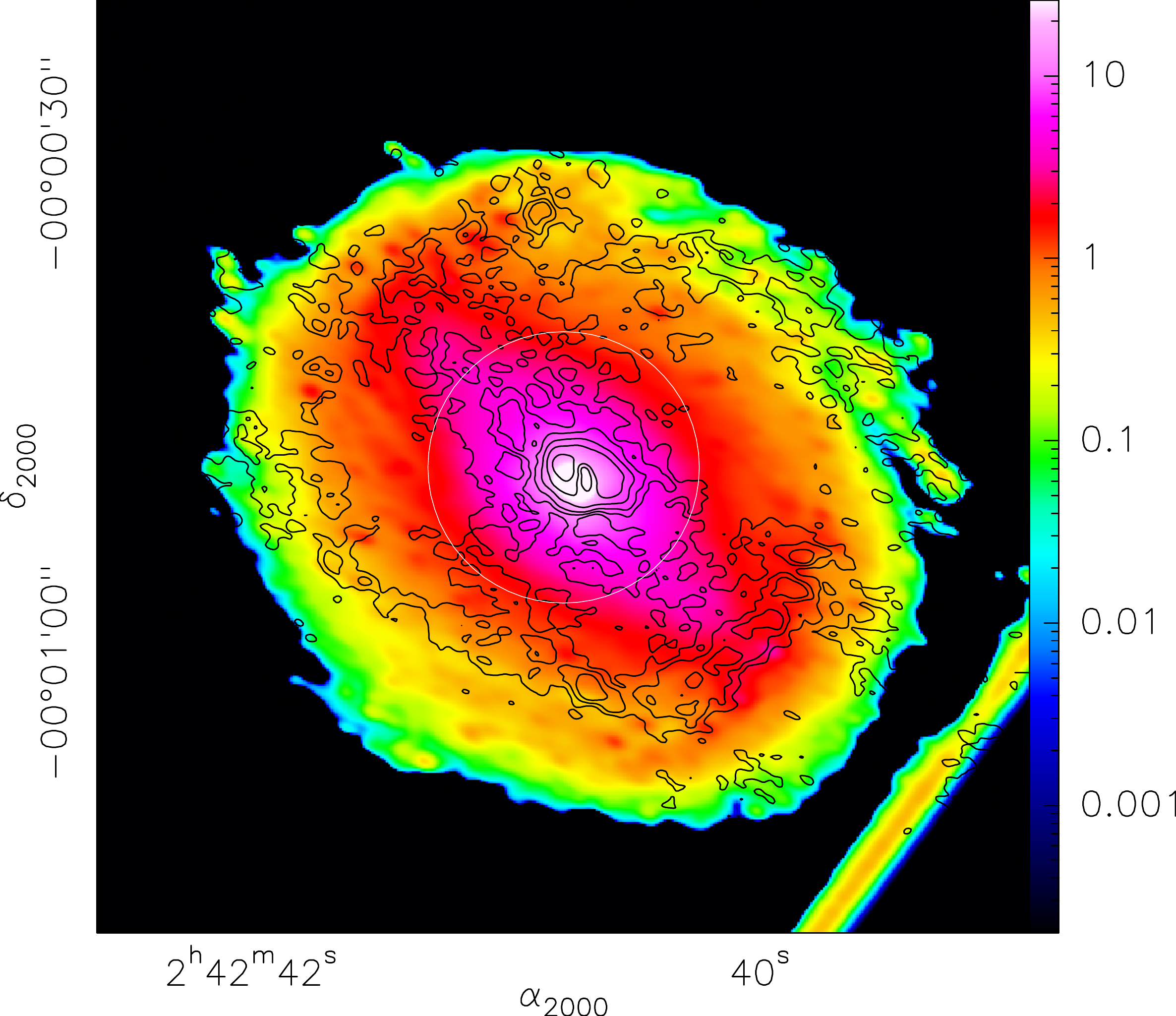}
   \caption{Overlay of the ALMA HCN(1--0) map (contours) on the HST/NICMOS F190N continuum image at 1.9 $mu$m HST (colour scale as shown in Jy). Contour spacing: 3$\sigma$, 7$\sigma$, 12$\sigma$, 24$\sigma$, 50$\sigma$ and 120$\sigma$, where 1$\sigma$=0.028 Jy beam$^{-1}$ km s$^{-1}$. The white circle of r = 8$\arcsec$ locates the inner of the galaxy, excluded of our analysis. The HST/NICMOS F190N image has been degraded to the spatial resolution of HCN(1--0) (1$\arcsec$ $\times$ 0$\arcsec$.6 $\approx$ 50 pc).}
          
             \label{Fig_20}%
   \end{figure}

Near-infrared emission is commonly used to trace the stellar mass in nearby galaxies, since light at these wavelengths mainly comes from old stars and is less affected by extinction \citep{Quillen1994}.
With this aim, we used the HST near-infrared (NIR) continuum narrow-band image of NGC~1068 at 1.9~$\mu$m, obtained with the F190N filter on the NICMOS~3 camera,  to trace the distribution of the stellar mass in the disk of the galaxy.
We assumed a constant mass-to-light ratio $M/L_{1.9 \mu m} = 0.2 M_{\odot}$/$L_{\odot}$  \citep[e.g., see][]{Querejeta2015} to obtain the local stellar mass surface density. Figure \ref{Fig_20} overlays the HCN(1--0) integrated intensity contours on the HST/NICMOS F190N image. The stellar bar feature is easily identified in the NIR image. Judging from the overall distribution of $\Sigma_{\rm star}$ in the disk, it appears that the bar-ring interface is characterised by higher  $\Sigma_{\rm star}$ values compared to the regions located elsewhere in the SB ring.
 


 \begin{figure*}[tb!]
 \centering
   	\includegraphics[width=0.61\linewidth]{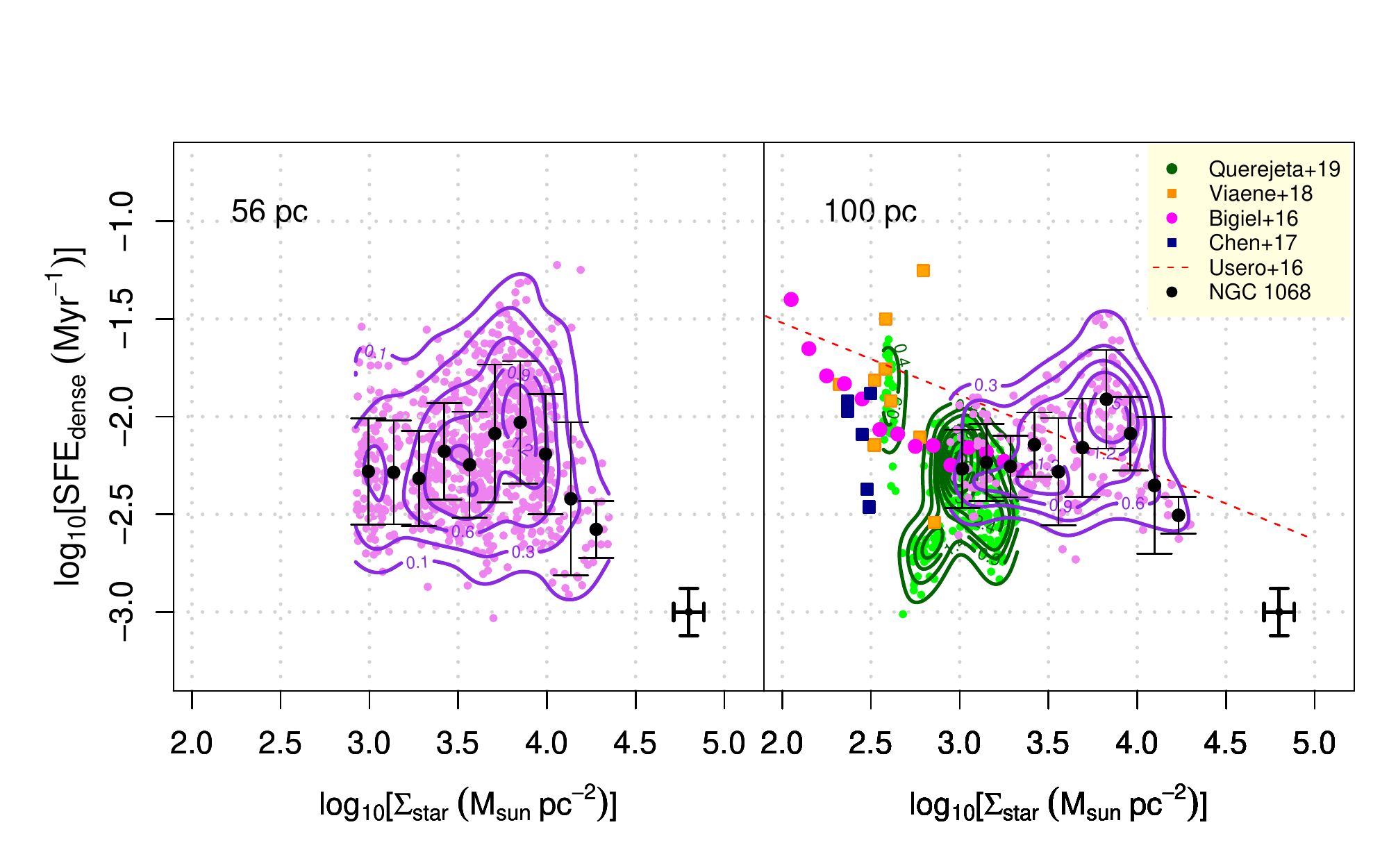}
   		\includegraphics[width=0.38\linewidth]{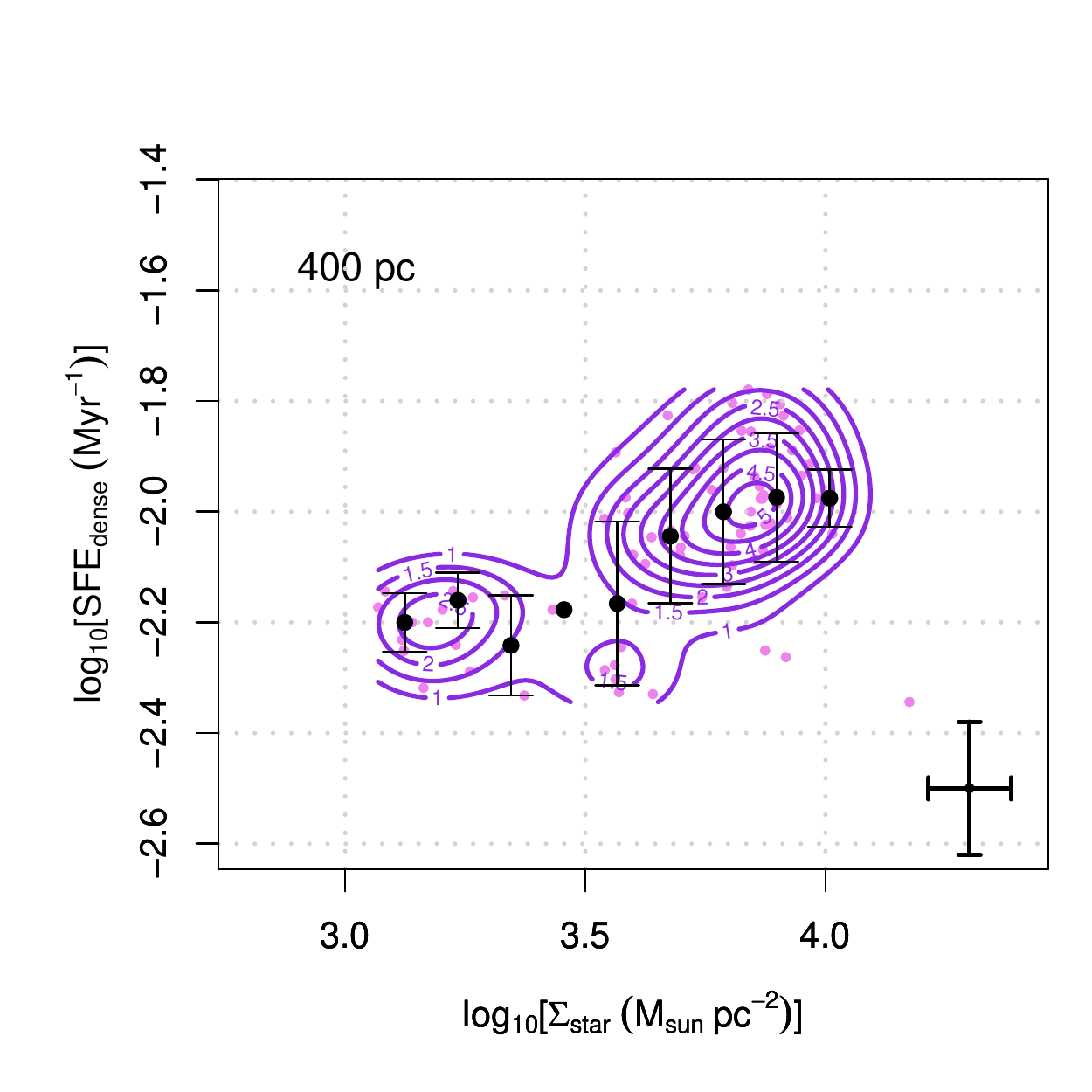}
   \caption{Star formation efficiency of the dense gas (SFE$_{\rm dense}$) as a function of the  stellar mass surface density ($\Sigma_{\rm star}$) in the SB ring of NGC~1068 at the "initial resolution" (56~pc) ({\it left panel}), 
and at two averaging scales: 100~pc ({\it middle panel}) and 400~pc ({\it right panel}).  Magenta circles and isodensity contours correspond to data measurements evaluated at the Nyquist-sampled grid of points  after convolution with the appropriate Gaussian for each aperture.  Black circles show median SFE$_{\rm dense}$ values  and errorbars indicate the rms scatter in the bins. Symbols and isodensity contours in the {\it middle panel} identify the data obtained in other galaxies.The dashed orange line represents the fit of \citet{Usero2015}.  Vertical and horizontal errorbars at the lower right corner of each panel account for the typical uncertainties, which amount to $\pm0.13$~dex and $\pm0.09$~dex, respectively.}
              \label{Fig_21}%
 \end{figure*}



 \begin{figure*}[bt!]
   \centering
   		\includegraphics[width=0.9\linewidth]{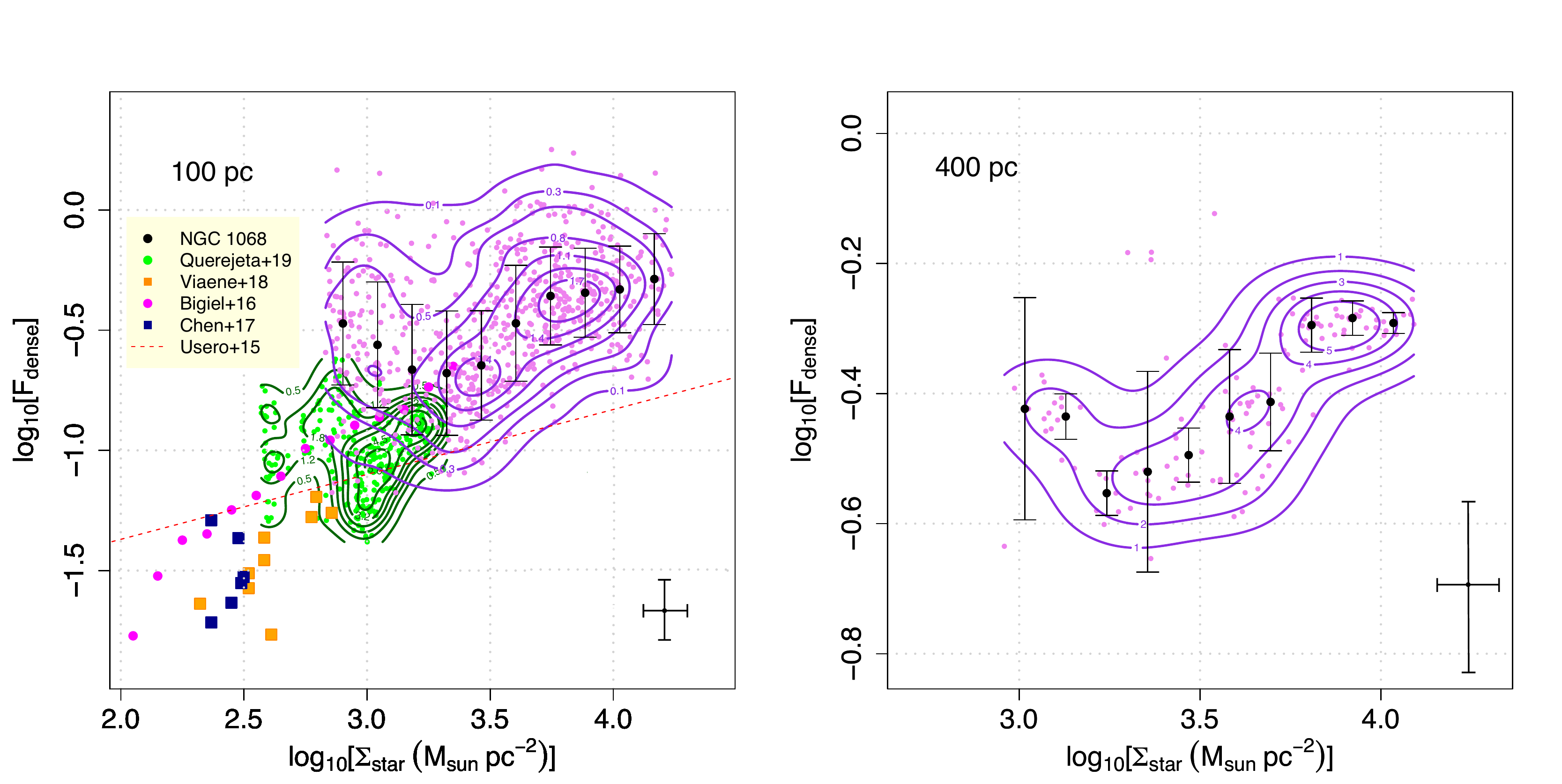}
   \caption{Fraction of dense gas ($F_{\rm dense}$) as a function of the  stellar mass surface density ($\Sigma_{\rm star}$) in the SB ring of NGC~1068 at the two averaging scales: 100~pc ({\it left panel}) and 400~pc ({\it right panel}). Symbols as in Fig.~\ref{Fig_21}.  Vertical and horizontal errorbars at the lower right corner of each panel account for the typical uncertainties, which amount to $\pm0.13$~dex and $\pm0.09$~dex, respectively.}
              \label{Fig_22}%
   \end{figure*}
   

Figure \ref{Fig_21} represents  SFE$_{\rm dense}$ as a function of $\Sigma_{\rm star}$ across the SB ring of NGC~1068 for the three working apertures adopted in Sect.~\ref{Tdep-b}. 
SFE$_{\rm dense}$ does not show any significant correlation with  $\Sigma_{\rm star}$  at the "initial resolution"  ($\rho_{\rm sp}=+0.10$, with a $p$-value~$>1\%$), or at the 100~pc averaging scale 
($\rho_{\rm sp}=+0.15$, with a $p$-value~$>1\%$). On the contrary,  SFE$_{\rm dense}$ shows a statistically significant positive correlation  with $\Sigma_{\rm star}$ in the SB ring for $\Delta A=400$~pc ($\rho_{\rm sp}=+0.59$, with a $p$-value~$<1\%$). We compare the location of NGC~1068  in the  SFE$_{\rm dense}$-$\Sigma_{\rm star}$ parameter space with the position occupied by the galaxies studied by the references listed in Table \ref{tabinfinite} in the middle panel of Fig.~\ref{Fig_21}.  NGC~1068 clearly deviates from the overall anti-correlation trend followed by other galaxies. The positive correlation observed in the NGC~1068 SB ring is in stark contrast with the anticorrelation trends identified in the  galaxies studied by  \citet{Usero2015} ($\rho_{\rm sp}=-0.50$),  \citet{Querejeta2019} ($\rho_{\rm sp}=-0.57$), and  \citet{Gallagher2018}  ($\rho_{\rm sp}=-0.66$).  In this context it is noteworthy that \citet{Querejeta2019} noticed that for a given value of $\Sigma_{\rm star}$ the  data points of M~51 span a significant 1~dex range  of  SFE$_{\rm dense}$,  namely similar to the spread of values seen in the NGC~1068 SB ring. This is an indication of the wide range of dynamical environments probed in both galaxies. 


Figure \ref{Fig_22} represents  $F_{\rm dense}$ as a function of $\Sigma_{\rm star}$ across the SB ring of NGC~1068 for the two averaging scales adopted in Sect.~\ref{Tdep-b}.
$F_{\rm dense}$  shows already a significant correlation with  $\Sigma_{\rm star}$  at $\Delta A=100$~pc ($\rho_{\rm sp}=+0.40$, with a $p$-value~$<1\%$). The correlation is further reinforced when we use  400~pc
as averaging scale ($\rho_{\rm sp}=+0.67$, with a $p$-value~$<1\%$). This is qualitatively and quantitatively  similar to the trends identified in the data of the galaxies shown in the left panel of Figure \ref{Fig_22} at scales of 100 pc, for which
\citet{Querejeta2019} derived a Spearman rank parameter of $\rho_{\rm sp}\sim+0.45$, and $\rho_{\rm sp}\sim+0.7$ when they added the datapoints from \citet{Chen2017} and \citet{Gallagher2018}.


  \begin{figure*}[tb!]
  \centering
    \includegraphics[width=0.61\linewidth]{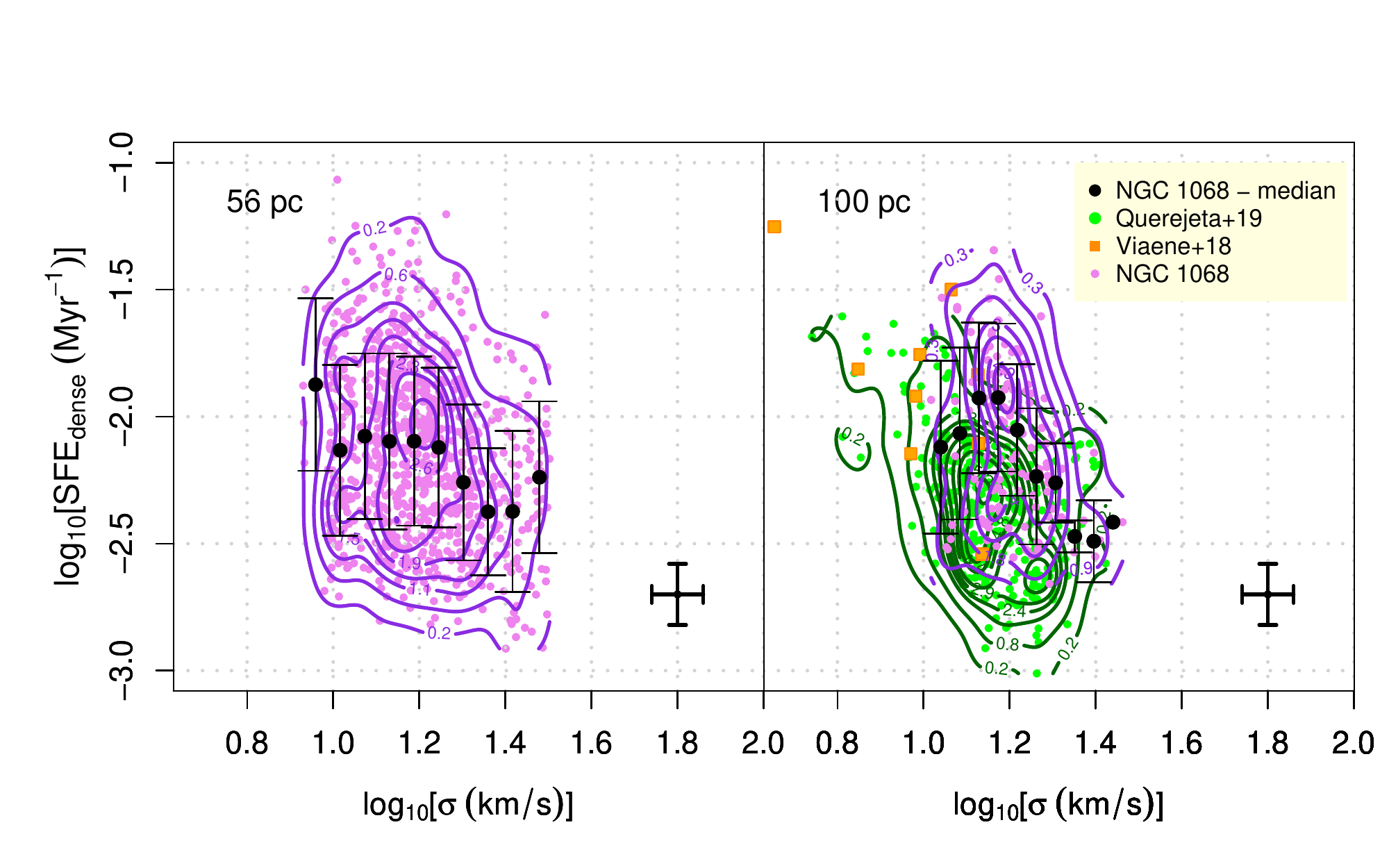}   	
   	\includegraphics[width=0.38\linewidth]{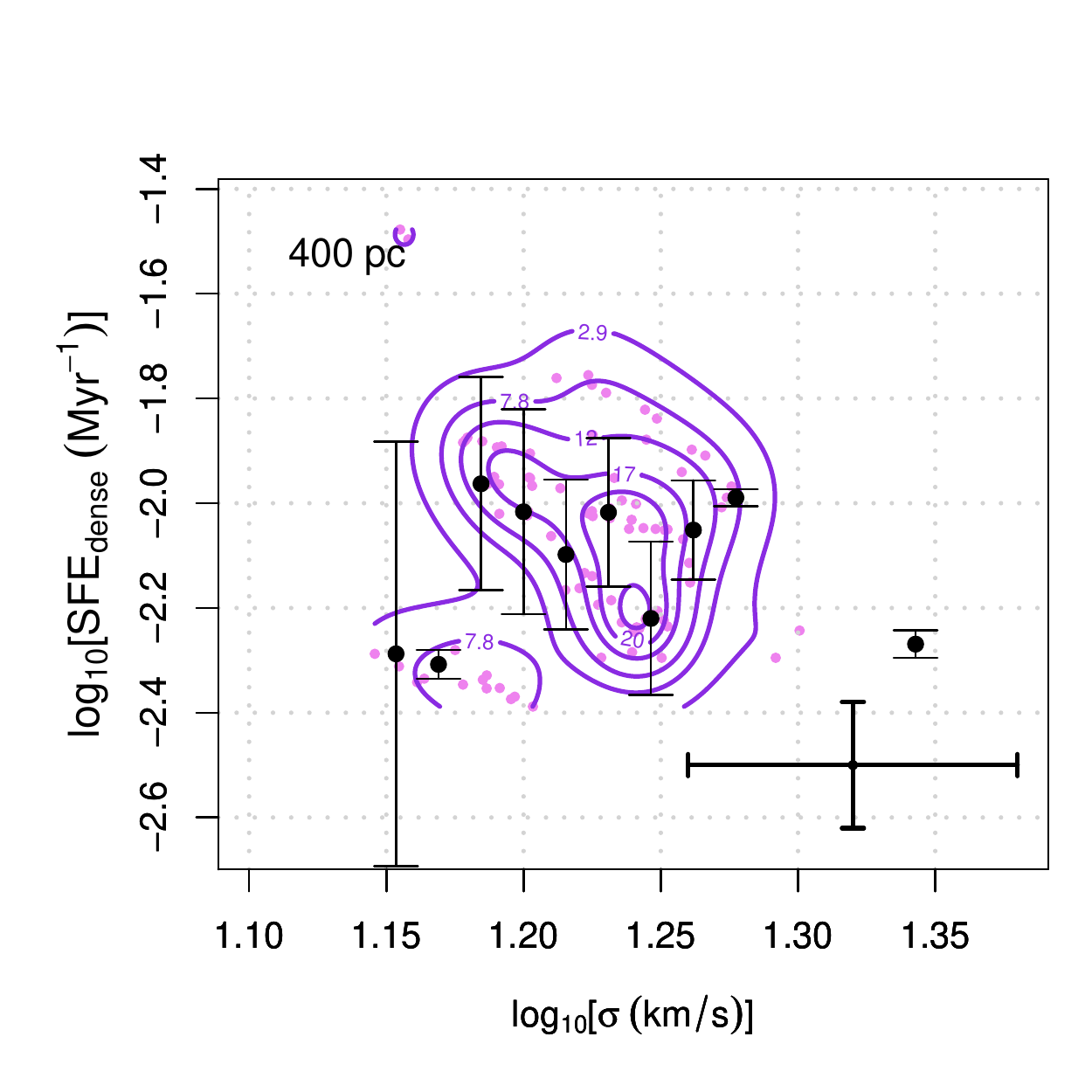}
   \caption{Star formation efficiency of the dense gas (SFE$_{\rm dense}$) as a function of the velocity dispersion of HCN ($\sigma$) in NGC~1068 for three averaging scales: 56~pc ({\it left panel}), 100~pc ({\it middle panel}) and 400~pc ({\it right panel}). Measurements at 100~pc scales are compared to literature data for other galaxies. Symbols as in Fig.~\ref{Fig_21}. Vertical and horizontal errorbars at the lower right corner of each panel account for the typical uncertainties, which amount to $\pm0.13$~dex and $\pm0.06$~dex, respectively.}
              \label{Fig_17}%
   \end{figure*}       
 


  \begin{figure}[tb!]
   \centering
   \includegraphics[width=.9\linewidth]{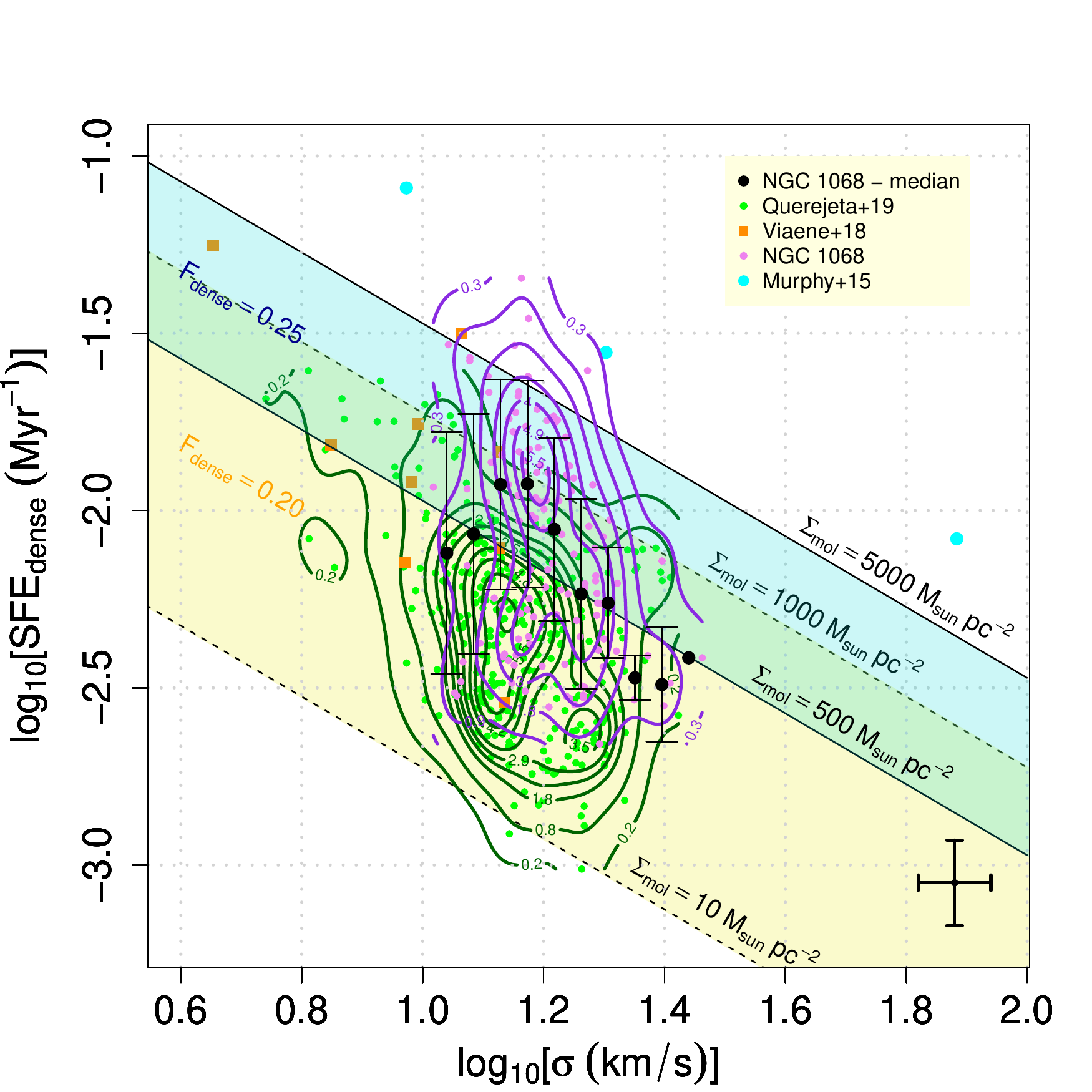}
   \caption{Same as the {\it middle panel} of Fig.~\ref{Fig_17} but adding a comparison with the trends of SFE$_{\rm dense}$   as a function of  the velocity dispersion of the gas ($\sigma$) predicted by the model of \citet{Meidt2020}. This model foresees different locations of galaxies in the  SFE$_{\rm dense}-\sigma$ parameter space depending on the fraction of dense gas ($F_{\rm dense}$) and, also, on the total molecular gas column densities ($\Sigma_{\rm mol}$). Symbols as in Fig.~\ref{Fig_21}. Vertical and horizontal errorbars at the lower right corner  account for the typical uncertainties, which amount to $\pm0.13$~dex and $\pm0.06$~dex, respectively.}
              \label{Figmodel}%
  \end{figure}    
   


  \begin{figure*}[htp]
   \centering
   		\includegraphics[width=.95\linewidth]{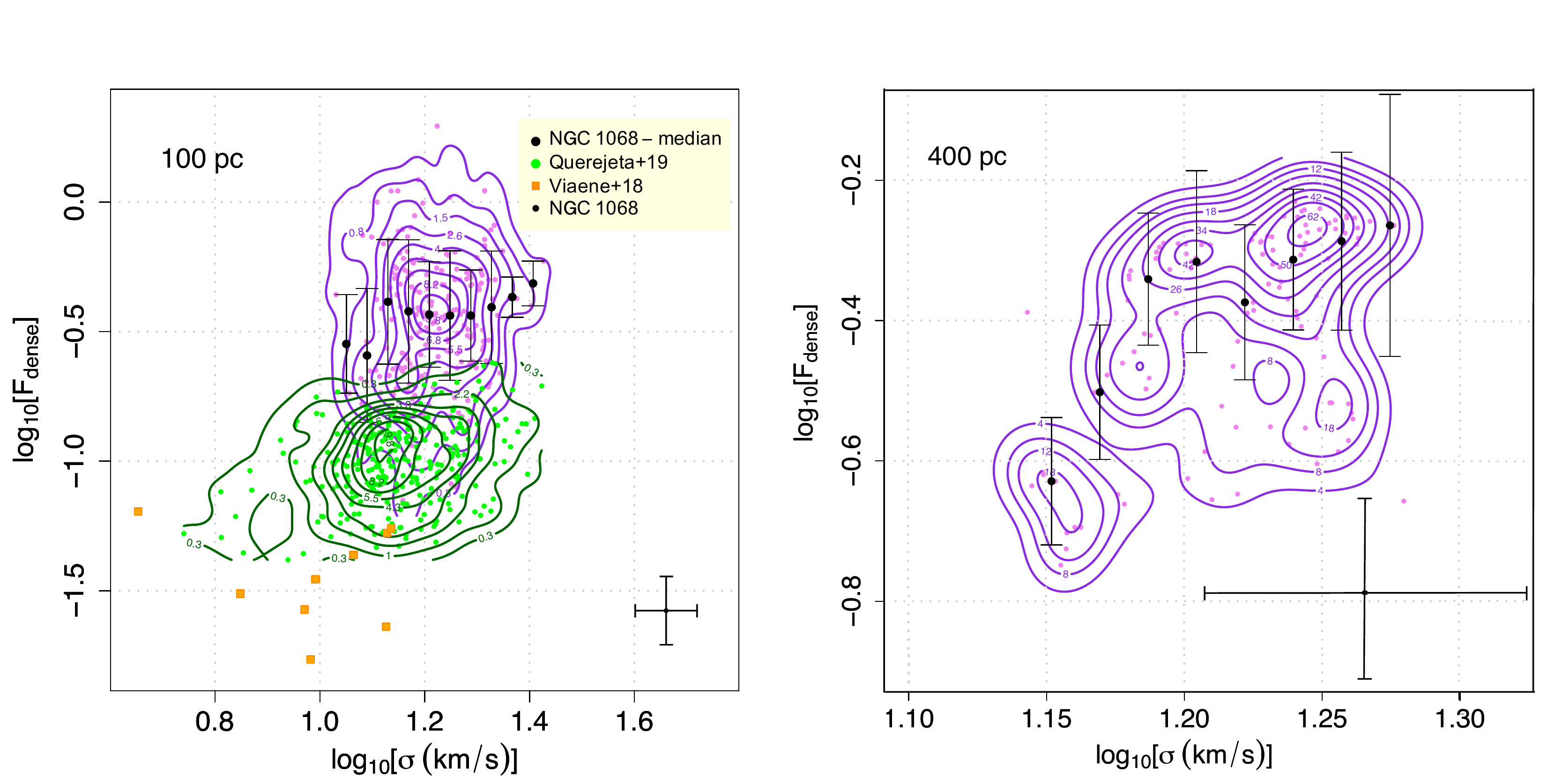}

   \caption{Dense gas fraction ($F_{\rm dense}$) as a function of velocity dispersion in the SB ring of NGC~1068 at the two averaging scales: 100~pc ({\it left panel}) and 400~pc ({\it right panel}). Measurements at 100~pc scales are compared to literature data for other galaxies. Symbols as in Fig.~\ref{Fig_21}. Vertical and horizontal errorbars at the lower right corner of each panel account for the typical uncertainties, which amount to $\pm0.13$~dex and $\pm0.06$~dex, respectively. 
            }
              \label{Fig_19}%
  \end{figure*}    
   


    \begin{figure}[htp]
      \centering
   \includegraphics[width=1\linewidth]{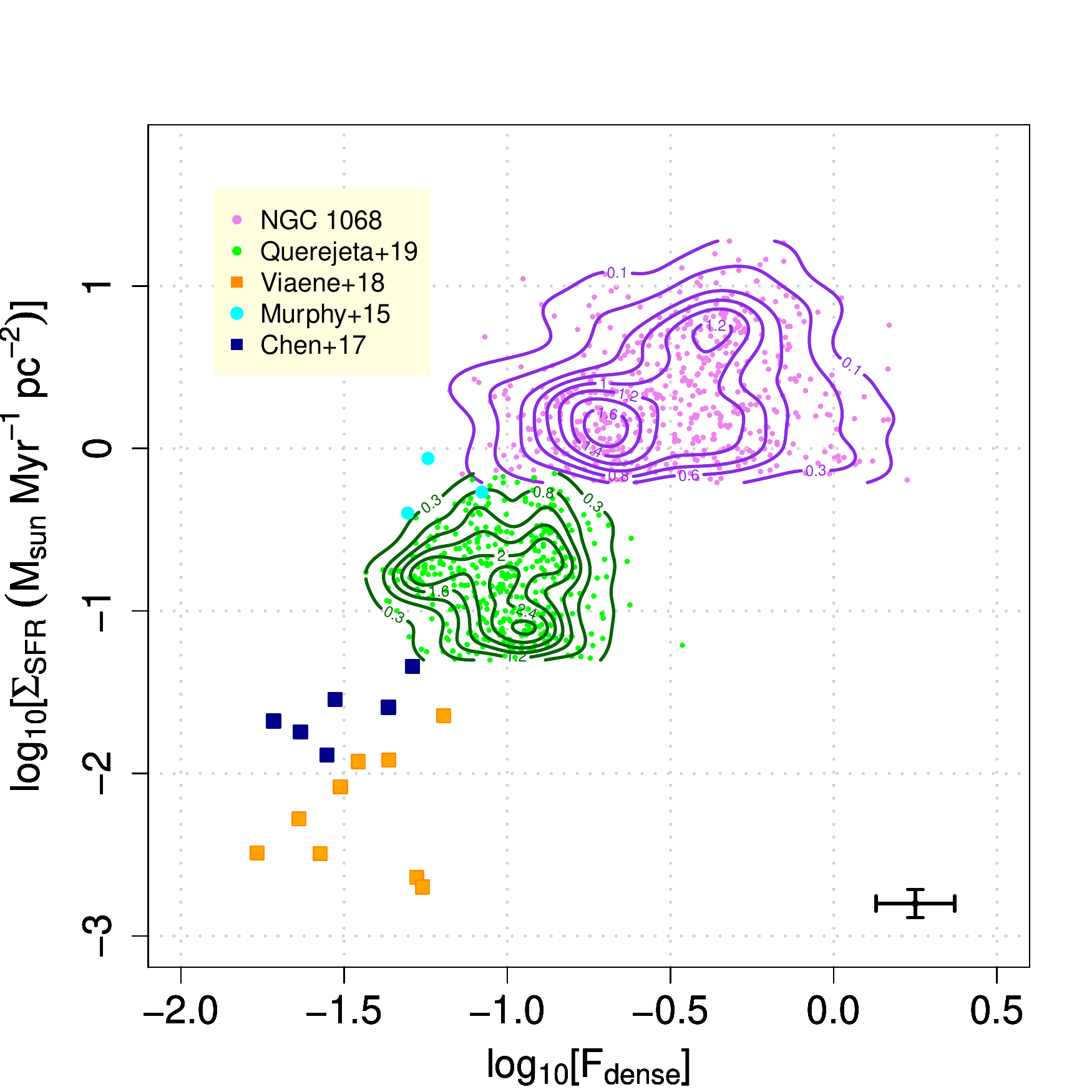}

   \caption{Star formation rate surface density ($\Sigma_{\rm SFR}$)  as a function of dense gas fraction ($F_{\rm dense}$) in NGC~1068 measured at 100~pc scales compared to the  data obtained in other galaxies (symbols as in Fig.~\ref{Fig_8}).  Vertical and horizontal errorbars at the lower right corner account for the typical uncertainties, which amount to $\pm0.09$~dex and $\pm0.13$~dex, respectively}.
               \label{Fig_16}%
   \end{figure}

\subsection{Trends  as a function of velocity dispersion} \label{SFE-sigma}

The role of velocity dispersion in setting the efficiency of molecular gas at forming stars is central to current models of turbulence-driven SF via their dependence on the Mach number \citep[e.g.,][]{Krumholz2005, Hennebelle2012, Federrath2015}.
From the observational point of view, and under the hypothesis that the velocity dispersion measured at a given scale reflects mainly turbulent motions in molecular gas, identifying trends in SFE as a function 
of local velocity dispersion may indicate whether SF tends to be either enhanced or suppressed  by turbulence. The predictions of models on the expected  trends of SFE$_{\rm dense}$ as a function of the "observed" velocity dispersion ($\sigma$) differ depending on the role of turbulence versus large-scale bulk motions (e.g., streaming motions) that are driven by the background gravitational potential. The degree of coupling to the large-scale gravitational potential implies that the critical threshold for star formation is not universal and that SFE$_{\rm dense}$ is therefore not constant. Specifically, the recent model published by \cite{Meidt2020} foresees that SFE$_{\rm dense}$ should decrease with $\sigma$  at fixed values of $F_{\rm dense}$ and $\Sigma_{\rm mol}$.

 \citet{Leroy2017} found a decreasing trend between the efficiency of the bulk molecular gas (SFE$_{\rm mol}$) and $\sigma$ measured from CO(1--0) in M~51.  \citet{Querejeta2019} also found a significant anti-correlation between SFE$_{\rm dense}$ and $\sigma$  in M~51 ($\rho_{\rm sp}=-0.62$).  A comparison of the observations of M~31 \citep{Viaene2018} and M~51  \citep{Querejeta2019} with  the predictions of \citet{Meidt2020}'s model shows a fair agreement for apertures of 100~pc.

 Figure~\ref{Fig_17} examines the trends in SFE$_{\rm dense}$ as a function of $\sigma$ across the SB ring of NGC~1068 for the  three apertures adopted in Sect.~\ref{Tdep-b}. We note that the value of $\sigma$ derived on scales of 56~pc is not a good proxy  for the "internal" turbulence of the dense cores probed by HCN, which in all likelihood  have sizes $\sim$a few pc.  Instead, $\sigma$  likely encapsulates a mix of the "macroscopic" turbulence between the cores and the residual gradient of large-scale bulk motions within the ALMA beam. The correlation shown in Fig.~\ref{Fig_17} is not statistically significant for the two extreme values of the averaging apertures (56~pc and 400~pc). However, there is a weak yet significant trend at 100~pc scales, for which we derive a  $\rho_{\rm sp}=-0.28$, with a $p$-value $=0.3\%$. Overall, the results obtained in NGC~1068 show a very marginal agreement with the results of  \citet{Viaene2018} and \citet{Querejeta2019} and, therefore, with the predictions of \citet{Meidt2020}'s model. As illustrated in  Fig.~\ref{Figmodel},  a  fraction of the data corresponding to the SB ring in NGC~1068  lies in the region predicted by \citet{Meidt2020} for  $F_{\rm dense}\sim0.25$ and a range for $\Sigma_{\rm mol}\sim 500-5000~M_{\sun}$pc$^{-2}$. Although these values are in rough agreement with the ones estimated for a high fraction of the SB ring positions, the decreasing trend in SFE$_{\rm dense}$ as a function of $\sigma$ is seen to be much shallower than the one predicted by \citet{Meidt2020}. However, as noted by \citet{Meidt2020},  a gravitational potential linked to a strong density wave (bar or spiral) may tend to degrade  the strength of the expected decreasing  trend,  which is estimated in their model using a purely axisymmetric disk.

Figure \ref{Fig_19} explores the trends of  $F_{\rm dense}$ as a function of $\sigma$ in the SB ring for the two averaging scales adopted in Sect.~\ref{Tdep-b}.
In contrast with the trends identified in M~31 and M~51 by  \citet{Viaene2018} and \citet{Querejeta2019} on 100~pc scales, $F_{\rm dense}$  fails to show any significant correlation
 with  $\sigma$  in the SB ring of NGC~1068 for $\Delta A=100$~pc ($\rho_{\rm sp}=+0.10$, with a $p$-value~$>1\%$), However, a significant positive trend appears 
when we use  400~pc as averaging scale ($\rho_{\rm sp}=+0.50$, with a $p$-value~$<1\%$).

\subsection{Relation between the star formation rate and the dense gas fraction} \label{SFR-F}

 \citet{Viaene2018} observed a significant  correlation ($\rho_{sp}=0.63$) between the SFR and $F_{\rm dense}$ in a number of regions in the disk of M~31 observed with  a spatial resolution of 100~pc.  \citet{Querejeta2019} 
 explored the correlation between $\Sigma_{\rm SFR}$ and $F_{\rm dense}$ in M~51 after incorporating the data obtained by  \citet{Chen2017} in M~51 on similar spatial scales and derived a similar positive trend: $\rho_{\rm sp}=0.63$, with a 
 $p$-value~$<1\%$. This correlation improved when  \citet{Querejeta2019} included the M~31 and NGC~3627 data of \citet{Viaene2018} and \citet{Murphy2015}, respectively ($\rho_{\rm sp}=0.66$).
 
Figure \ref{Fig_16} shows the region occupied by the NGC~1068 data in the $\Sigma_{\rm SFR}$--$F_{\rm dense}$ plot for an adopted  averaging scale $\Delta A=100$~pc.
We do not find a strong correlation between   $\Sigma_{\rm SFR}$ and $F_{\rm dense}$ ($\rho_{\rm sp}$=0.22) when we consider the  NGC~1068 data alone. In particular, this correlation is weaker than the one found between  $\Sigma_{\rm SFR}$
and $\Sigma_{\rm dense}$ for the same spatial scales ($\rho_{\rm sp}$=0.35; see Table~\ref{tab}). This result is an indication that the dense gas fraction is not a  better predictor of star formation than the dense gas content in the SB ring. However, if we include the data published for the galaxies displayed in Figure \ref{Fig_16}, the overall correlation improves significantly ($\rho_{sp}$=0.72).
\citet{Querejeta2019} used all the available data at 100~pc scales and similarly concluded that the dense gas fraction does not seem to be a better predictor of the SFR surface density than the dense gas surface density.

\color{black}

\section{A scenario for star formation in the SB ring} \label{discussion}

The results of our work support the relevance of dynamical environment in setting the efficiency of SF of the dense molecular gas in galaxy disks. Specifically, we find that  SFE$_{\rm dense}$ is comparatively boosted by up to  a factor of three to four in the bar-ring interface of NGC~1068 relative to the regions located elsewhere in the ring.

\begin{figure}[tbp!]
      \centering
   \includegraphics[width=1\linewidth]{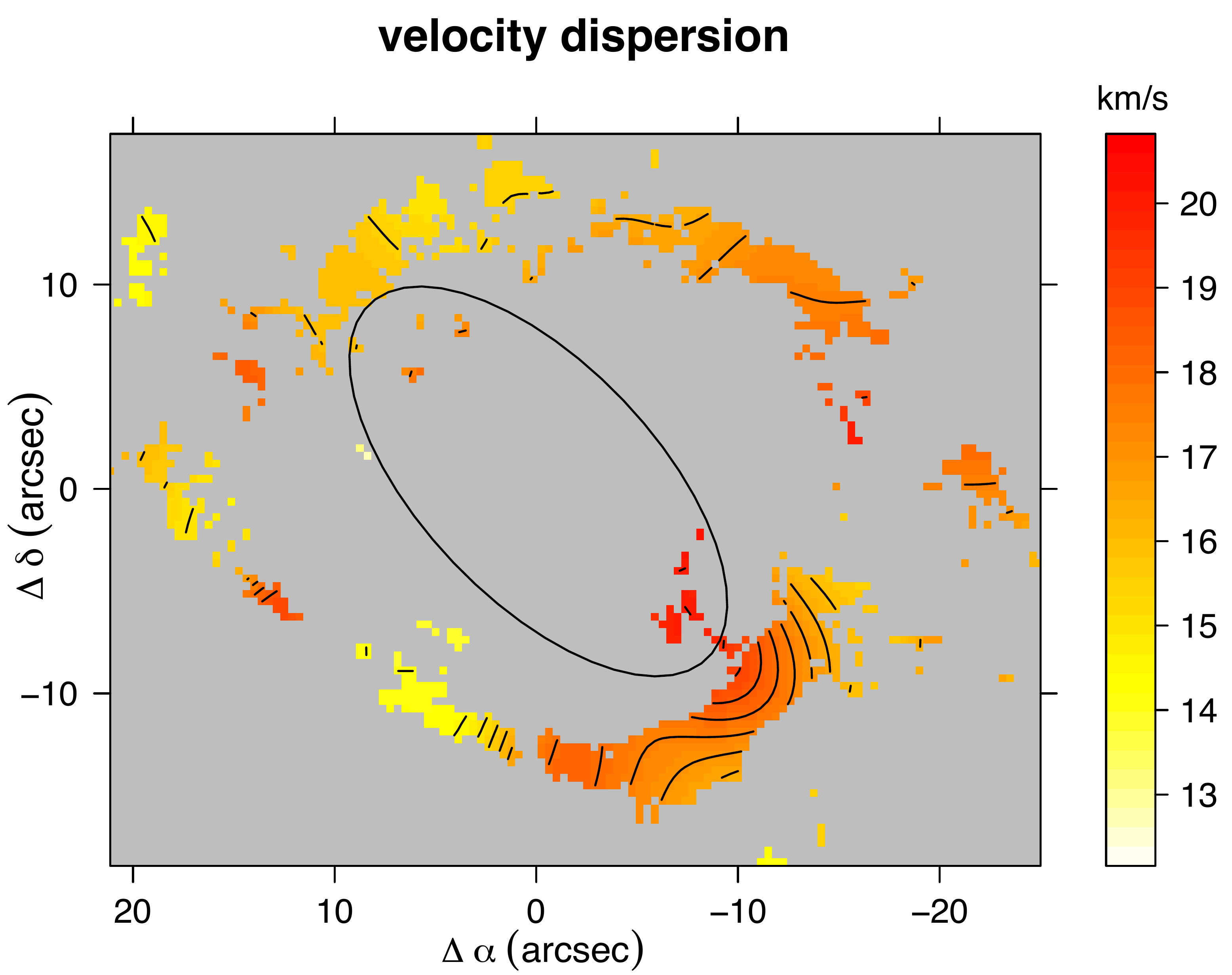}

   \caption{Same as Fig. \ref{Fig_12} but showing the map of the velocity dispersion ($\sigma$) obtained using 400 pc as averaging scale. Contour levels for $\sigma$ go from 14.0 to 19.0 in steps of 0.5~km~s$^{-1}$. } 
               \label{map_sigma}%
   \end{figure}

The velocity dispersion of the dense gas as derived from the HCN(1--0) line  at the "initial resolution" of ALMA  shows little variations over the SB ring.  For an averaging scale of 400~pc, Fig.~\ref{map_sigma} shows that $\sigma$  changes from $\sim14$ to $\sim19$~km~s$^{-1}$ with an estimated mean value of about 16~km~s$^{-1}$~\footnote{A similar result is obtained using the HCO$^+$(1--0) line as an alternative  tracer of the kinematics of the dense gas.}. Specifically, $\sigma$ shows only a moderate $\sim30\%$ increase in the southern region of the bar-ring interface (see Fig.~\ref{map_sigma}). This is in agreement with the picture drawn from the analysis of the velocity dispersion estimated  at 42~pc resolution from the CO(3--2) line, as illustrated by  Fig~10 of \citet{Burillo2014}.  Furthermore, the right panel of Fig.~\ref{Fig_19} shows that, while  $F_{\rm dense}$ increases by $\sim 0.6$~dex, the corresponding increase in $\sigma$ ($\sim 0.12$~dex) is a factor of three smaller. Although $F_{\rm dense}$ and $\Sigma_{\rm dense}$ are not identical quantities, the superlinear trend of  $F_{\rm dense}$ as a function $\sigma$ implies that the trends in the boundedness parameter described in Sect.~\ref{Tdep-b} are mostly driven by the observed boost in $F_{\rm dense}$  in the bar-ring interface. All in all, these results suggest that molecular gas undergoes an efficient compression likely as a result of an enhanced rate of cloud-cloud collisions in the bar-ring interface. However, the shallow trends in $\sigma$ suggest that cloud-cloud collisions have not increased to any significant level the "macroscopic" turbulence between the cores of dense gas probed on scales of 56~pc  in this region.

The SB ring is formed by a tightly wound two-arm spiral structure. The gas accumulates in a region where two 
density-wave resonances are thought to be overlapping: the inner Lindblad resonance (ILR) of the $\sim17$~kpc-size outer stellar oval and the corotation of the $\sim2.6$~kpc-diameter  stellar bar \citep{BH1997, Schinnerer2000,Emsellem2006}. 
The accumulation of gas at the SB ring suggests that the barrier imposed by the corotation of the (nuclear) stellar bar has been overcome \footnote{Rings are expected to form either at the outer Lindblad resonance (OLR) or at the ILR of the bar.}. This could reflect either the influence of the outer stellar oval, which produces gas inflow down to its ILR, or alternatively, the influence of a decoupled (lower pattern speed) spiral mode, which would also induce gas inflow inside its own corotation.  Based on a Fourier decomposition of streaming motions derived from CO(3--2), \citet{Burillo2014} found the signature of systematic gas inflow across the SB ring, an indication that  the two-arm spiral   
constitutes an independent wave feature characterised by a lower pattern speed. The inward motions detected are particularly strong at the region connecting the bar with the spiral \citep[see Fig.~14 of][]{Burillo2014}. In either case, orbital crowding and the intersection of molecular gas on the orbits of the bar and the spiral are seen to enhance SFE$_{\rm dense}$ in their interface region. \citet{Rico-Vilas21} recently analysed the continuum emissions at 147 and 350~GHz observed with ALMA, together with the Pa$\alpha$ HST/NICMOS image of NGC~1068, and identified  14 super star clusters (SSC) in the SB ring. In agreement with the scenario described above, most of  the SSCs (11 out of 14) are seen to be located at the bar ends.


NGC~1068 is not by any means an isolated case that illustrates how the complex molecular gas dynamics in a bar-spiral arm interface can trigger SF activity in a  galaxy disk. In particular,   \citet{Beuther2012} found evidence that the particular dynamical environment of the W43 mini-starburst complex in the Milky Way has likely increased cloud interactions and the subsequent SF activity in this region, which is located near the Galactic bar-spiral arm interface.  Furthermore,  \citet{Beuther2017}, and more recently, \citet{Beslic21} found an increase of SF activity in the bar-spiral arm interface region of the strongly barred galaxy NGC~3627, as a result of the accumulation of gas where the two orbit families (related to the bar and the spiral) intersect, triggering cloud-cloud collisions. Specifically,  the work of \citet{Beslic21} reveals that, compared to other regions in the disk of NGC~3627, the efficiency of the dense molecular gas derived from H$\alpha$/HCN ratios is significantly  enhanced in the bar-spiral arm interface region. Similarly to the scenario described by \citet{Beuther2017} for  NGC~3627, we conclude that a configuration where the bar and spiral may rotate at two different pattern speeds can explain the intense star formation at the bar-ring interface of NGC~1068.

Several hydrodynamical numerical simulations have investigated the role of large-scale bars in AGN fueling and in triggering SF in galaxy disks \citep[e.g.;][]{Ren13, Ren15, Ems15}. In particular,  \citet{Ren15} showed that the SFE and the formation of massive stellar associations are  enhanced at the extremities of the bar to a level comparable to the one observed in galaxy-galaxy interactions as a result of cloud-cloud collisions. This scenario seems to account for the observed pattern of SFE$_{\rm dense}$ in NGC~1068.

\section{Summary and conclusions}\label{summary}

We used ALMA to image the emission of dense molecular gas in the $r\sim1.3$~kpc SB ring of the Seyfert~2 galaxy NGC\,1068 with a 
resolution of $\sim 56$~pc  using  the 1--0 transitions of HCN and HCO$^{+}$.  We also used ancillary data of CO (1--0), as well as CO(3--2) and its underlying continuum 
emission at the resolutions of $\sim100$~pc and $\sim40$~pc, respectively. These observations allow us 
to probe a wide range of molecular gas densities ($n_{\rm H_2}\sim10^{3-5}$cm$^{-3}$). The SF rate is derived from Pa$\alpha$ line emission imaged by HST/NICMOS. 
We analysed the influence of the dynamical environment on different formulations of SF relations in the SB ring and compared our results with the general predictions of 
density-threshold and turbulent SF models.

The main results of this paper are summarized as follows:

\begin{itemize}

\item

We derived  spatially resolved KS laws for a set of seven spatial resolutions, ranging from $\sim40$~pc  up to  $\sim700$~pc. We studied how these relations change depending on the adopted aperture sizes and on the 
choice of molecular gas tracer. For a given spatial resolution the correlation parameters  derived from the high density tracers (CO(3--2), HCN(1--0) and HCO$^+$(1--0)) are about a factor of two to three larger than that derived from CO(1--0). 

\item

The KS correlations lose statistical significance below a critical spatial scale $\approx$ 300-400~pc common for all gas tracers. For spatial scales $\geq300-400$~pc the correlation improves monotonically as a function of the aperture size for all gas tracers excluding CO(1--0). While dust continuum emission shows a behaviour similar to the rest of high density tracers, the dust-based KS correlation is significant already at 100~pc scales.

\item

NGC~1068  lies within the general  $\Sigma_{\rm SFR}$--$\Sigma_{\rm dense}$ linear KS relationship  derived from a compilation of data obtained in other galaxies.
In particular, the location of NGC 1068 in the KS plot is intermediate between the one of normal galaxies and (U)LIRG, a result that underlines the relatively extreme conditions in the SB ring. 

\item

The efficiency of SF of the dense molecular gas, defined as SFE$_{\rm 
dense}\equiv\Sigma_{\rm SFR} / \Sigma_{{\rm dense}}$, shows a scattered distribution as a function of the HCN luminosity at the "initial resolution" of ALMA  ($\sim 56$~pc) around a mean value of $\simeq0.01$Myr$^{-1}$. 
However, we find evidence of a significant environmental dependence of SFE$_{\rm dense}$ reflected in the existence of systematic trends across the different regions of the SB ring, which are inconsistent with the 
predictions of density-threshold models.

\item

 With the aim of resolving the degeneracy associated with the SFE$_{\rm dense}$-$L'$(\rm HCN) plot, we explored an  alternative prescription for SF relations, which includes the 
dependence of SFE$_{\rm dense}$ on the boundedness  of the gas, measured by the parameter $b$  defined as $b\equiv\Sigma_{\rm dense}/\sigma^{2} \propto \alpha_{\rm vir}^{-1}$.
We identified two branches in the version of the SFE$_{\rm dense}$--$b$ plot derived for an averaging scale of 400~pc. 
The two branches correspond to two dynamical environments defined by their proximity to the region where the SB ring is connected to the stellar bar of NGC~1068. 
\item

 We studied the trends in  SFE$_{\rm dense}$ as a function of the dense gas fraction ($F_{\rm dense}$), the stellar mass surface density ($\Sigma_{\rm star}$), and the velocity dispersion ($\sigma$) in the SB ring using different
 averaging scales.  We find that SFE$_{\rm dense}$ correlates both with $F_{\rm dense}$ and $\Sigma_{\rm star}$. 
 However, we find no significant correlation of SFE$_{\rm dense}$ with  $\sigma$. 
 These results differ to a large extent from those derived from previous kpc-scale studies of galaxy disks and high-resolution ($\sim100$~pc) observations of M~51.

\item
 The trends in the boundedness parameter in the SB ring are
mostly driven by the observed boost in $F_{\rm dense}$  in the bar-ring interface region. The results reflect a significant compression of molecular gas as a result of an enhanced rate of cloud-cloud collisions in the bar-ring interface on the one hand, and an efficient dissipation of turbulence on the other.

\end{itemize}

The outcome of our work emphasizes the relevance of different dynamical environments to modulating  SFE$_{\rm dense}$ in galaxy disks. Taken at face value,  this result is inconsistent with the 
predictions of density-threshold models, which foresee the existence of a nearly constant canonical value for SFE$_{\rm dense}$. There is mounting observational evidence that SFE$_{\rm dense}$ can show 
systematic trends based on studies carried out both in our Galaxy and nearby galaxies. However, the theoretical scenarios advanced to explain the variations of  SFE$_{\rm dense}$ reported in the literature are diverse. 
This diversity reflects the variety of trends observed in   SFE$_{\rm dense}$ as a function of key physical variables  such as $F_{\rm dense}$,  $\Sigma_{\rm star}$, and $\sigma$ in galaxy disks.   
Similarly to the case of the barred galaxy NGC~3627 discussed by \citet{Beuther2017}, the enhanced SF activity in the bar-ring interface of NGC~1068 could be explained by a dynamical configuration in which the bar and the two-arm spiral feature are independent $m=2$ modes rotating at different pattern speeds. Further studies are required to explore the frequency of this type of dynamical decoupling between bars and spirals and their role at setting the SF efficiency
of dense molecular gas in galaxy disks in general and in galactic rings in particular.

\begin{acknowledgements}
We thank the referee for the useful comments and suggestions. This paper makes use of the following ALMA data: ADS/JAO.ALMA$\#$2013.1.00055.S and $\#$2011.0.00083.S.
ALMA is a partnership of ESO (representing its member states), NSF (USA), and NINS (Japan), together with NRC (Canada) and NSC and ASIAA (Taiwan),
in cooperation with the Republic of Chile. The Joint ALMA Observatory is
operated by ESO, AUI/NRAO, and NAOJ. The National Radio Astronomy
Observatory is a facility of the National Science Foundation operated under cooperative
agreement by Associated Universities, Inc. 
We used observations made
with the NASA/ESA {\it Hubble} Space Telescope, and obtained from the {\it Hubble}
Legacy Archive, which is a collaboration between the Space Telescope Science
Institute (STScI/NASA), the Space Telescope European Coordinating Facility
(ST-ECF/ESA), and the Canadian Astronomy Data centre (CADC/NRC/CSA). MSG acknowledges support from the Spanish Ministerio de Econom\'{\i}a y  
Competitividad through the grants BES-2016-078922 and ESP2015-68964-P. SGB and MQ acknowledge support from the  research project PID2019-106027GA-C44 of the Spanish Ministerio de Ciencia e Innovaci\'on. AF acknowledges support from the  research project PID2019-106235GB-I00. 
AAH, SGB and AU work was funded by grant PGC2018-094671-B-I00 funded by MCIN/AEI/ 10.13039/501100011033 and by ERDF A way of making Europe. AAH's  work was done under project No. MDM-2017-0737 Unidad de Excelencia "Mar\'{\i}a de Maeztu"- Centro de Astrobiolog\'{\i}a (INTA-CSIC). MPS acknowledges support from the Comunidad de Madrid through the Atracci\'on de Talento Investigador Grant 2018-T1/TIC-11035 and PID2019-105423GA-I00 (MCIU/AEI/FEDER,UE). LC and MSG acknowledge support from  the research project PID2019-106280GB-100.
\end{acknowledgements}

\bibliographystyle{aa}
\bibliography{NGC1068_SFrelations}

\begin{appendix} 
\section{Determination of the missing flux in HCN and CO maps } \label{app1}


 \begin{figure*}[htp]
   \centering
   \includegraphics[width=0.9\linewidth]{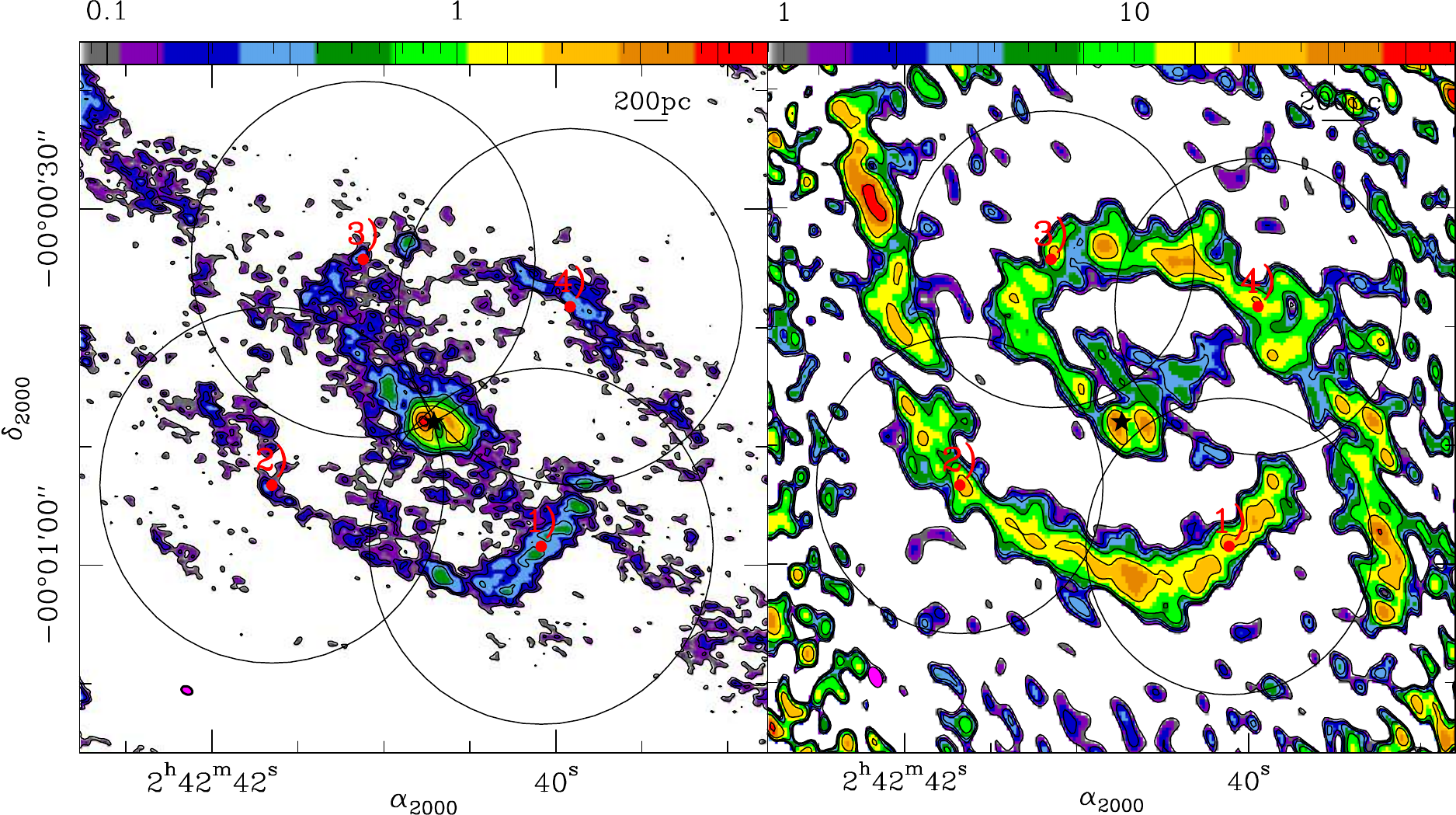}
   \caption{{\it Left panel}: Overlay of the four $\sim30\arcsec$-size apertures (black circles, identified by the numerical labels 1-4) selected to cover the SB ring of NGC~1068  on the ALMA HCN(1--0) map (colour scale and contours as in Fig.~\ref{Fig_1}), used in Appendix~\ref{app1} to estimate the missing flux correction. {\it Right panel}: same as {\it left panel} but showing the overlay of the four $\sim12\arcsec$-size apertures (black circles, identified by the numerical labels 1-4) selected to cover the SB ring of the galaxy on the PdBI CO(1--0) map of \citet{Schinnerer2000}. Colour scale is given in Jy~km~s$^{-1}$~beam$^{-1}$ units and the position of the AGN is identified by the grey star markers in both maps. The (magenta) filled ellipses at the bottom left in the two panels represent the beam sizes.}
              \label{anexo1}%
   \end{figure*}    
    


\begin{figure}[tb!]
   \centering
   \includegraphics[width=.48\linewidth]{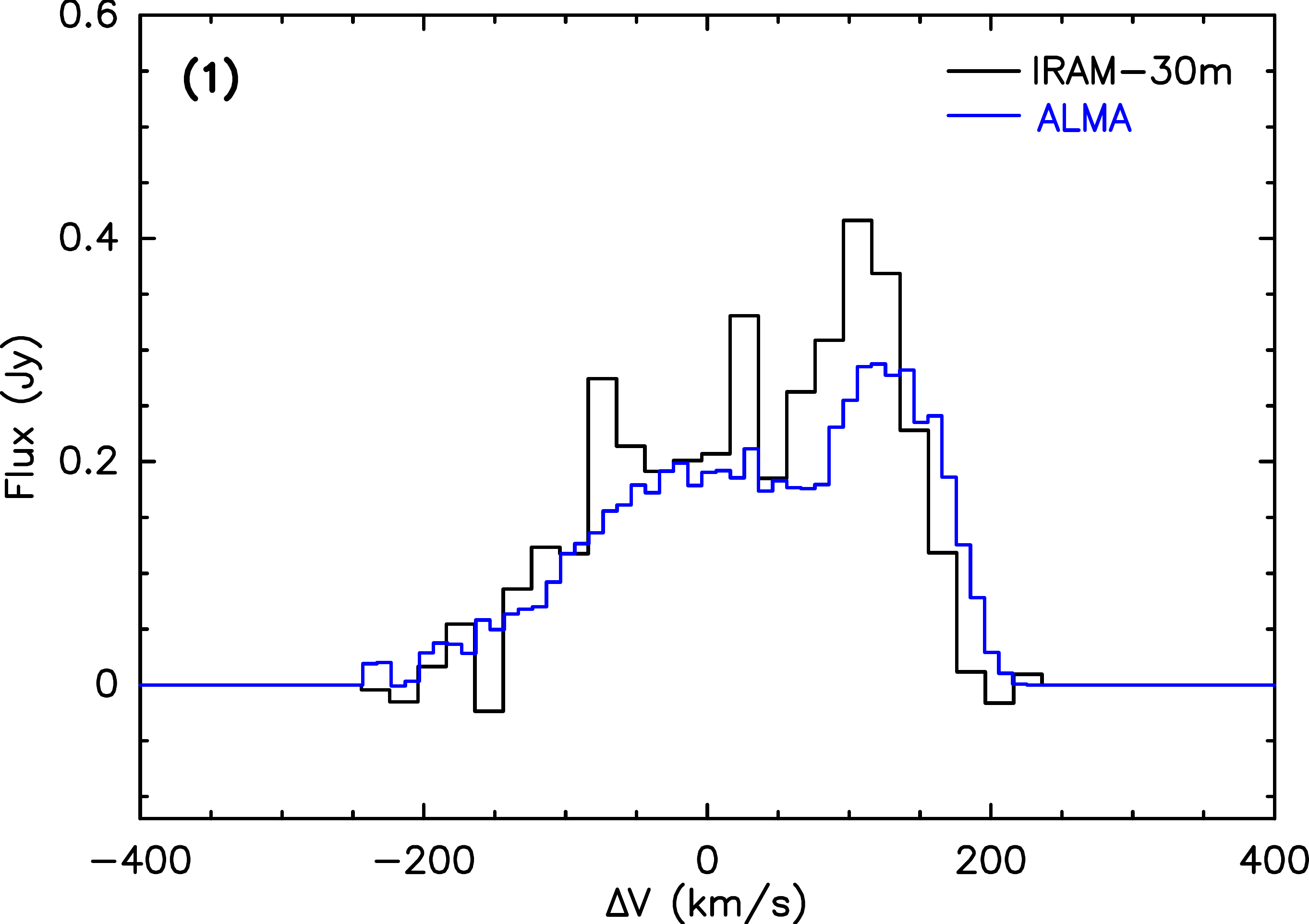}
      \includegraphics[width=.48\linewidth]{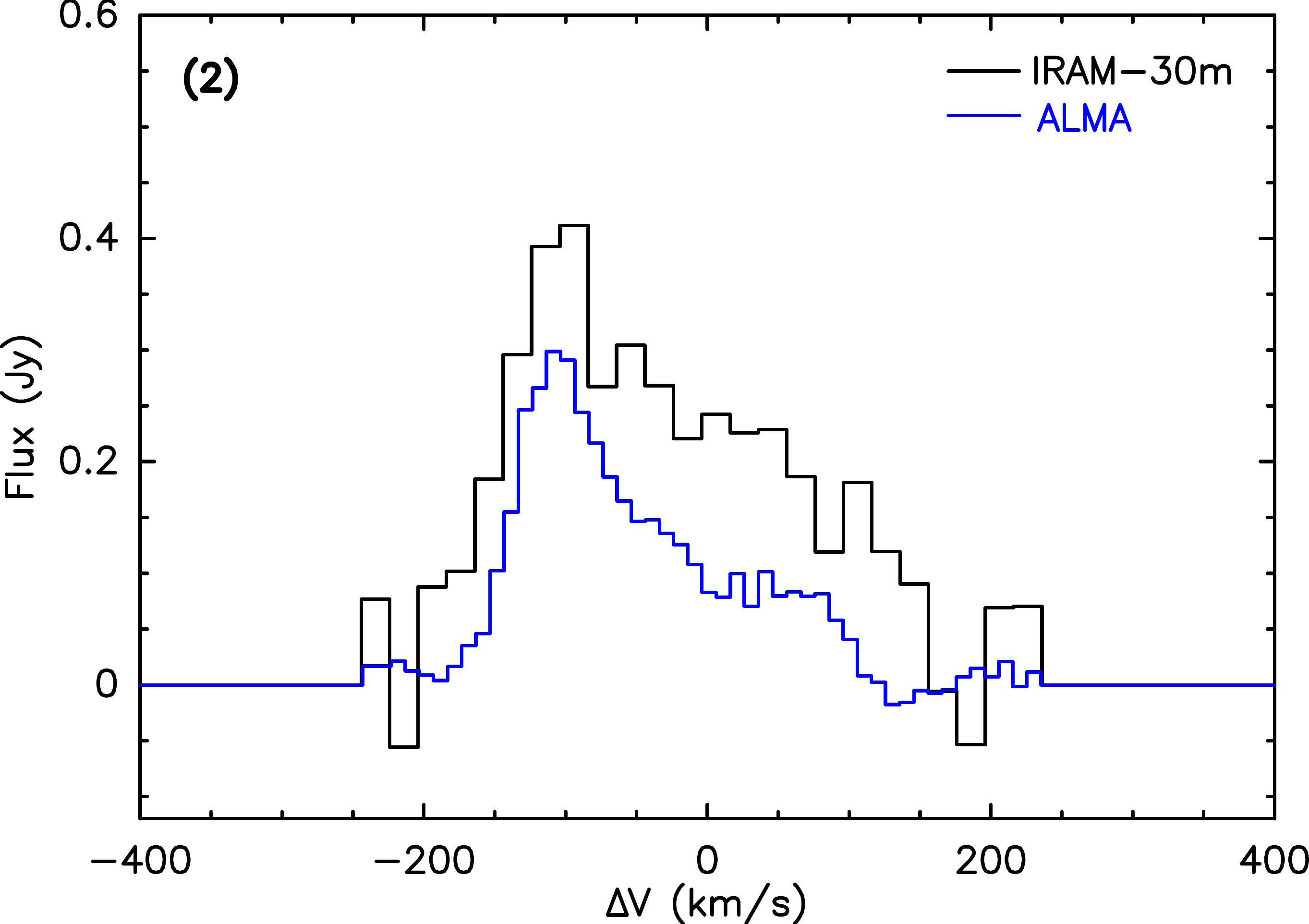}
       \includegraphics[width=.48\linewidth]{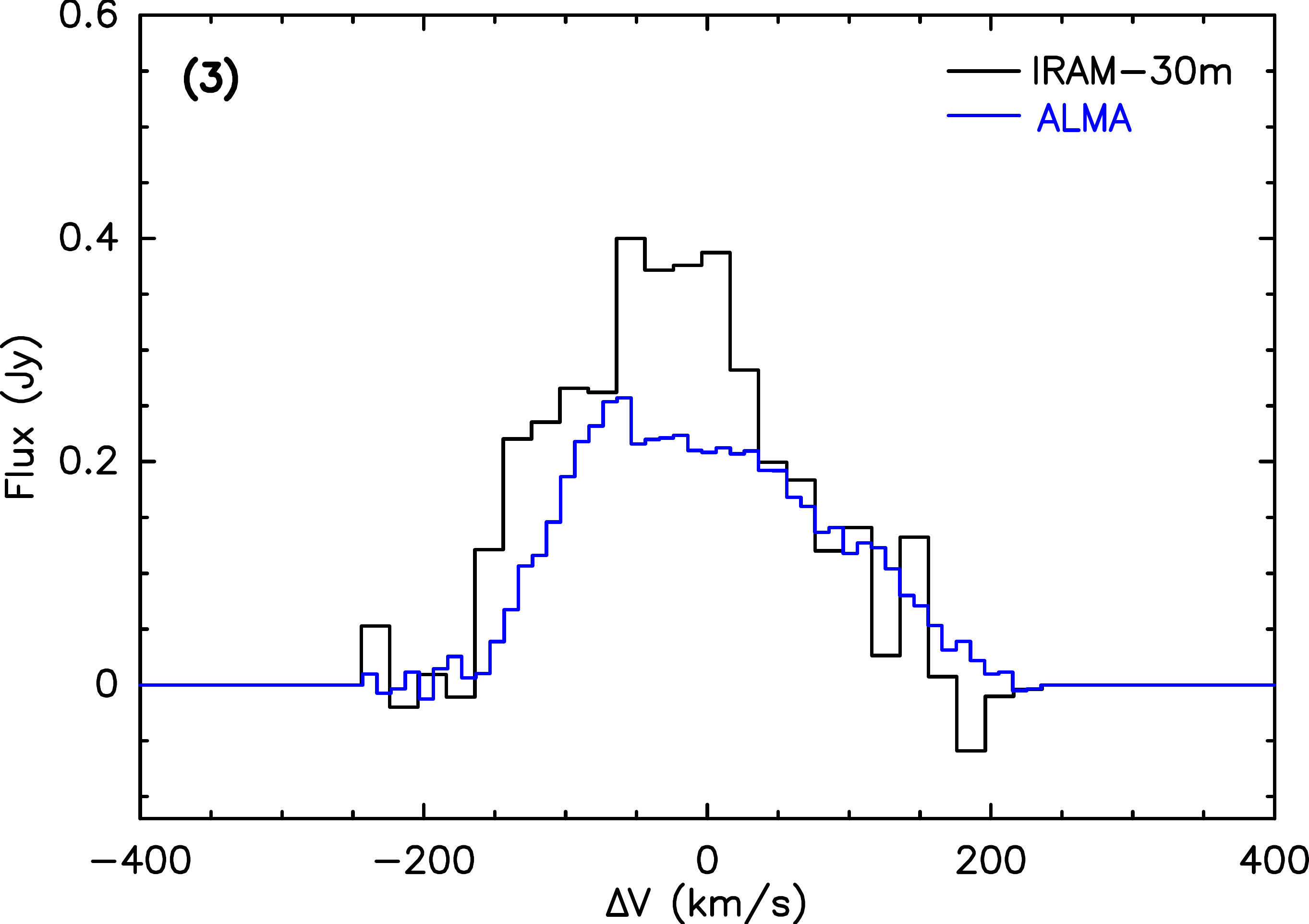}
        \includegraphics[width=.48\linewidth]{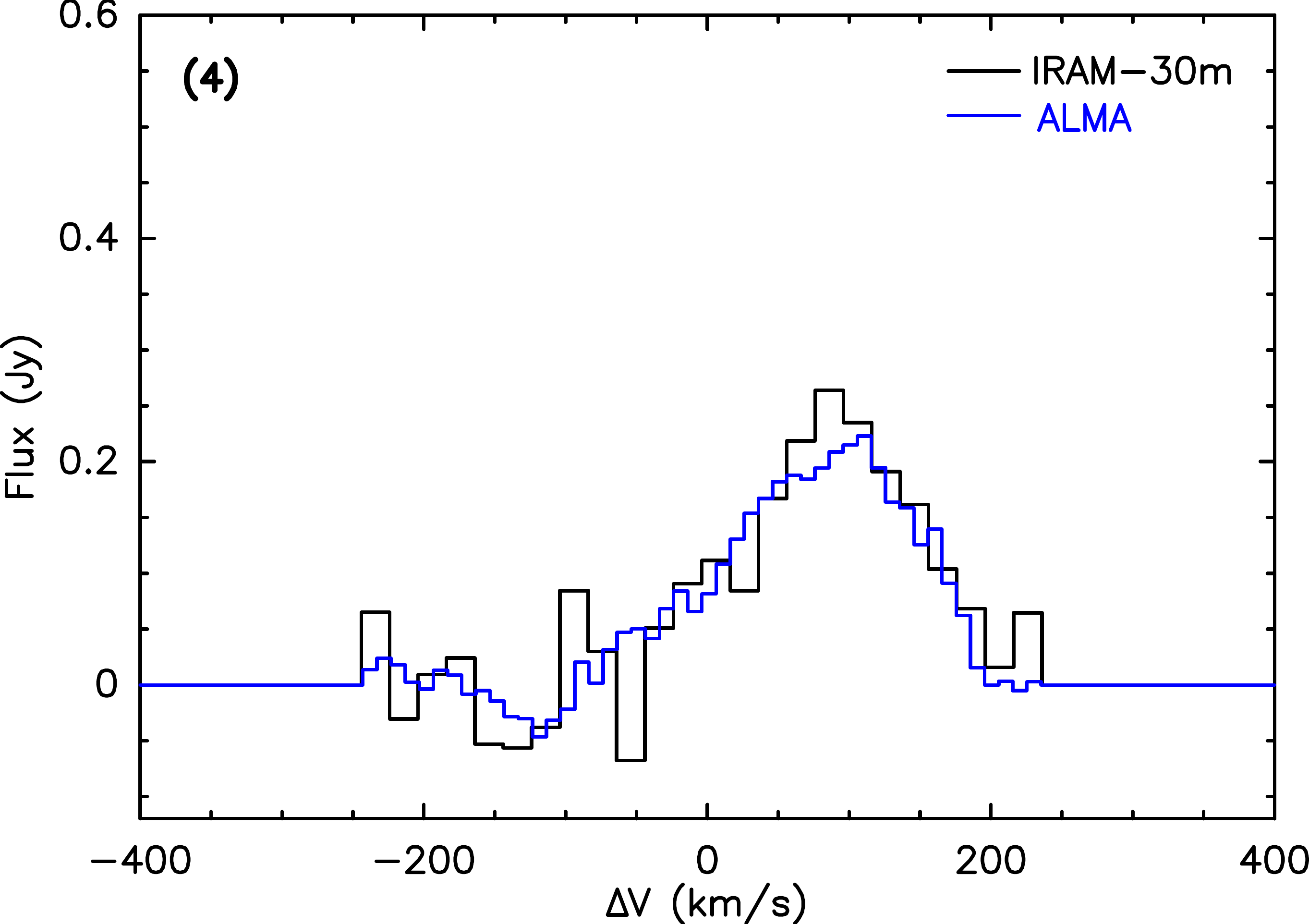}
   \caption{Comparison of the HCN(1--0) emission observed by ALMA (blue histograms) and the IRAM-30m telescope (black histograms) inside the four selected regions  identified in Fig.~\ref{anexo1} (labelled as 1, 2, 3, and 4 in each panel).}
              \label{figanexo1}
   \end{figure}


\begin{figure}[tb!]
   \centering
   \includegraphics[width=.48\linewidth]{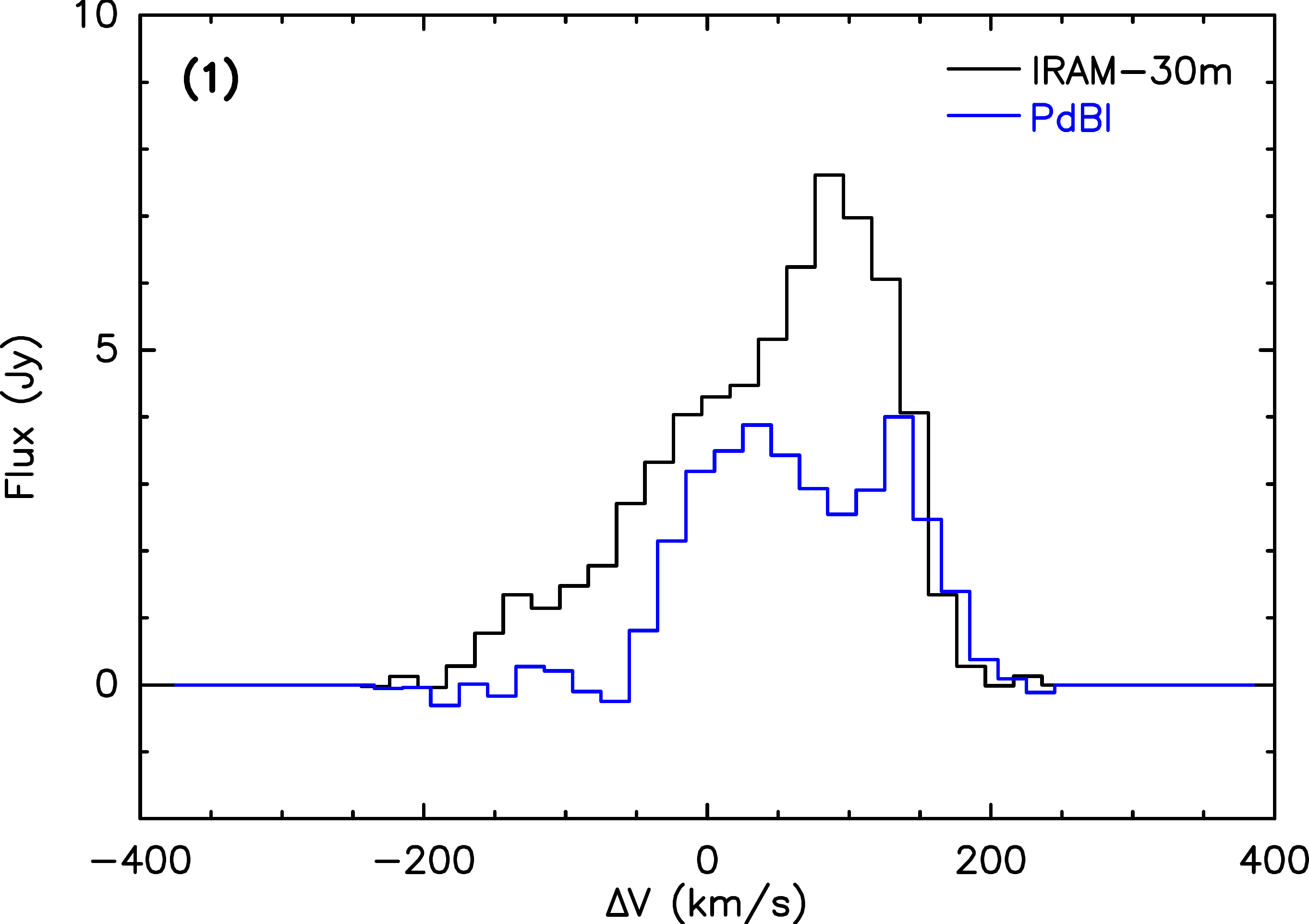}
      \includegraphics[width=.48\linewidth]{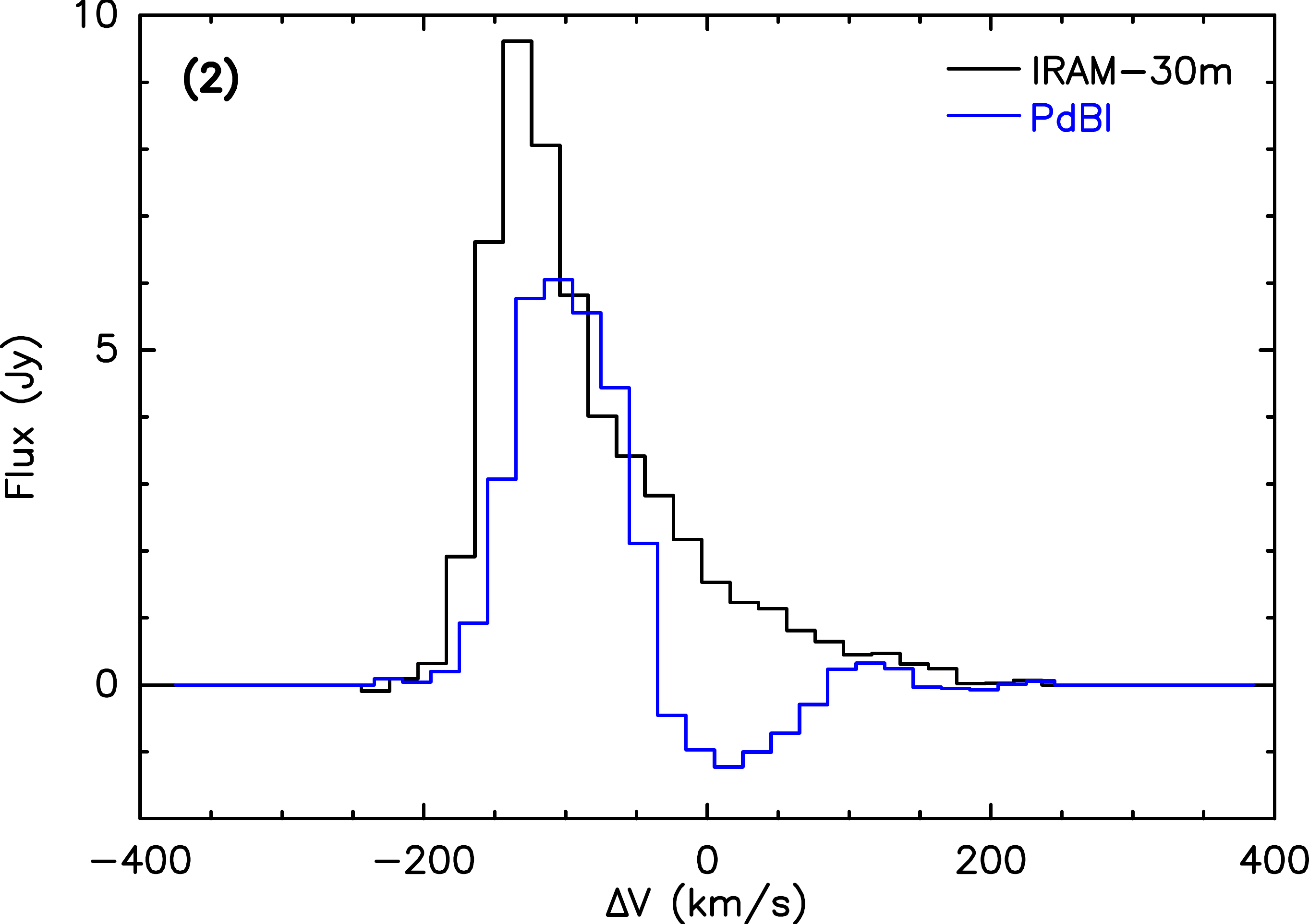}
       \includegraphics[width=.48\linewidth]{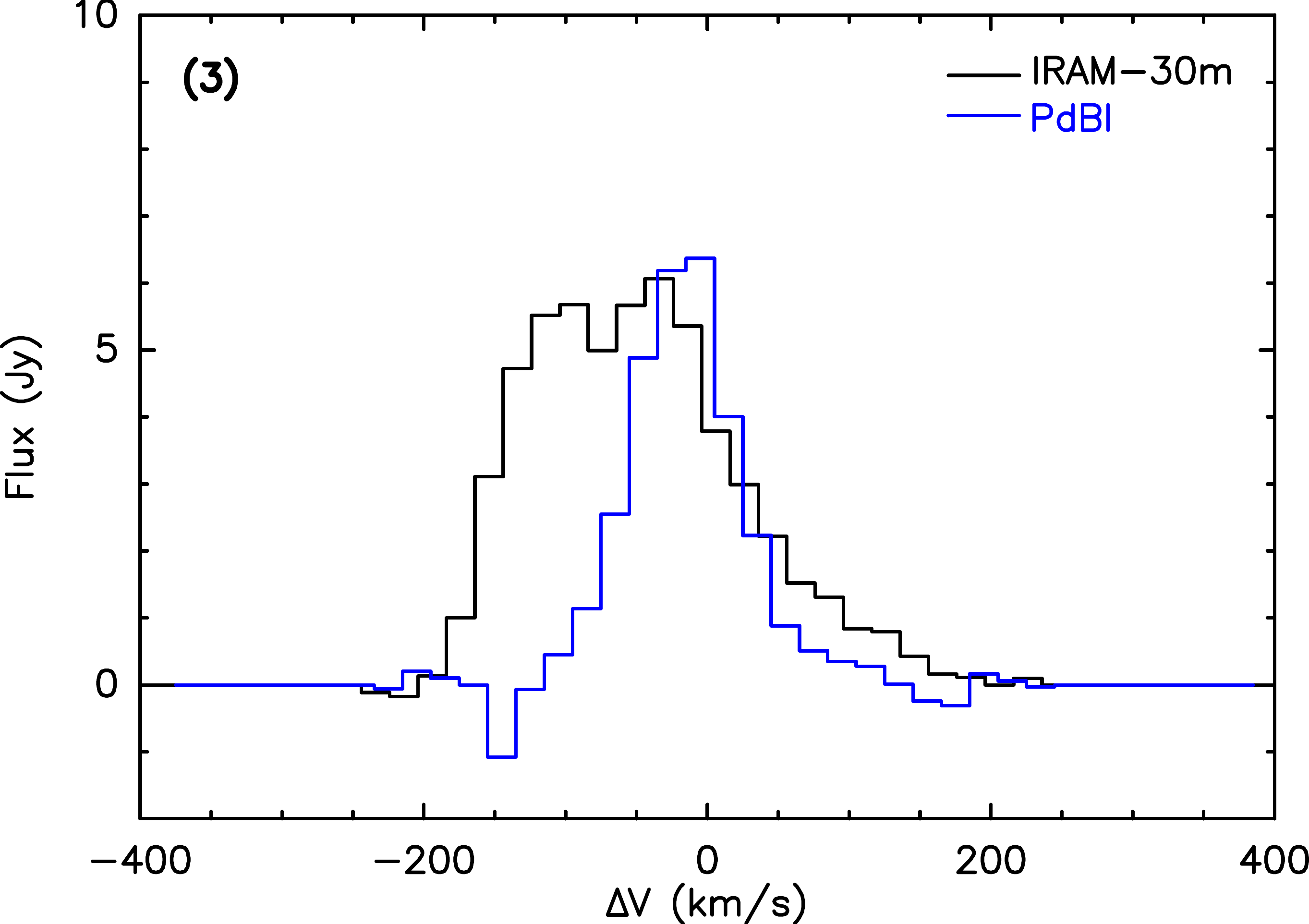}
        \includegraphics[width=.48\linewidth]{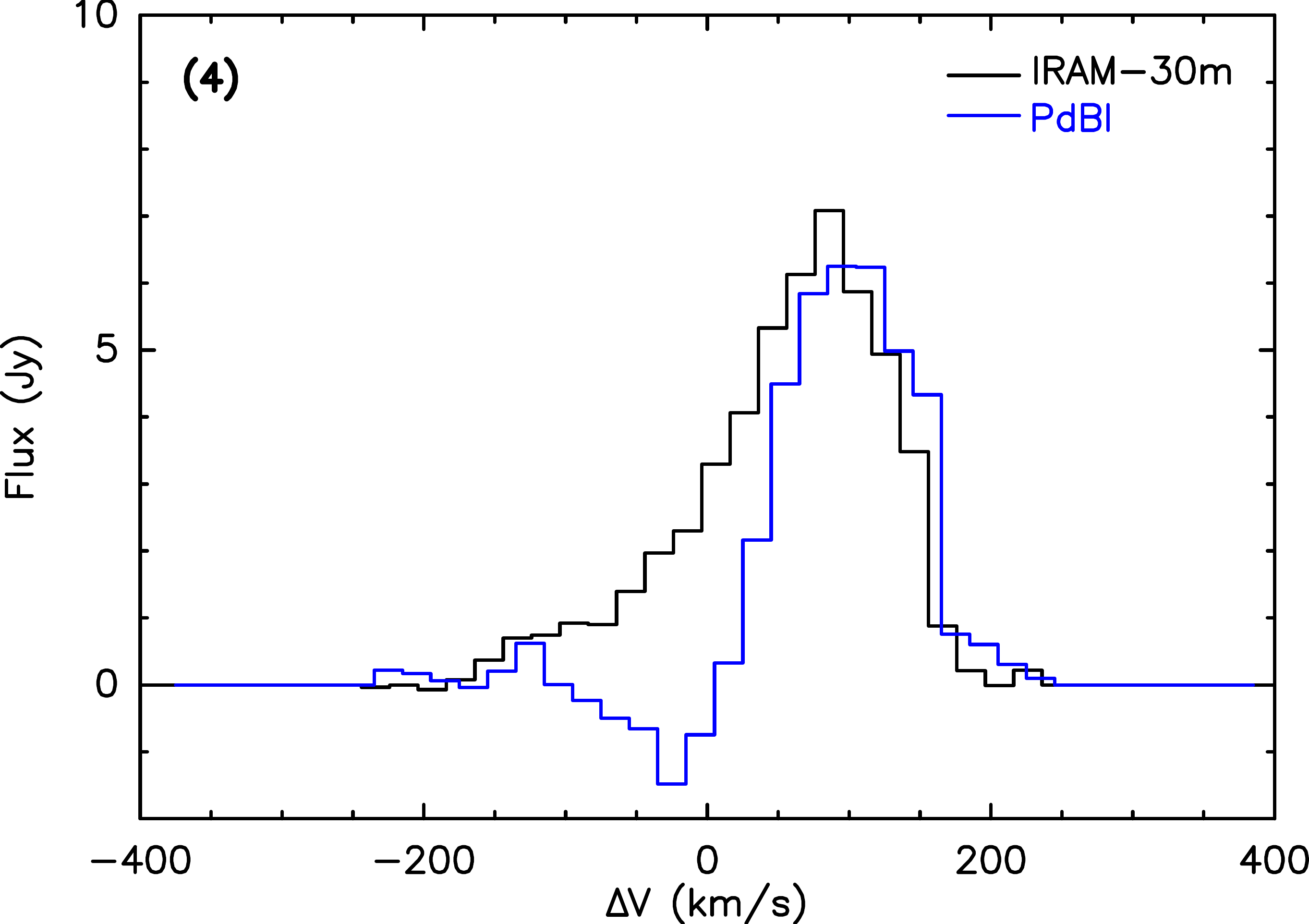}
   \caption{Same as Fig.~\ref{figanexo1}, but showing the comparison of the CO(1--0) emission observed by the IRAM-PdBI (blue histograms) and the IRAM-30m telescope (black histograms) inside the four selected regions  identified in Fig.~\ref{anexo1} (labelled as 1, 2, 3, and 4 in each panel).}
              \label{figanexo2}
   \end{figure}

We estimated the missing flux in the HCN(1--0) map obtained by ALMA and in the CO(1--0) map of the IRAM array (PdBI) \citep{Schinnerer2000}. To this aim, we measured the spatially-integrated fluxes in both  lines using a set of four apertures covering  the region occupied by the SB ring of NGC~1068. The interferometer fluxes are  compared with the corresponding single-dish fluxes measured from observations obtained by the IRAM-30m telescope (Usero et al., private communication). 
Figure \ref{anexo1} illustrates the apertures chosen overlaid on the HCN ({\it left panel}) and CO ({\it right panel}) interferometer images of the galaxy. The sizes of the apertures used to derive the spatially-integrated fluxes of the interferometer maps correspond to the single-dish beams of the IRAM-30m telescope, which are  $\sim30\arcsec$ and $\sim24\arcsec$ for HCN and CO, respectively. 

Figures \ref{figanexo1} and  \ref{figanexo2} compare  the  spectra of HCN and CO obtained  by integrating the emission of the lines inside the four selected regions shown in Fig.~\ref{anexo1} with the IRAM-30m spectra. Based on this comparison we estimate that on scales of $\sim 30\arcsec$ ($\sim2$~kpc) approximately $20-25\%$ of the total flux gets filtered in the HCN ALMA map. A similar estimate for the CO map yields a  flux filtering of about 40-45$\%$ on scales $\sim 24\arcsec$ ($\sim1.7$~kpc). This is about a factor of two larger than the value estimated by \citet{Schinnerer2000}, who compared the fluxes measured with the PdBI with 
 those measured with the combined BIMA array and Kitt Peak single-dish telescope datasets on scales $\sim 55\arcsec$ (3.8~kpc).

As in this paper we analyse spatial scales that are much smaller than the critical scales reported above ($\thickapprox$ 2-4 kpc), we can foresee that the missing flux factors will be significantly less than 25-45$\%$  for the scales that are relevant to our study.

\section{Pa$\alpha$ analysis as SFR tracer} \label{anexo-pa}

\begin{figure}[h!]
   \centering
   \includegraphics[width=.9\linewidth]{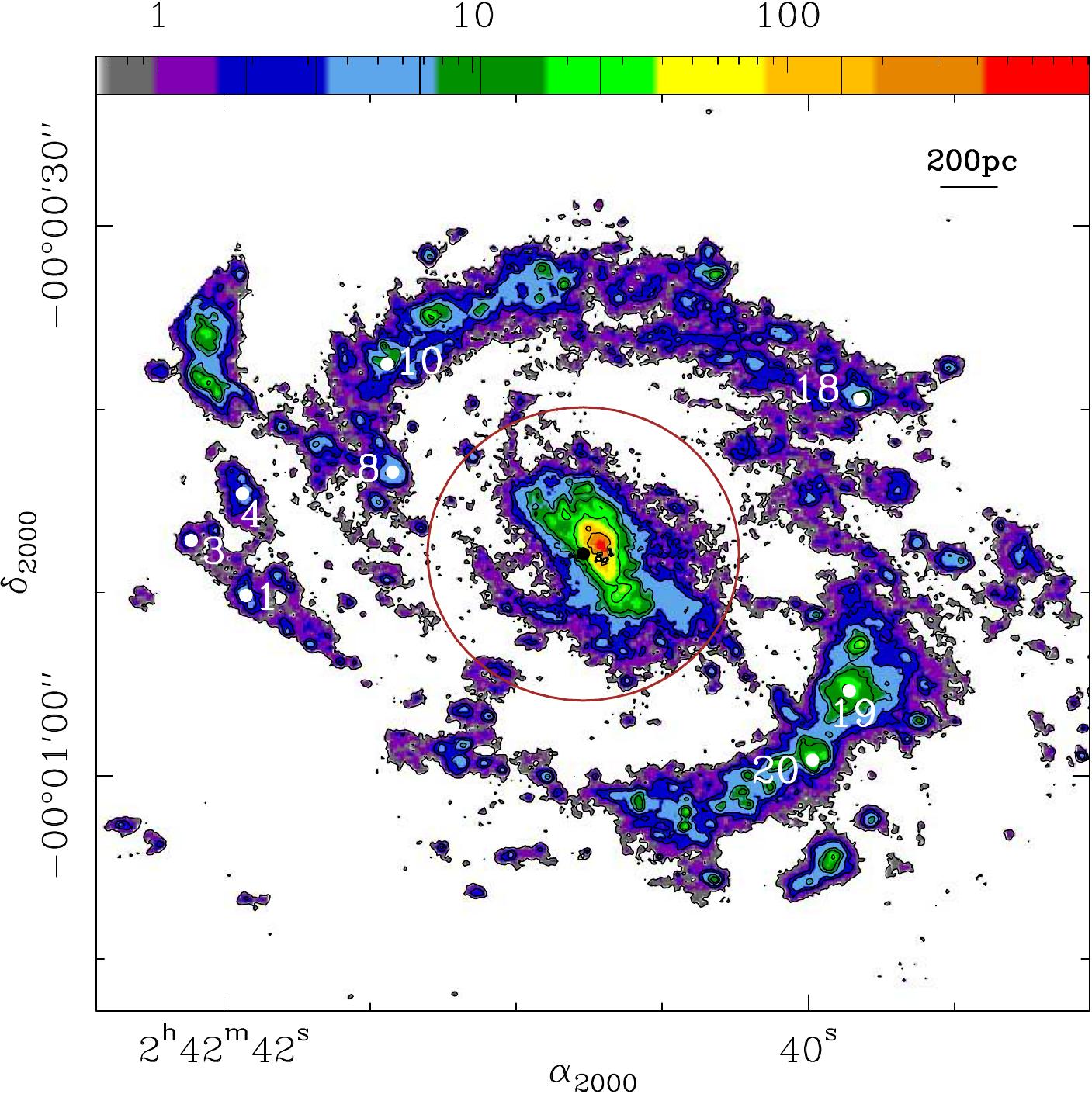}
   \caption{HST/NICMOS Pa$\alpha$ image of NGC 1068. The (white) labels identify the regions selected to estimate the extinction of Pa$\alpha$, according to the nomenclature used by \citet{angeles2000}. } 
              \label{anexo2}%
   \end{figure}

To validate our working assumption, namely that the Pa$\alpha$ image used in this work to derive the distribution of recent SF in NGC~1068 does not require a significant correction by dust extinction,    
we compared the Pa$\alpha$ fluxes measured in the HST/NICMOS image of the galaxy with the H$\alpha$ fluxes measured using the ground-based image of \citet{angeles2000} 
over a representative number of hot spots of the SB ring. Figure \ref{anexo2} shows the position of the eight regions considered in this comparison selected from Table~4 of \citet{angeles2000}.
 We derived the Pa$\alpha$  spatially-integrated fluxes inside the eight selected regions using the sizes of  H$\alpha$ SF knots listed in Table~4 of \citet{angeles2000} and obtained the H$\alpha$/Pa$\alpha$ flux ratios for each aperture (listed in Table~\ref{tabla-anexo}). 
 To estimate the colour excess ($E(B-V)$) and extinction (A$_{Pa\alpha}$) values we used the prescriptions of  \citet{calzetti2001} \citep[see also][]{calzetti1994, calzetti2000} and the standard equation:
 
 \begin{equation}
F_{\rm obs}(\lambda)=F_{\rm int}(\lambda)10^{-0.4k(\lambda)E(B-V)} , 
\end{equation}
where $F_{\rm obs}$ and $F_{\rm int}$ are the observed and intrinsic fluxes for the lines, respectively, and $E(B-V)$ is the colour excess. We assumed that the differential extinction between the two lines is determined by  $k(H\alpha)-k(Pa\alpha)=2.104$ \citep{calzetti2001}. Table \ref{tabla-anexo} lists  the H$\alpha$/Pa$\alpha$ line ratios  as well as the values estimated for $E(B-V)$ and  A$_{Pa\alpha}$  for the eight selected regions.

The median value for the distribution of H$\alpha$/Pa$\alpha$ ratios listed in Table~\ref{tabla-anexo} is $\sim6.8$. The latter is close to the expected value corresponding to the canonical case B recombination: $\sim 7.8$. 
The implied mean extinction correction at 1.875 $\mu$m is A$_{\rm Pa\alpha}$=0.03 
mag, which translates into very low dust opacities for the ensemble of the studied SF knots. Although in 
regions 18,19 and 20,  A$_{\rm Pa\alpha}$ is seems to reach values $\sim0.1-0.2$, the implied opacities  are still compatible with optically thin emission for  Pa$\alpha$. 
Moreover, assuming that the flux scale is uncertain to $\pm$ 20$\%$ due to absolute calibration errors, we can therefore conclude that Pa$\alpha$ emission is not significantly affected by dust extinction 
in the SB ring of NGC 1068.

\begin{table}[htbp]
\begin{center}
\begin{tabular}{cccc} 
\hline
\textbf{Region} & \textbf{H$\alpha$/Pa$\alpha$} & \textbf{E(B-V) [mag]} & \textbf{A$_{Pa\alpha}$ [mag]} \\
\hline 
1 & 7.7 & 0.01 & $\leq$ 0.01 \\ 
3 & 9.9 & -0.12 & -0.05 \\ 
4 & 7.7 & 0.01 & $\leq$ 0.01 \\ 
8 & 9.0 & -0.07 & -0.03 \\
10 & 5.9 & 0.15 & 0.05 \\ 
18 & 3.8 & 0.37 & 0.14 \\ 
19 & 2.1 & 0.67 & 0.24 \\ 
20 & 3.5 & 0.41 &  0.15 \\ \hline
median & 6.8 & 0.08 & 0.03 \\ \hline
\end{tabular}
\caption{Observed line ratios and derived extinction  parameters for the SB ring knots of NGC~1068.} \
\tablefoot{We identify the SB ring knots according to the label convention used in Table~4 of  \citet{angeles2000} in Column (1). Columns (2), (3), and (4)  list the H$\alpha$/Pa$\alpha$ ratios, the colour excess ($E(B-V)$) and the extinction (A$_{Pa\alpha}$) values for each region, respectively. The median values of columns (2), (3), and (4)  are listed at the last row.}
\label{tabla-anexo}
\end{center}
\end{table}


\section{Kennicutt-Schmidt plots} \label{AppD}
\begin{figure*}[h!]
   \centering
    \includegraphics[width=.97\linewidth]{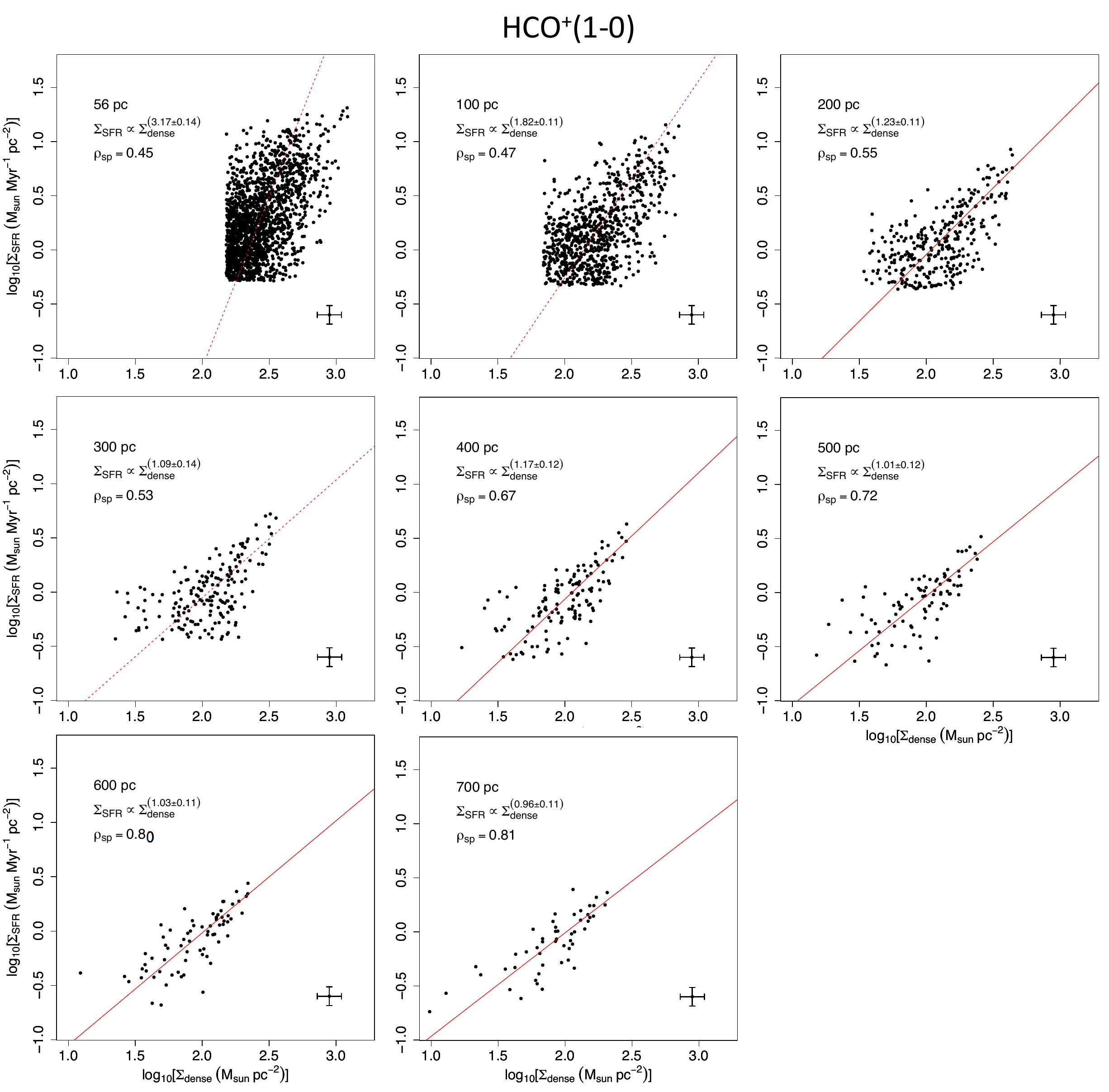}
     \caption{Same as Fig. \ref{Fig_5}, but showing all the spatial scales used in this work for HCO$^{+}$(1-0). Vertical and horizontal errorbars at the lower right corner of each panel account for the typical uncertainties, which amount to $\pm$0.09 dex on both axes.}
              \label{Fig_29}
   \end{figure*}

\begin{figure*}[h!]
   \centering
    \includegraphics[width=.97\linewidth]{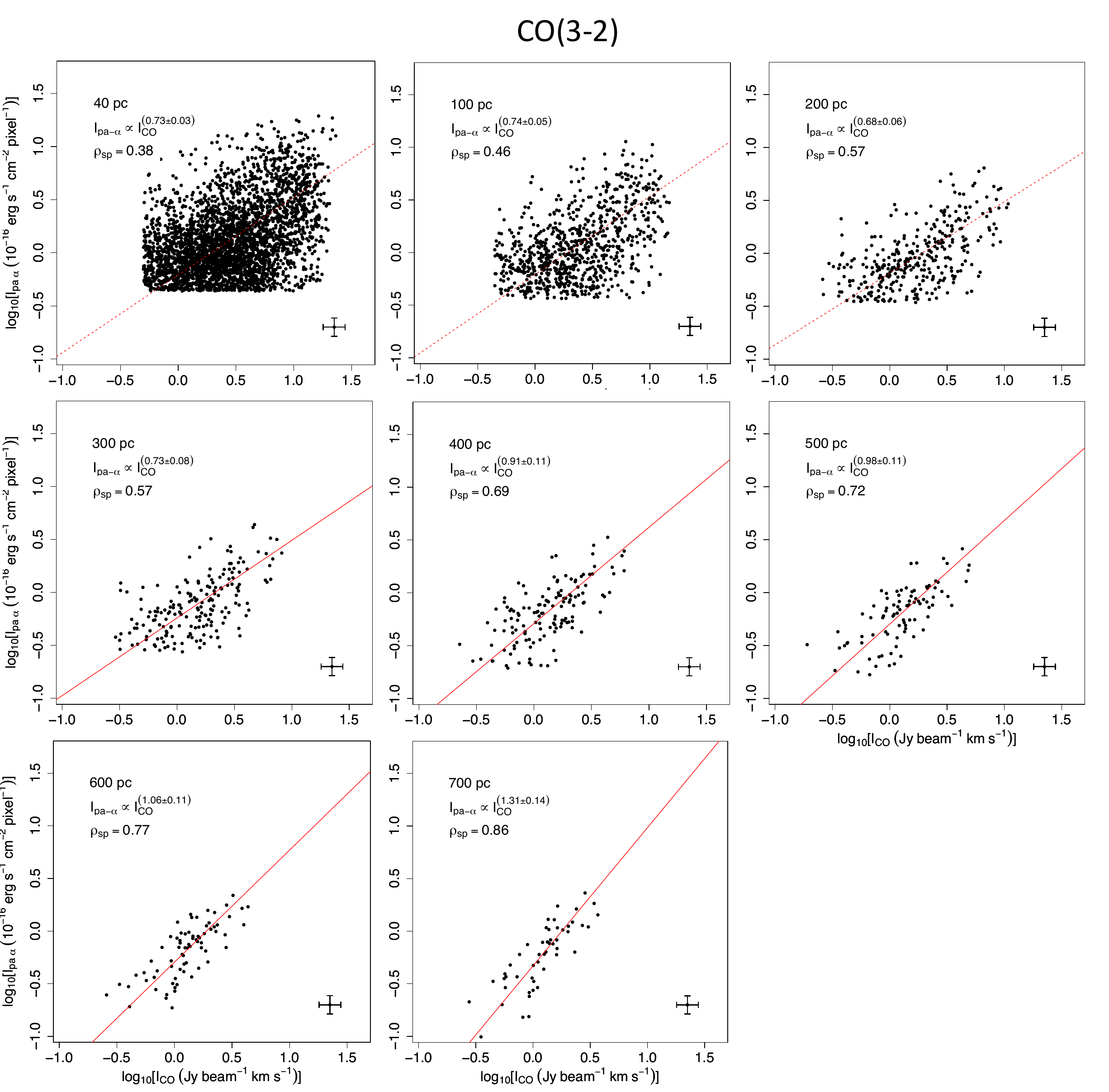}
    \caption{Same as Fig. \ref{Fig_5}, but showing all the spatial scales used in this work for CO(3-2). Vertical and horizontal errorbars at the lower right corner of each panel account for the typical uncertainties, which amount to $\pm$0.09 dex on both axes.}
              \label{Fig_30}
   \end{figure*}

\begin{figure*}[h!]
   \centering
    \includegraphics[width=.97\linewidth]{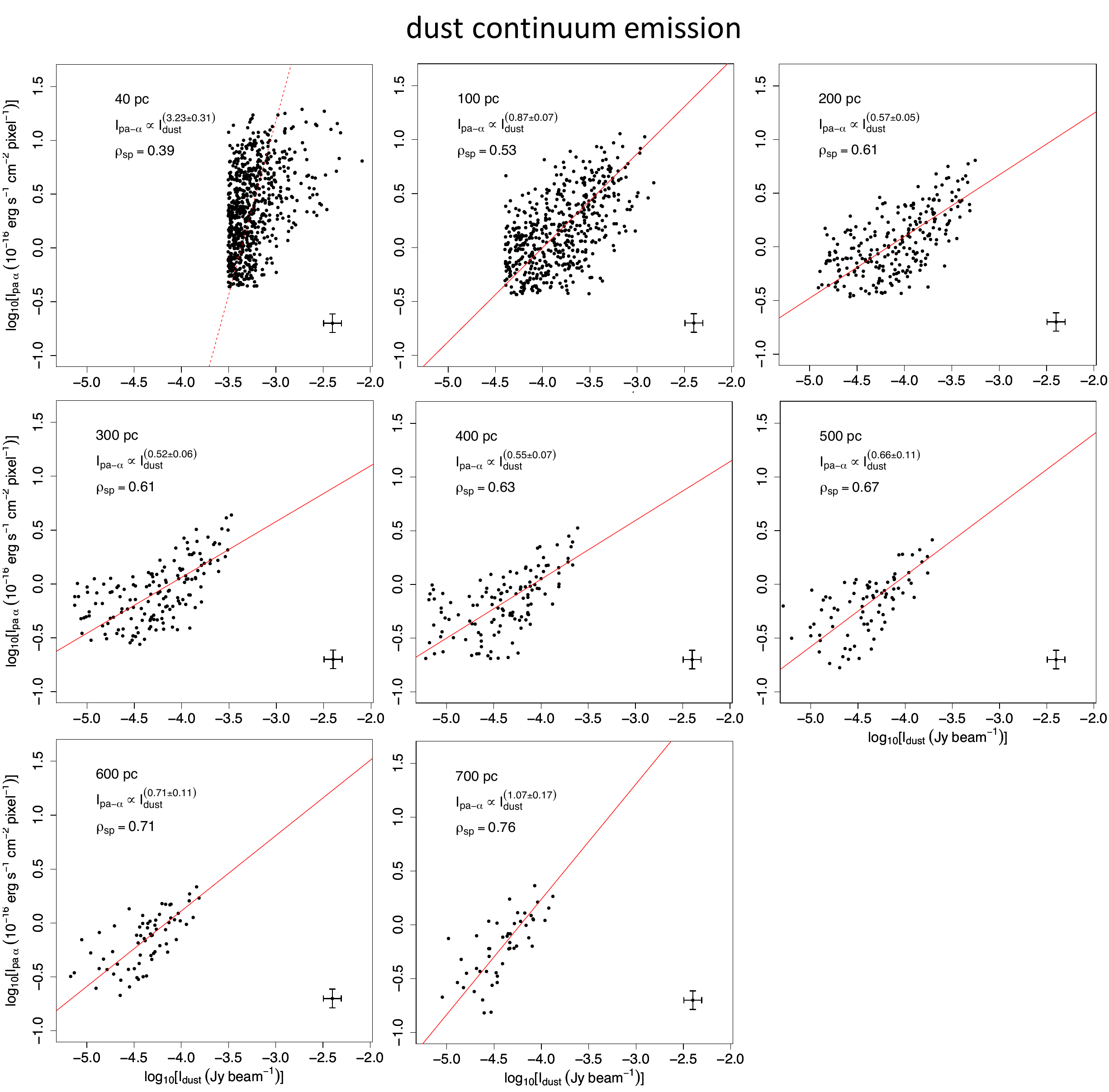}
    \caption{Same as Fig. \ref{Fig_5}, but showing all the spatial scales used in this work for dust continuum emission. Vertical and horizontal errorbars at the lower right corner of each panel account for the typical uncertainties, which amount to $\pm$0.09 dex on both axes.}
              \label{Fig_31}
   \end{figure*}

\begin{figure*}[h!]
   \centering
    \includegraphics[width=.97\linewidth]{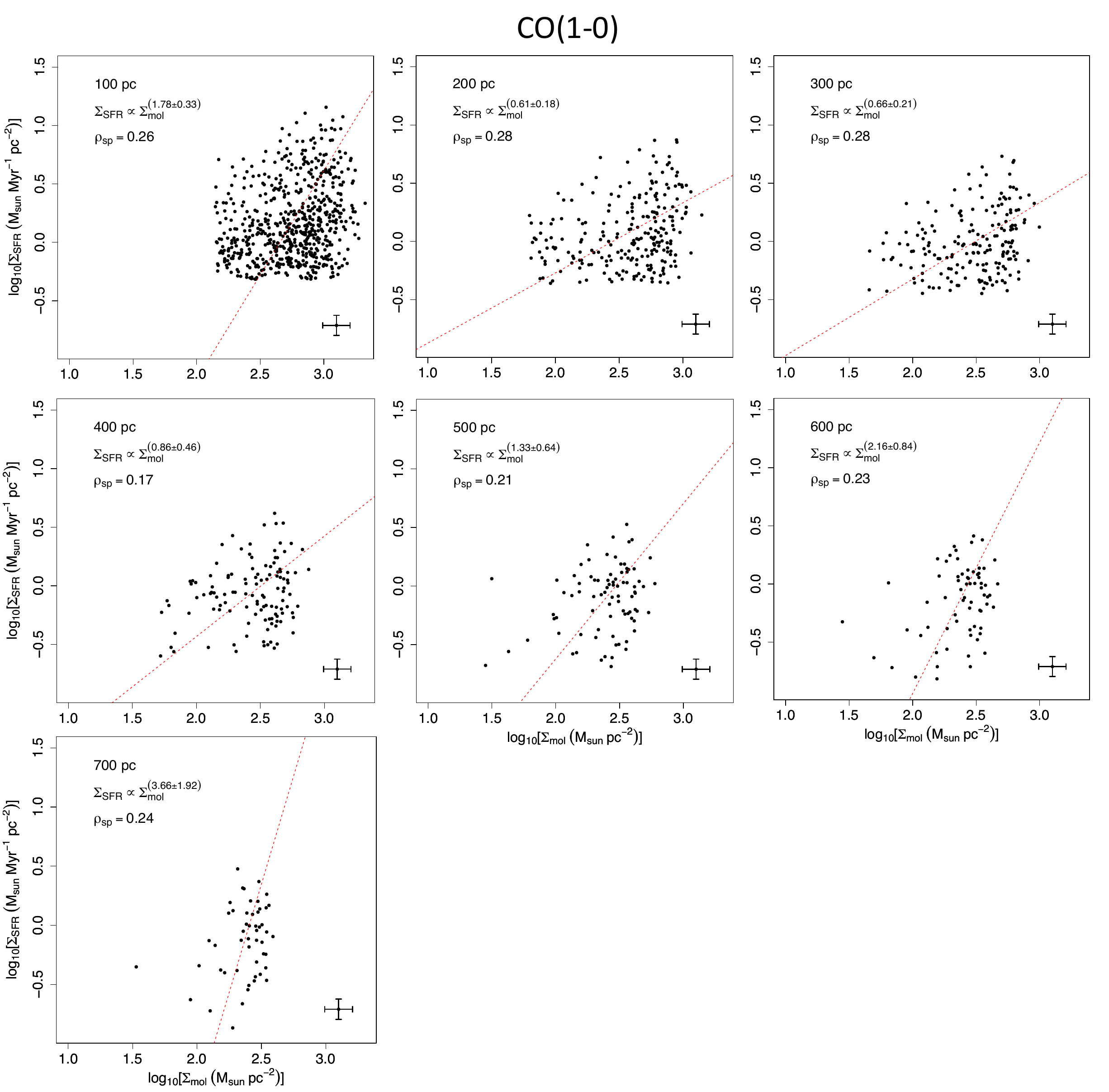}
    \caption{Same as Fig. \ref{Fig_5}, but showing all the spatial scales used in this work for CO(1-0). Vertical and horizontal errorbars at the lower right corner of each panel account for the typical uncertainties, which amount to $\pm$0.09 dex and $\pm$0.10 dex, respectively.}
              \label{Fig_32}
   \end{figure*}

We represent the Kennicutt-Schmidt plots derived from the different tracers for the range of scales used in this work (Figures \ref{Fig_29},  \ref{Fig_30}, \ref{Fig_31} and \ref{Fig_32}).

%
%
%
%
%


\section{Alternative formulation of the intensity-weighted average of the $b$ parameter} \label{AppC}

\begin{figure*}[h!]
   \centering
    	\includegraphics[width=.61\linewidth]{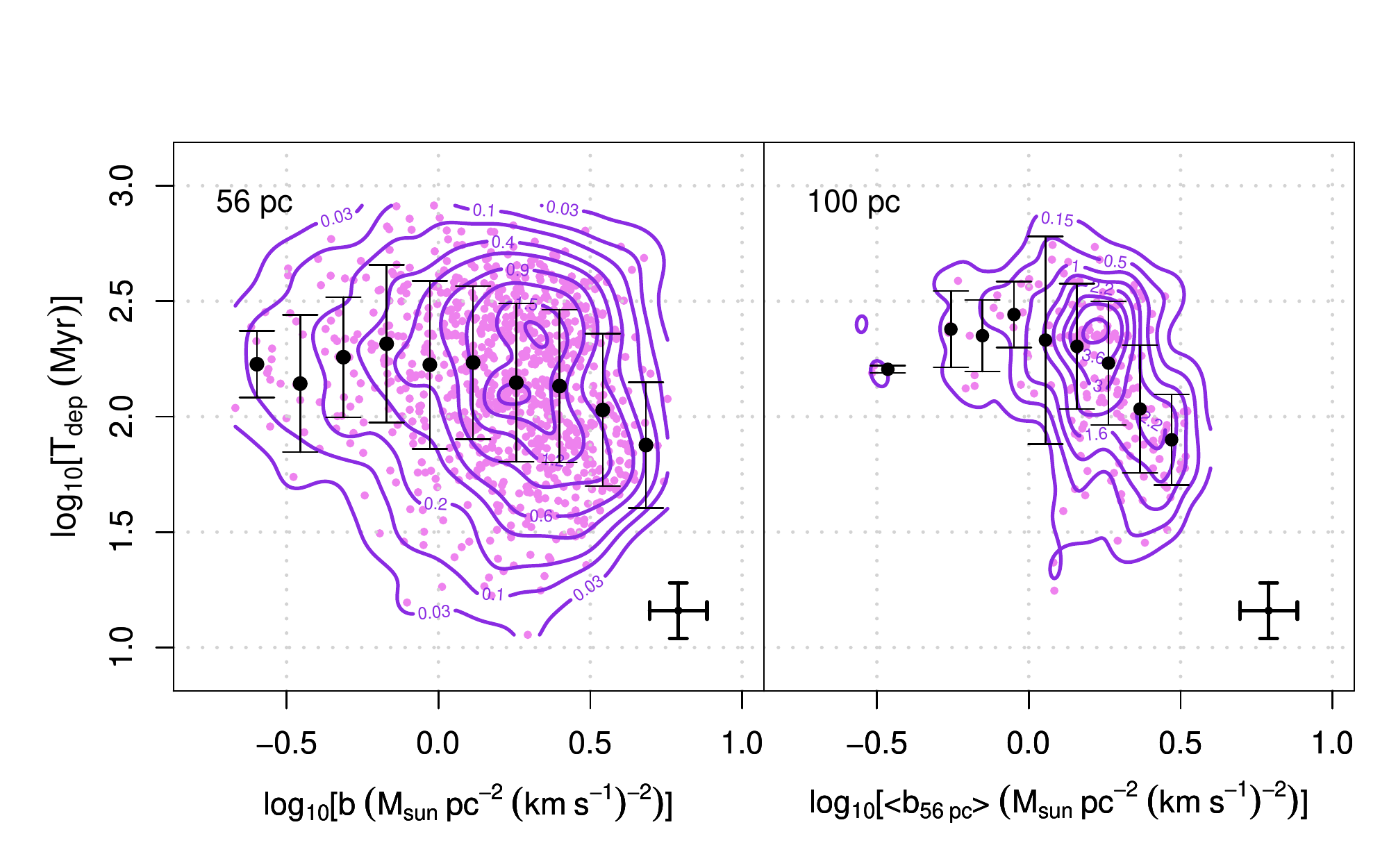}
    	\includegraphics[width=.38\linewidth]{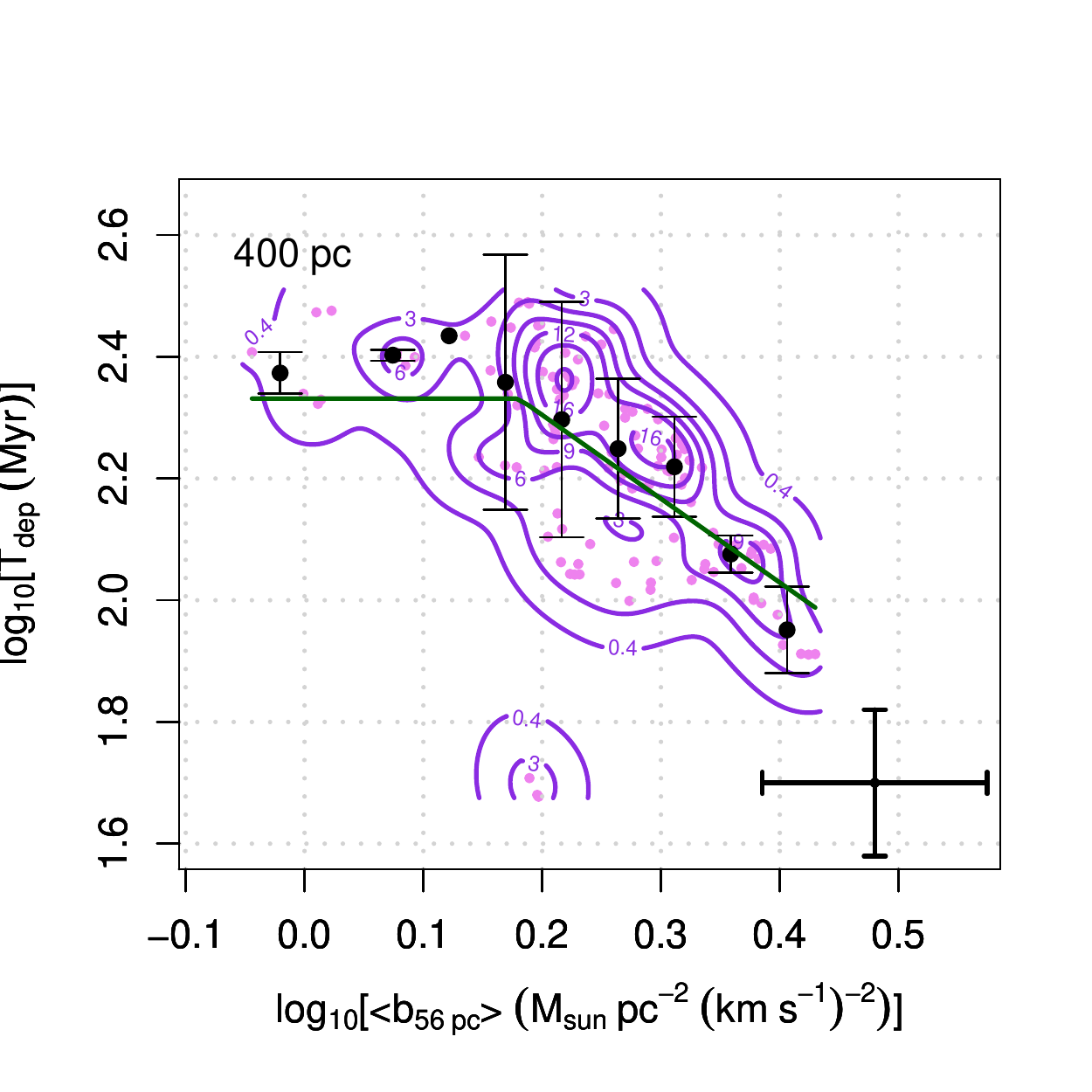}
   \caption{Same as Fig.~\ref{Fig_11} but derived using the alternative formulation of the intensity-weighted average of the $b$ parameter used by \citet{Leroy2017}.}
              \label{Fig_27}
   \end{figure*}

\citet{Leroy2017} used a definition of the intensity-weighted average of the $b$ parameter different to the one used in Sect~\ref{Tdep-b}. In particular, instead of deriving the averages of $b$ as defined in 
Eq.~\ref{Eq8}, they  performed  the averages of $\Sigma_{\rm gas}$ and $\sigma^2$ separately and defined : $\langle b \rangle \equiv \langle \Sigma_{\rm gas} \rangle /  \langle \sigma^{2} \rangle$. We have derived a new version of Fig.~\ref{Fig_11} using the definition of \citet{Leroy2017} particularized for $\Sigma_{\rm dense}$ and our set of two apertures (100~pc and 400~pc).  The results obtained with this definition, shown in Fig.~\ref{Fig_27} are virtually identical to those shown in Fig.~\ref{Fig_11}. In particular,  we obtain (anti) correlation Spearman rank parameters $\rho_{\rm sp}=-0.42$ and $-0.60$ for  $\Delta A=100$~pc and $400$~pc, respectively, and associated two-sided $p$-values $<1\%$.
Furthermore, after applying the {\tt MARS} routine to the 400~pc-scale scatter-plot we find a turnover  at log$_{\rm10}$($\langle b \rangle$)$=0.18~M_{\sun}$pc$^{-2}$(km s$^{-1}$)$^{-2}$ and  a slope $\simeq-1.4$ for $\langle b \rangle$ values beyond the turnover.


\section{Trends as a function of the boundedness of the gas from HCO$^{+}$} \label{AppE}

\begin{figure*}[h!]
   \centering
    	\includegraphics[width=.61\linewidth]{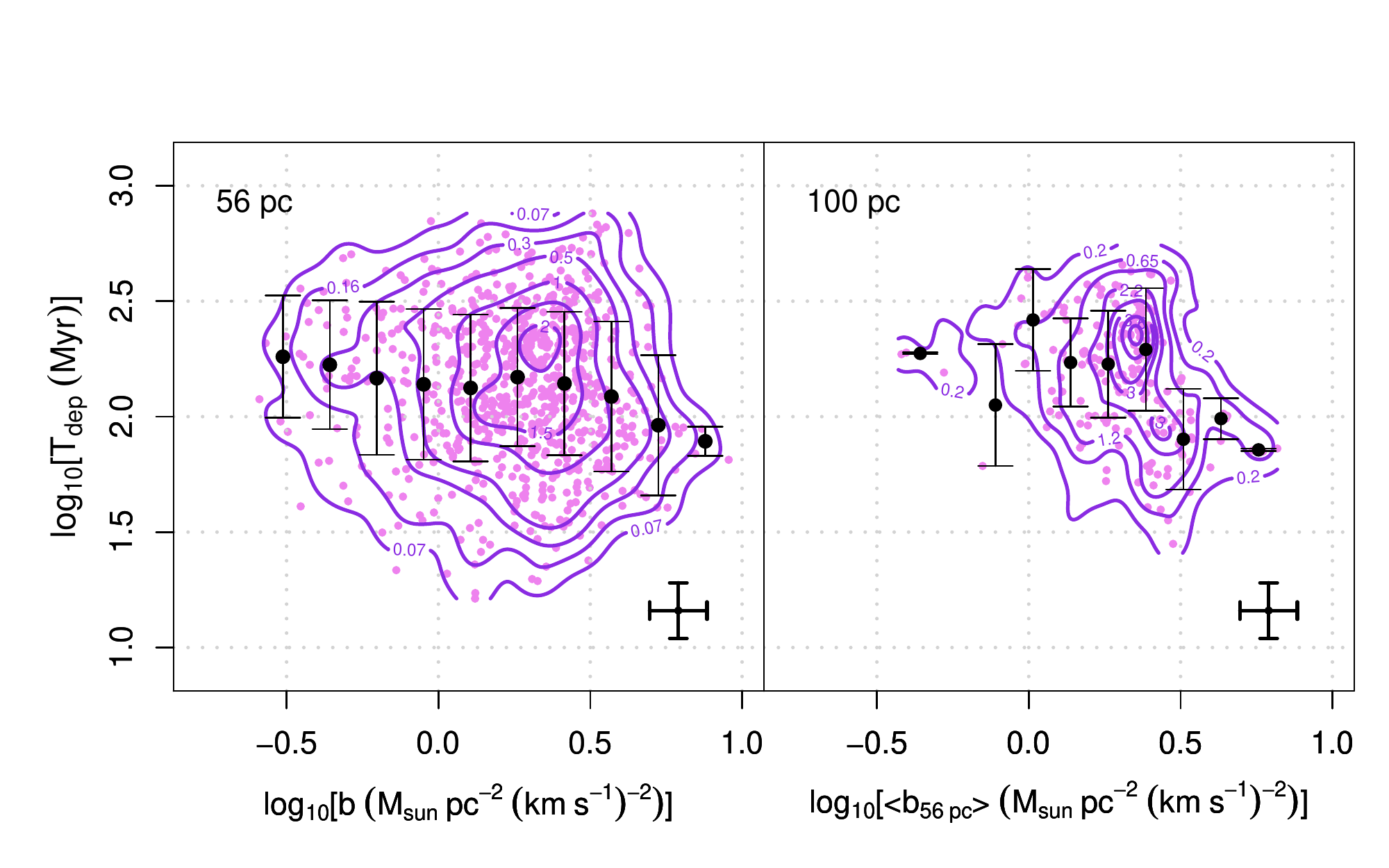}
    	\includegraphics[width=.38\linewidth]{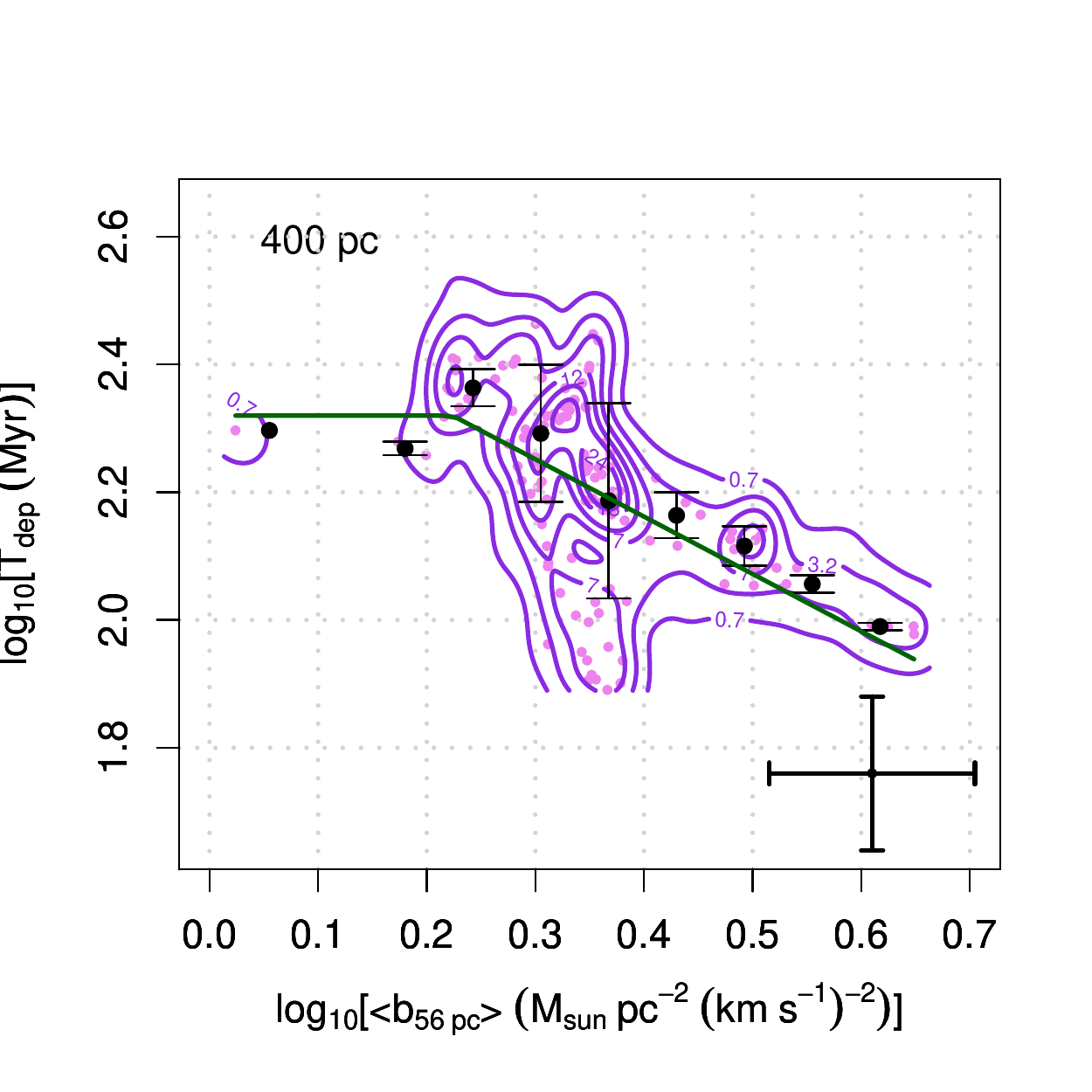}
   \caption{Same as Fig.~\ref{Fig_11} but estimated from HCO$^{+}$.}
              \label{Fig_33}
   \end{figure*}

We have studied in Sect. \ref{Tdep-b} how the depletion time of the dense molecular gas estimated from HCN changes as a function of the self-gravity of the gas ($b$ $paramater$). We now analyse this with the other dense gas tracer that we use in this work, HCO$^{+}$. Figure \ref{Fig_33} represents the depletion time as a function of the $b$ parameter at 56~pc, 100~pc and 400~pc. We observe a similar trend to Figure \ref{Fig_11}, obtaining (anti) correlation Spearman rank paramaters $\rho_{sp}$ = -0.07 (p-value = 0.06), -0.32 (<1$\%$) and -0.65 (<1$\%$) for 56~pc, 100~pc and 400~pc, respectively. At 400~pc scales, we identify a turnover in the plot located around log$_{10}$($\langle b \rangle$) $\simeq$ 0.2 $M_{\sun}$pc$^{-2}$(km s$^{-1}$)$^{-2}$ and a slope $\simeq$ -0.91 for $\langle b \rangle$ values beyond the turnover.

Figure \ref{Fig_33} shows the spatial distribution of $T_{dep}^{dense}$ and $\langle b \rangle$ derived for $\Delta A$ = 400~pc in the SB ring of NGC\,1068 from HCO$^{+}$. We observe that higher (lower) SFE$_{dense}$ ($T_{dep}^{dense}$) values are located closer to the bar-ring interface. We find that the HCN and HCO$^{+}$ show similar correlations in the SB ring of NGC\,1068.

\begin{figure*}[h!]
   \centering
    	\includegraphics[width=.49\linewidth]{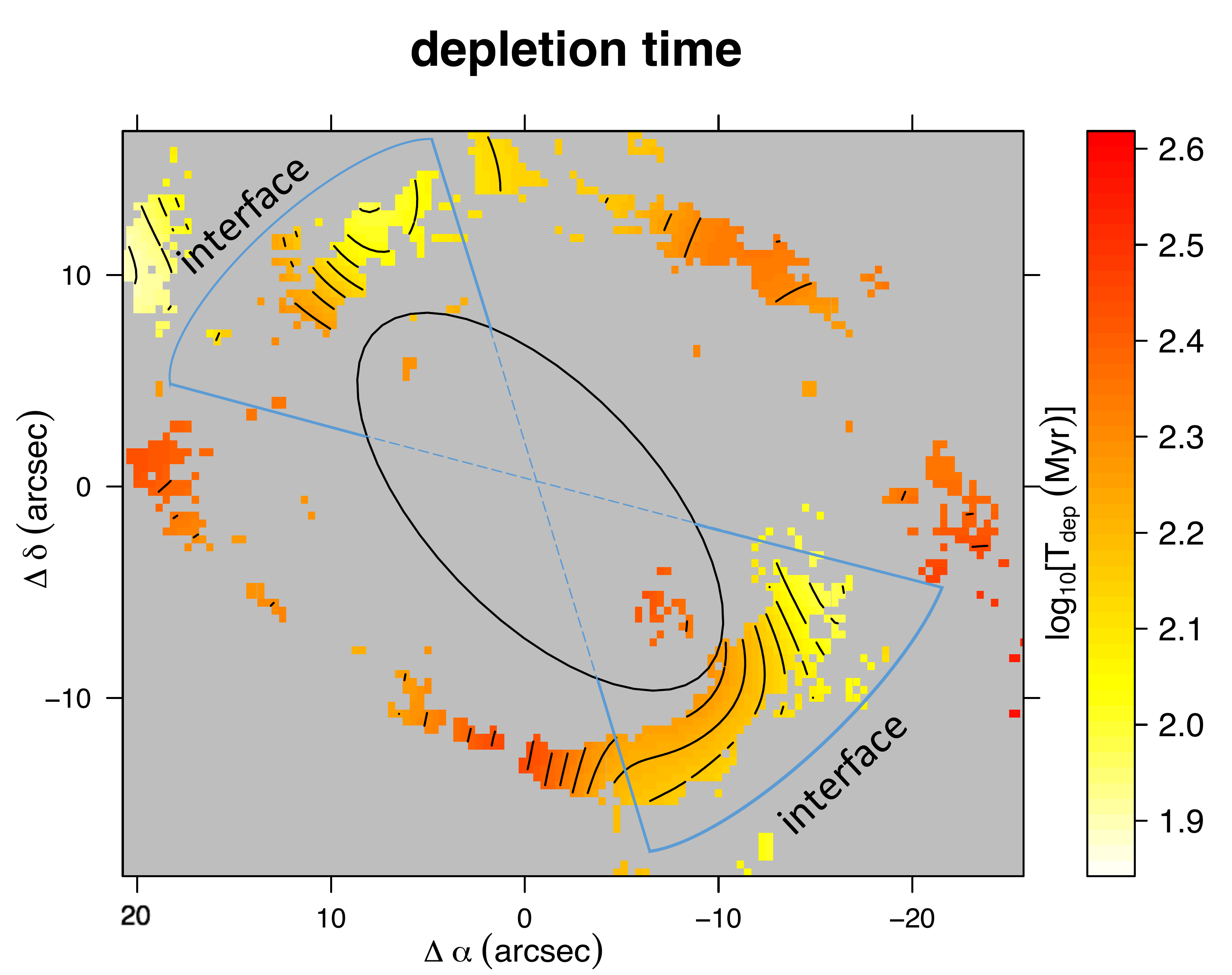}
    	\includegraphics[width=.49\linewidth]{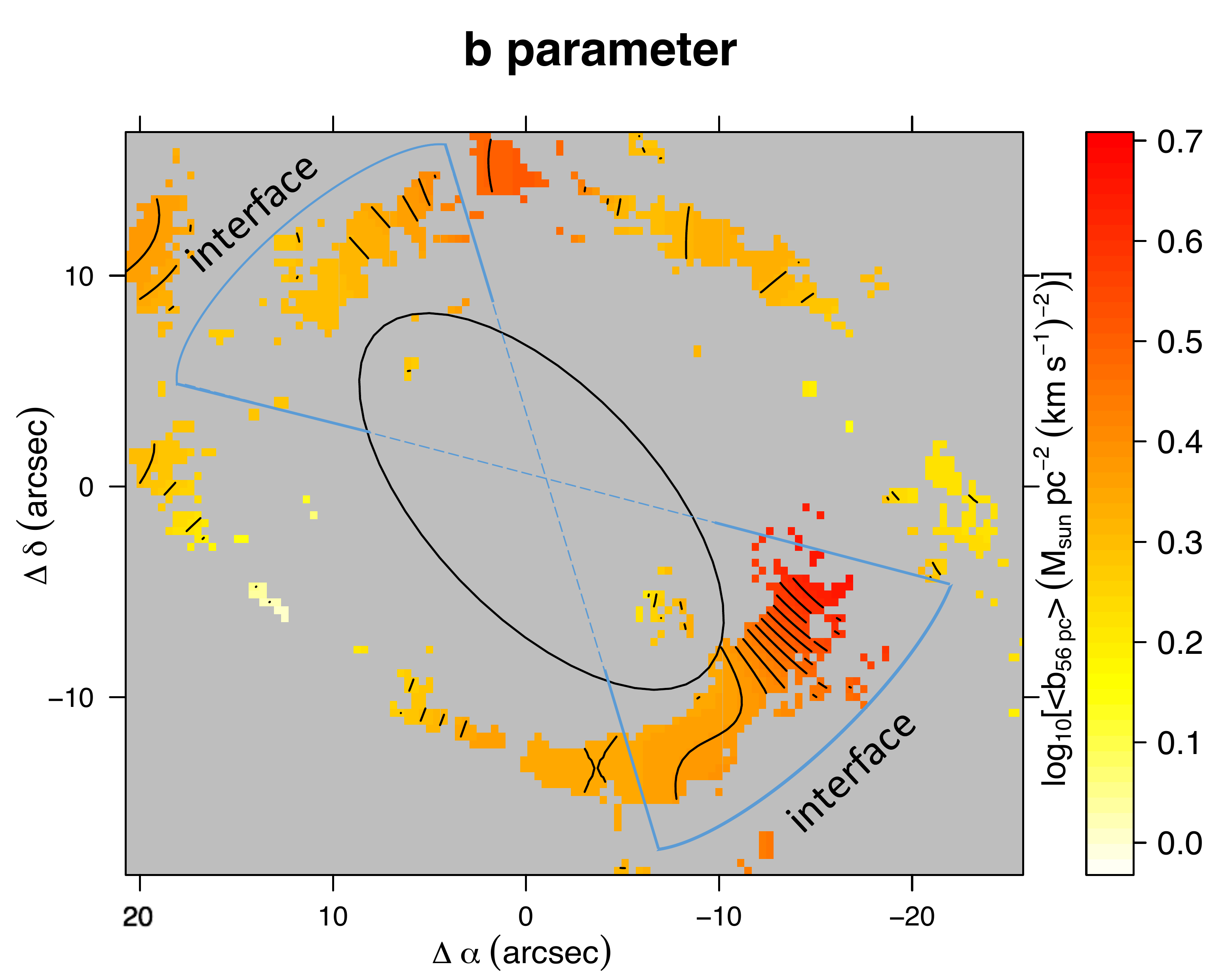}
   \caption{Same as Fig.~\ref{Fig_12} but estimated from HCO$^{+}$. $Left$ $panel$: contour levels for log$_{10}$($T_{dep}^{dense}$) go from 1.90 to 2.50 in steps of 0.045 (in Myr). $Right$ $panel$: contour levels for log$_{10}$($\langle b \rangle$) go from 0.02 to 0.64 to 0.025 (in M$_{\odot}$pc$^{-2}$(km~s$^{-1}$)$^{-2}$).}
              \label{Fig_34}
   \end{figure*}


\end{appendix}


\end{document}